\documentclass[fleqn,usenatbib]{mnras}

\usepackage{amsfonts}
\usepackage[T1]{fontenc}
\usepackage{ae,aecompl}
\usepackage[position=top]{subfig}
\usepackage{float}
\usepackage{comment}
\usepackage{xcolor}
\usepackage{grffile}% A pacakge which changes the algorithm to check for known extensions, so that we can insert pdf figures into this text.
\usepackage{etoolbox}
\makeatletter % Workaround to only color the year for cite commands
  \patchcmd{\NAT@citex}
    {\@citea\NAT@hyper@{%
      \NAT@nmfmt{\NAT@nm}%
      \hyper@natlinkbreak{\NAT@aysep\NAT@spacechar}{\@citeb\@extra@b@citeb}%
      \NAT@date}}
    {\@citea\NAT@nmfmt{\NAT@nm}%
    \NAT@aysep\NAT@spacechar\NAT@hyper@{\NAT@date}}{}{}

  \patchcmd{\NAT@citex}
    {\@citea\NAT@hyper@{%
      \NAT@nmfmt{\NAT@nm}%
      \hyper@natlinkbreak{\NAT@spacechar\NAT@@open\if*#1*\else#1\NAT@spacechar\fi}%
        {\@citeb\@extra@b@citeb}%
      \NAT@date}}
    {\@citea\NAT@nmfmt{\NAT@nm}%
    \NAT@spacechar\NAT@@open\if*#1*\else#1\NAT@spacechar\fi\NAT@hyper@{\NAT@date}}
    {}{}
\makeatother

\usepackage{etoolbox}
 \makeatother
\usepackage{graphicx}	% Including figure files
\usepackage{amsmath}	% Advanced maths commands
\usepackage{amssymb}	% Extra maths symbols
\usepackage{pifont}
\newcommand{\cmark}{\ding{51}}%
\newcommand{\xmark}{\ding{55}}%
\newcommand\HII{{H\,\textsc{ii}}} % ionized hydrogen
\title[First star formation with PBHs]{Effects of stellar-mass primordial black holes on first star formation}

\author[B. Liu, S. Zhang \& V. Bromm]{Boyuan Liu\textsuperscript{\href{https://orcid.org/0000-0002-4966-7450}{\includegraphics[width=2.5mm]{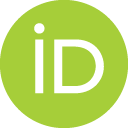}}\,}\thanks{E-mail: boyuan@utexas.edu}$^{1}$, 
Saiyang Zhang\textsuperscript{\href{https://orcid.org/0000-0003-1541-177X}{\includegraphics[width=2.5mm]{orcid.png}}\,}$^{2}$ and Volker Bromm\textsuperscript{\href{https://orcid.org/0000-0003-0212-2979}{\includegraphics[width=2.5mm]{orcid.png}}}$^{1}$
\\
$^{1}$Department of Astronomy, University of Texas, Austin, TX 78712, USA\\
$^{2}$Department of Physics, University of Texas, Austin, TX 78712, USA\\ 
}

\date{Accepted XXX. Received YYY; in original form ZZZ}

\pubyear{2022}

\begin{document}
\label{firstpage}
\pagerange{\pageref{firstpage}--\pageref{lastpage}}
\maketitle

\begin{abstract}
We use cosmological hydrodynamic zoom-in simulations and semi-analytical models to study the effects of primordial black holes (PBHs) on first star formation. Our models self-consistently combine two competing effects: initial (isocurvature) perturbations induced by PBHs and BH accretion feedback. %The former accelerates structure formation, while the latter increases the halo mass threshold above which stars can form by efficient cooling. 
Focusing on PBHs with masses $\sim 30\ \rm M_{\odot}$, we find that the standard picture of first star formation in molecular-cooling minihaloes is not changed by PBHs, as the simulated star-forming gas clouds in the central parsec are very similar to those in the $\rm \Lambda CDM$ case when PBHs make up $f_{\rm PBH}\sim 10^{-4}-0.1$ of dark matter. With a dynamical friction timescale of $\sim 2-10\ \rm Myr$ when the central gas density reaches $10^{5}\ \rm cm^{-3}$, it is also unlikely that PBHs can sink into star-forming discs and affect the evolution of protostars, although they may interact with the stars during the main-sequence stage. At larger scales, PBHs tend to shift star formation to more massive haloes, and accelerate structure formation. The latter effect is stronger in regions with higher initial overdensities. %In the simulated over-dense regions, star formation is accelerated for $f_{\rm PBH}\lesssim 10^{-3}$, but delayed for $f_{\rm PBH}\gtrsim 0.01$. 
For $f_{\rm PBH}\sim 10^{-4}-0.01$ (allowed by observational constraints), the collapsed mass fraction of haloes hosting Population~III stars is similar (within a factor of $\sim2$ at $z\lesssim 30$) to that in $\rm \Lambda CDM$, implying that the impact of stellar-mass PBHs on the cosmic star formation history at $z\gtrsim 10$ is small. We also find that the Lyman-Werner photons from PBH accretion in atomic-cooling haloes may facilitate the formation of direct-collapse BHs.
\end{abstract}
\begin{keywords}
early universe -- dark ages, reionization, first stars -- black hole physics -- dark matter
\end{keywords}

%%%%%%%%%%%%%%%%%%%%%%%%%%%%%%%%%%%%%%%%%%%%%%%%%%

%%%%%%%%%%%%%%%%% BODY OF PAPER %%%%%%%%%%%%%%%%%%

\section{Introduction}
The detection of gravitational waves (GWs) from mergers of black holes (BHs) has triggered renewed interest in primordial black holes (PBHs, \citealt{Zeldovich1966,Hawking1971,Carr1974}), particularly those with masses of $m_{\rm PBH}\sim 10-100\ \rm M_{\odot}$, as a bona fide dark matter (DM) candidate \citep{Bird2016,Sasaki2016,Clesse2017,DeLuca2020pbh,DeLuca2021}. However, it is inferred from recent GW data \citep{abbott2020population} that such stellar-mass PBHs can only make up a small (mass) fraction ($f_{\rm PBH}\sim 0.2-1\%$) of DM \citep{Ali-Haimoud2017,Raidal2017,Hutsi2021,Wong2021}, and that the entire population of observed BH mergers is unlikely to be explained by PBHs \citep{Hall2020}.

Formation of PBHs in the very early Universe is well motivated theoretically from a variety of mechanisms {(see e.g. \citealt{Khlopov2010})}, such as collapse of adiabatic (curvature) density perturbations \citep{Escriva2022} and collapse of domain walls formed by quantum fluctuations of a scalar field during inflation \citep{Belotsky2019}. Moreover, even if PBHs do not constitute the entire dark sector, their existence has interesting implications for a broad range of astrophysical phenomena across cosmic history (reviewed in e.g. {\citealt{Belotsky2014,Sasaki2018,Carr2020,Carr2021}}), and is implicated by a variety of observations {\citep{Clesse2018,Hawkins2022}}. For instance, in the local Universe, accretion onto PBHs can heat the interstellar medium (ISM) in dwarf galaxies, whose properties can be used to constrain PBH parameters \citep{Takhistov2022,Lu2021}. Besides, the dynamical heating of (particle) cold DM (CDM) by PBHs can induce a cusp-to-core transition in the DM density profile, providing a solution to the cusp-core problem \citep{Boldrini2020}. 

%At higher redshifts, particularly d
During the Cosmic Dark Age ($z\gtrsim 30$) and Cosmic Dawn ($z\sim 5-30$), the PBH accretion produces various cosmic radiation backgrounds \citep{Hasinger2020} that can alter the thermal and chemical history of the intergalactic medium (IGM). As a result, the abundance of PBHs with $m_{\rm PBH}\gtrsim 1\ \rm M_{\odot}$ are constrained by the cosmic microwave background (CMB, e.g. \citealt{Poulin2017}) and the 21-cm signal from neutral hydrogen (e.g. \citealt{Ricotti2008,Bernal2018,Hektor2018,Mena2019,Yang2021}). Beyond the evolving radiation backgrounds, PBHs also affect cosmic structure formation via the `seed' and `Poisson' effects \citep{Carr2018}. The former dominates when the abundance of PBHs is very small ($f_{\rm PBH}\rightarrow 0$), such that nonlinear, bound DM structures form around individual PBHs, which hardly interact with each other. The latter effect acts in the opposite limit of $f_{\rm PBH}\rightarrow 1$, where the large-scale behavior is still dominated by the adiabatic mode as in the standard CDM model, but a discreteness noise is introduced at small scales. 
Such effects have been evaluated with \textit{semi-analytical} models in previous studies (e.g. {\citealt{Kashlinsky1983,Kashlinsky2016,Gong2017,Cappelluti2022}}), showing that minihaloes with masses $M_{\rm h}\sim 10^{5}-10^{6}\ \rm M_{\odot}$, capable of hosting Population~III (Pop~III) stars within the standard picture of first star formation (see e.g. \citealt{bromm2013}), collapse earlier due to the enhancement of small-scale density perturbations from PBHs. The radiation from stars and accretion discs around PBHs in such star-forming haloes can explain the observed cosmic infrared and X-ray backgrounds, as well as the correlation between them (e.g. {\citealt{Kashlinsky2012,Mitchell-Wynne2016,Kashlinsky2018,Li2018}}).

However, the semi-analytical approach is only accurate in the two limits, while the interplay between the `seed' and `Poisson' effects can be complex in the intermediate cases with $f_{\rm PBH}\sim 10^{-4}-0.1$, which happens to be the most promising range for stellar-mass PBHs ($m_{\rm PBH}\sim 10-100\ \rm M_{\odot}$) according to current observational constraints (see e.g. \citealp{Poulin2017,Hektor2018,Carr2021,Hutsi2021}). In this regime, the DM structures around individual PBHs do interact with each other, forming larger structures, and the nonlinear dynamics can only be captured with ($N$-body) simulations \citep{Inman2019}. It is shown by \citet{Inman2019} with (DM-only) simulations (at $z\ge 99$) for $m_{\rm PBH}\sim 30\ \rm M_{\odot}$ that there is significant clustering of PBHs in large haloes for $f_{\rm PBH}\gtrsim z\times 10^{-4}$, and that PBHs also affect the internal structure of DM haloes. Furthermore, in addition to the enhancement of density perturbations, the feedback from PBH accretion can change the thermodynamics and chemistry of gas in minihaloes, which may delay or even prevent star formation with efficient heating of gas, similar to the scenario in dwarf galaxies at low $z$ \citep{Takhistov2022,Lu2021}. Previous studies have not looked into PBH accretion feedback in the nonlinear regime quantitatively. Therefore, how exactly PBHs impact first star formation is still unclear.

In this paper, we use cosmological hydrodynamic (zoom-in) simulations to study the effects of stellar-mass PBHs ($m_{\rm PBH}\sim 10-100\ \rm M_{\odot}$) on first star formation, which take into account both the enhancement of initial density perturbations and accretion feedback from PBHs for the first time. Assuming a monochromatic\footnote{The mass spectrum of PBHs can also be broad or contain multiple spikes, depending on the formation mechanism (see e.g. \citealt{Carr2018,Tada2019,Carr2019,Carr2021qcd}). As we are mostly concerned with stellar-mass PBHs in a relatively narrow mass range ($m_{\rm PBH}\sim 10-100\ \rm M_{\odot}$), we adopt a monochromatic mass function for simplicity. {Note that constraints on PBH abundance for monochromatic mass distributions can also be converted to those for extended mass distributions \citep{Bellomo2018}.}} mass function, our fiducial PBH model has a PBH mass $m_{\rm PBH}=33\ \rm M_{\odot}$ and a fraction of PBHs in the dark sector $f_{\rm PBH}=10^{-3}$. We choose this particular mass because it is the location of the Gaussian peak in the best-fit Power-Law + Peak model, expressing the mass distribution of BHs detected by the LIGO-Virgo-KAGRA Scientific Collaboration \citep{abbott2020population}. The fraction $f_{\rm PBH}=10^{-3}$ is approximately the highest value allowed by (the tightest) constraints from GWs, 21-cm cosmology, and CMB observations  (\citealp{Poulin2017,Hektor2018,Carr2021,Hutsi2021}). Considering the uncertainties in these constraints and for the sake of theoretical exploration, we further consider 
%two cases with $f_{\rm PBH}=0.01$ and $10^{-4}$. 
three cases with $f_{\rm PBH}=0.1$, 0.01 and $10^{-4}$. 
To evaluate the effect of PBH mass, we also investigate the case with $m_{\rm PBH}=100\ \rm M_{\odot}$ and $f_{\rm PBH}=10^{-3}$. 

In Sec.~\ref{s2}, we describe the numerical details of our simulations, including the zoom-in setup, initial conditions and BH physics. The simulation results are presented in Sec.~\ref{s3}. In Sec.~\ref{s4}, we provide interpretations of the simulation results with semi-analytical arguments, and discuss the possible roles played by PBHs in later stages of halo evolution, beyond the period covered by our simulations. Finally, we summarize our main findings and discuss potential caveats, as well as directions for future work in Sec.~\ref{s5}.

%PBH and GWs \citep{Bird2016,Sasaki2016,Clesse2017,DeLuca2021}, 
%review \citep{Sasaki2018}, constraints \citep{Ali-Haimoud2017,Raidal2017,Hutsi2021,Wong2021,Hall2020}, 
%hints for PBH DM \citep{Clesse2018}, 
%PBH formation: collapse of domain walls formed by quantum fluctuations of a scalar field during inflation \citep{Belotsky2019}, collapse of adiabatic (curvature) density perturbations \citep{Escriva2022}. 
%Effects of PBHs \citep{Carr2020} and observational constraints \citep{Carr2021}:
%dynamical heating of the DM halo (cusp-to-core transition) \citep{Boldrini2020},
%radiation backgrounds from PBH accretion \citep{Hasinger2020}, constraints from 21-cm signals for BH accretion , 
%enhancement of structure formation \citep{Carr2018}: abundance of minihaloes, CIB \& CXB, seeds of SMBHs . 

%Clustering of PBHs \citep{Belotsky2019}, \citep{Desjacques2018,Bringmann2019,DeLuca2020}

\begin{table*}
    \centering
    \caption{Simulation parameters and key properties (see Sec.~\ref{s3} and Appendices~\ref{aa}-\ref{ac} for the definitions of runs). $\sigma_{8}$ reflects the amplitude of initial (adiabatic) density perturbations in the parent box. $m_{\rm PBH}$ is the (initial) PBH mass (assuming that all PBHs have the same mass initially). $f_{\rm PBH}$ is the mass fraction of PBHs in the dark sector. $\epsilon_{r}$ is the efficiency of radiation-thermal coupling %, which can be written as $\epsilon_{r}\equiv f_{h}f_{\rm abs}$, where $f_{\rm abs}\simeq 0.66$ is the fraction of radiation energy absorbed by the ISM, and $f_{h}\sim 1/3$ is the fraction of energy deposited as heat 
    (see Sec.~\ref{s2.3.3}). 
    PBHP is a flag that shows whether the initial (isocurvature) perturbations of DM (and gas) particles by PBHs are included (\cmark) or not (\xmark), enhanced ($\uparrow$) or suppressed ($\downarrow$). 
    $v_{\rm b\chi}$ is the magnitude of streaming velocity between DM and gas, expressed in units of the root-mean-square streaming velocity (see Sec.~\ref{s2.2}). 
    $z_{\rm col}$ is the redshift at which the maximum hydrogen number density reaches $10^{5}\ \rm cm^{-3}$, $t_{\rm col}$ is the corresponding cosmic age, and $M_{\rm h}$ is the halo (virial) mass at this moment, derived from the \href{https://bitbucket.org/gfcstanford/rockstar/src}{\textsc{rockstar}} halo finder \citep{behroozi2012rockstar}. $N_{\rm PBH}$ is the number of PBHs in the halo (within $R_{\rm vir}$). $T_{\rm IGM}$ is the IGM temperature at $z_{\rm col}$, estimated with the volume-weighted temperature of gas in the density range of $\rho_{\rm gas}\sim 0.1-10\bar{\rho}_{\rm gas}$, given $\bar{\rho}_{\rm gas}$ as the cosmic average gas density. The value in the bracket is the reference IGM temperature in the standard CDM cosmology from the fitting formula in \citet{Tseliakhovich2010}. }
    \begin{tabular}{ccccccccccccc}
    \hline
        Run & $\sigma_{8}$ & $m_{\rm PBH}\ [\rm M_{\odot}]$ & $f_{\rm PBH}$ & $\epsilon_{r}$ & PBHP & $v_{\rm b\chi}/\sigma_{\rm b\chi}$ & $z_{\rm col}$ & $t_{\rm col}\ [\rm Myr]$ & $M_{\rm h}\ [\rm M_{\odot}]$ & $N_{\rm PBH}$ & $T_{\rm IGM}\ [\rm K]$ & \\
    \hline
        \texttt{CDM\_A} & 2.0 & - & - & - & - & 0 & 30.3 & 97.9 & $2\times 10^{5}$ & - & 45.1 (19.7) \\%& \cmark\\
        %\texttt{CDM\_A\_ST} & 2.0 & - & - & - & 0.8 & - & ? & ? \\
        \texttt{PBH3\_A} & 2.0 & 33 & $10^{-3}$ & 0.22 & \cmark & 0 & 37.6 & 71.0 & $4.3\times 10^{5}$ & 19 & 37.5 (29.4) \\
        \texttt{PBH3\_NP\_A} & 2.0 & 33 & $10^{-3}$ & 0.22 & \xmark & 0 & 28.2 & 108.6 & $3.7\times 10^{5}$ & 14 & 64.8 (17.2) \\
        \texttt{PBH3\_NF\_A} & 2.0 & 33 & $10^{-3}$ & 0 & \cmark & 0 & 42.7 & 58.9 & $1.3\times 10^{5}$ & 9 & 37.9 (37.1) \\
        \texttt{PBH1\_A} & 2.0 & 33 & $10^{-1}$ & 0.22 & \cmark & 0 & 22.5 & 150.2 & $9.2\times 10^{5}$ & 3195 & 260.3 (11.4) \\
        \texttt{PBH2\_A} & 2.0 & 33 & $10^{-2}$ & 0.22 & \cmark & 0 & 27.0 & 115.5 & $7.5\times 10^{5}$ & 254 & 104.1 (15.9) \\
        \texttt{PBH4\_A} & 2.0 & 33 & $10^{-4}$ & 0.22 & \cmark & 0 & 31.6 & 91.8 & $1.6\times 10^{5}$ & 0 & 39.3 (21.3) \\
        \texttt{PBH3\_M100\_A} & 2.0 & 100 & $10^{-3}$ & 0.22 & \cmark & 0 & 42.4 & 59.6 & $2.5\times 10^{5}$ & 5 & 44.0 (36.6) \\
    \hline 
        \texttt{CDM\_ST\_A} & 2.0 & - & - & - & - & 0.8 & 26.9 & 116.3 & $4.4\times 10^{5}$ & - & 85.8 (15.8)\\
        \texttt{PBH3\_ST\_A} & 2.0 & 33 & $10^{-3}$ & 0.22 & \cmark & 0.8 & 34.8 & 79.6 & $5.8\times 10^{5}$ & 25 & 41.3 (25.5) \\
        \texttt{PBH3\_WF\_A} & 2.0 & 33 & $10^{-3}$ & 0.02 & \cmark & 0 & 39.8 & 65.5 & $2.4\times 10^{5}$ & 13 & 36.3 (32.6) \\
        \texttt{PBH3\_SF\_A} & 2.0 & 33 & $10^{-3}$ & 1 & \cmark & 0 & 37.0 & 72.7 & $4.7\times 10^{5}$ & 20 & 40.3 (28.6) \\
        \texttt{PBH3\_WP\_A} & 2.0 & 33 & $10^{-3}$ & 0.22 & \cmark$\downarrow$ & 0 & 30.4 & 97.1 & $4.0\times 10^{5}$ & 17 & 39.5 (36.2) \\
        \texttt{PBH3\_SP\_A} & 2.0 & 33 & $10^{-3}$ & 0.22 & \cmark$\uparrow$ & 0 & 42.1 & 60.2 & $3.1\times 10^{5}$ & 14 & 50.6 (19.8) \\
    \hline
        \texttt{CDM\_B} & 1.6 & - & - & - & - & 0 & 22.3 & 152.7 & $3.8\times 10^{5}$ & - & 34.5 (11.2) \\%& \cmark\\
        %\texttt{CDM\_B\_ST} & 1.6 & - & - & - & 0.8 & - & ? & ? \\
        \texttt{PBH3\_B} & 1.6 & 33 & $10^{-3}$ & 0.22 & \cmark & 0 & 23.6 & 140.6 & $2.7\times 10^{5}$ & 13 & 36.6 (12.4)\\
        \texttt{PBH3\_NP\_B} & 1.6 & 33 & $10^{-3}$ & 0.22 & \xmark & 0 & 21.6 & 160.2 & $4.5\times 10^{5}$ & 9 & 41.5 (10.5)\\
        \texttt{PBH3\_NF\_B} & 1.6 & 33 & $10^{-3}$ & 0 & \cmark & 0 & 29.4 & 102.1 & $1.2\times 10^{5}$ & 3 & 20.9 (18.6) \\
        \texttt{PBH1\_B} & 1.6 & 33 & $10^{-1}$ & 0.22 & \cmark & 0 & 16.3 & 239.2 & $1.7\times 10^{6}$ & 5628 & 224.1 (6.3) \\
        \texttt{PBH2\_B} & 1.6 & 33 & $10^{-2}$ & 0.22 & \cmark & 0 & 18.7 & 196.5 & $5.9\times 10^{5}$ & 209 & 122.5 (8.1) \\
        \texttt{PBH4\_B} & 1.6 & 33 & $10^{-4}$ & 0.22 & \cmark & 0 & 23.8 & 138.6 & $2.1\times 10^{5}$ & 1 & 29.0 (12.6) \\
        \texttt{PBH3\_M100\_B} & 1.6 & 100 & $10^{-3}$ & 0.22 & \cmark & 0 & 27.6 & 111.7 & $2.6\times 10^{5}$ & 3 & 31.3 (16.6) \\
    \hline
        \texttt{PBH3\_WP\_B} & 1.6 & 33 & $10^{-3}$ & 0.22 & \cmark$\downarrow$ & 0 & 22.1 & 154.5 & $4.3\times 10^{5}$ & 11 & 41.1 (11.0) \\
        \texttt{PBH3\_SP\_B} & 1.6 & 33 & $10^{-3}$ & 0.22 & \cmark$\uparrow$ & 0 & 30.6 & 96.3 & $4.9\times 10^{5}$ & 11 & 23.4 (20.1) \\
    \hline
    \end{tabular}
    \label{t1}
\end{table*}

\section{Methodology}
\label{s2}
In total we run 24 simulations combining different PBH parameters, implementations of PBH physics and initial conditions, whose key characteristics are given in Table~\ref{t1} and further explained below. 
Our cosmological hydrodynamic simulations are conducted with the \textsc{gizmo} code \citep{hopkins2015new} that uses the Lagrangian meshless finite-mass (MFM) hydro solver (with a number of neighbours $N_{\mathrm{ngb}}=32$), combined with the parallelization scheme and Tree+PM gravity solver from \textsc{gadget-3} \citep{springel2005cosmological}. The hydro and gravity solvers are coupled with a non-equilibrium primordial chemistry and cooling network for 12 species ($\rm H$, $\rm H^{+}$, $\rm H^{-}$, $\rm H_{2}$, $\rm H_{2}^{+}$, $\rm He$, $\rm He^{+}$, $\rm He^{2+}$, $\rm D$, $\rm D^{+}$, $\rm HD$, $\rm e^{-}$) detailed in \citet{bromm2002,johnson2006}. %In addition to the original network, we further consider 4 reactions\footnote{$\rm H^{-}+h\nu \rightarrow H+e^{-}$ \citep[taking into account the contribution of non-thermal photons]{Coppola2011}, $\rm H_{2}^{+}+h\nu \rightarrow H+H^{+}$, $\rm H_{2}^{+}+h\nu \rightarrow 2H^{+}+e^{-}$, $\rm H_{2}+h\nu \rightarrow H_{2}^{+}+e^{-}$ \citep{galli1998}.} between $\rm H^{-}$, $\rm H_{2}^{+}$, $\rm H_{2}$ and CMB photons from \citet{galli1998,Coppola2011}, such that the $\rm H_{2}$ abundance in the IGM predicted by our network is consistent with that in \citet{galli2013dawn} at $z\lesssim 50$. 

To resolve the cold, dense gas clouds (with temperature $T\lesssim 500\ \rm K$ and hydrogen number density $n_{\rm H}\gtrsim 10^{4}\ \rm cm^{-3}$) in minihaloes, the standard formation sites of Pop~III stars, we run two sets of zoom-in simulations targeted at two overdense regions at $z\gtrsim 20$ (see Table~\ref{t1}). %for a list of the zoom-in runs). 
In Sec.~\ref{s2.1}, we summarize the setups of the parent simulations, zoom-in regions and numerical parameters. For each set, we run a reference simulation for the standard $\Lambda$CDM case and modify the initial conditions of this CDM run for PBH models with different parameters, based on linear perturbation theory, which is explained in Sec.~\ref{s2.2}. Beyond the initial conditions regulated by PBHs, we also adopt sub-grid models for BH physics (dynamics, accretion and feedback) to model the effects of PBHs on the thermal and chemical evolution of the ISM, as described in Sec.~\ref{s2.3}.

\begin{comment}
\begin{table}
    \centering
    \caption{Initial abundances at $z_{\rm ini}=300$, defined as the ratio of the number density of a species and the number density of the representative nuclei (see fig.~1 and 2 in \citealt{galli2013dawn}).}
    \begin{tabular}{lllllll}
        \hline
        Species & $\quad\quad\quad\quad\quad\quad\quad\quad$ & Definition  & Abundance \\
        \hline
        $\rm H_{0}$ & & $n_{\rm H_{0}}/n_{\rm H}$ & 0.9995 \\
        $\rm H^{+}$ & & $n_{\rm H^{+}}/n_{\rm H}$ & $5\times 10^{-4}$ \\
        $\rm H^{-}$ & & $n_{\rm H^{-}}/n_{\rm H}$ & $2.5\times 10^{-19}$ \\
        $\rm H_{2}$ & & $n_{\rm H_{2}}/n_{\rm H}$ & $2\times 10^{-11}$ \\
        $\rm H_{2}^{+}$ & & $n_{\rm H_{2}^{+}}/n_{\rm H}$ & $3\times 10^{-16}$ \\
        $\rm e^{-}$ & & $n_{\rm e^{-}}/n_{\rm H}$ & $5\times 10^{-4}$ \\
        $\rm He_{0}$ & & $n_{\rm He_{0}}/n_{\rm He}$ & 1.0 \\
        $\rm He^{+}$ & & $n_{\rm He^{+}}/n_{\rm He}$ & $1.4\times 10^{-20}$ \\
        $\rm He^{++}$ & & $n_{\rm He^{++}}/n_{\rm He}$ & 0 \\
        $\rm D_{0}$ & & $n_{\rm D_{0}}/n_{\rm D}$ & 0.9995 \\
        $\rm D^{+}$ & & $n_{\rm D^{+}}/n_{\rm D}$ & $5\times 10^{-4}$ \\
        $\rm HD$ & & $n_{\rm HD}/n_{\rm D}$ & $8.4\times 10^{-11}$ \\
        \hline
    \end{tabular}
    \label{t2}
\end{table}
\end{comment}

\subsection{Simulation setups}
\label{s2.1}
We start with two parent simulations with a box size of $L\sim 200\ \rm kpc$ and $2\times 128^3$ particles (including both DM and gas) in $\Lambda$CDM cosmology with parameters: $\Omega_{m}=0.3089$, $\Omega_{b}=0.04864$, $n_{s}=0.96$, and $h=0.6774$ \citep{planck}. To accelerate structure formation in our small box, we enhance the initial density perturbations\footnote{{As shown in \citet{Park2020}, although this approach cannot fully capture the structure formation history in real over-dense regions, Pop~III star formation at the halo scale is not affected. }} by adopting $\sigma_{8}=2$ and 1.6 for Case A and B, respectively, while the cosmological mean is $\sigma_{8}=0.8159$ \citep{planck}. The initial conditions are generated with the \textsc{music} code \citep{hahn2011multi} at an initial redshift of $z_{\rm ini}=300$. %The initial abundances of the 12 elements at $z_{\rm ini}=300$ are obtained from the results in \citet[see their fig.~1 and 2]{galli2013dawn} for chemical evolution of the IGM, as specified in Table~\ref{t2}. 
The mass of a DM (gas) particle in the parent simulations is $\sim 140\ (26)\ \rm M_{\odot}$, such that typical star-forming minihaloes with masses $M_{\rm h}\gtrsim 10^{5}\ \rm M_{\odot}$ at $z\sim 20-30$ are well resolved. We stop the simulation when the densest gas particle has reached $n_{\rm H}\ge 10^{4}\ \rm cm^{-3}$, at which point gas clouds in a few minihaloes have entered the runaway-collapse phase with $n_{\rm H}\gtrsim 10^{2}\ \rm cm^{-3}$ and $T\lesssim 500\ \rm K$. Throughout this study, DM haloes are identified with the \href{https://bitbucket.org/gfcstanford/rockstar/src}{\textsc{rockstar}} halo finder \citep{behroozi2012rockstar}.

We choose one halo with collapsing gas at the final snapshot of each parent simulation. For Case A, the target halo has a virial mass $M_{\rm h}\sim 3.7\times 10^{5}\ \rm M_{\odot}$ and a (physical) virial radius $R_{\rm vir}\sim 80\ \rm pc$ at $z\sim 28$. While for Case B, we pick a halo with $M_{\rm h}\sim 6.1\times 10^{5}\ \rm M_{\odot}$ and $R_{\rm vir}\sim 130\ \rm pc$ at $z\sim 20$. In the zoom-in simulations, we increase the resolution by $\Delta_{\rm res}=2$ levels, i.e. a factor of 4 (64) for length (mass). To avoid contamination of low-resolution particles within the virial radius of the target halo, we define the Lagrangian region as comprising all DM particles within $R_{\rm L}=(1.5\Delta_{\rm res}+1)R_{\rm vir}=4R_{\rm vir}$, according to \citet{Onorbe2014}. These particles are traced back to their initial positions and a rectangular box enclosing all of them is defined as the high-resolution region, where we have mass resolution of $m_{\rm DM}\sim 2\ \rm M_{\odot}$ for DM and $m_{\rm gas}\sim 0.4\ \rm M_{\odot}$ for gas. We adopt a co-moving softening length of $\epsilon_{\rm gas}=\epsilon_{\rm DM}=0.01\ h^{-1}\rm kpc$ for both DM and gas in the high-resolution region. The zoom-in initial conditions for $\Lambda$CDM cosmology are also generated with the \textsc{music} code \citep{hahn2011multi} at $z_{\rm ini}=300$, where the co-moving volumes of the high-resolution regions are $\sim 10^{5}$ and $2\times 10^{5}\ \rm kpc^{3}$ for Case A and B, respectively.

\subsection{Initial conditions with PBHs}
\label{s2.2}
We follow the analysis in \citet{Inman2019} to produce initial conditions including PBHs, which assumes purely adiabatic primordial perturbations on all scales before the formation of PBHs and considers isocurvature perturbations introduced by the discreteness of PBHs at small scales. The overdensity of either (particle) DM %\footnote{If not specially annotated, the term `DM' refers to the part of the dark sector excluding PBHs, which is treated as ideal collision-less fluid.} 
and PBHs can be decomposed into an adiabatic term $\delta_{\rm ad}(a)=T_{\rm ad}(a)\delta_{\rm ad}^{0}$ and an isocurvature term:
\begin{align}
    \delta_{\rm DM}(a)&=\delta_{\rm ad}(a)+\left[T_{\rm iso}(a)-1\right]f_{\rm PBH}\delta_{\rm iso}(a)\ ,\label{e1}\\
    \begin{split}
    \delta_{\rm PBH}(a)&=\delta_{\rm DM}(a)+\delta_{\rm iso}^{0}= \delta_{\rm ad}(a)+T^{\rm PBH}_{\rm iso}(a)\delta_{\rm iso}^{0}\ ,
    \end{split}\label{e2}
\end{align}
where $T^{\rm PBH}_{\rm iso}(a)=1+\left[T_{\rm iso}(a)-1\right]f_{\rm PBH}$, $\delta_{\rm ad}^{0}$ is the primordial adiabatic perturbation, $\delta_{\rm iso}^{0}$ is the primordial isocurvature perturbation of PBHs which make up a mass fraction of $f_{\rm PBH}$ in the dark sector, $\delta_{\rm iso}(a)$ is the perturbation in CDM induced by PBHs, $T_{\rm ad}(a)$ and $T_{\rm iso}(a)$ are the linear transfer functions of the adiabatic and isocurvature modes, respectively. In our case, $a=1/(1+z_{\rm ini})\sim 0.003$. %Note that in Equ.~\ref{e2}, we have ignored the back reaction on PBHs from the DM overdensities induced by PBHs, as such effects are only expected to be significant at scales smaller than the average separation between PBHs $d_{\rm PBH}$ where the isocurvature mode dominates anyway when the adiabatic mode is still deep in the linear regime.

In the zoom-in initial conditions for $\Lambda$CDM cosmology described in the preceding subsection~\ref{s2.1}, the adiabatic mode $\delta_{\rm ad}(a)$ has already been encoded in the positions (and velocities) of simulation particles generated by \textsc{music}. We use this information to generate the initial positions and velocities of PBH particles (which only reside in the zoom-in region). We first apply a grid on the high-resolution region in the CDM initial condition, whose cell size is chosen such that on average each cell contains around one PBH. We calculate the local overdensity of DM in each cell $j$ as $\delta_{j}$ and draw the number of PBHs contained in this cell from a Possion distribution with parameter $(\delta_{j}+1)f_{\rm PBH}M_{\rm tot}/m_{\rm PBH}$, where $M_{\rm tot}$ is the total dark sector mass in the zoom-in region. The adiabatic mode (important at large scales) is realized by this process as cells of higher overdensities containing more PBHs. Next, within each cell, we place the PBHs randomly since they are expected to be Poisson distributed on such small scales\footnote{For simplicity, we have ignored any clustering of PBHs at birth that may arise from certain PBH formation mechanisms (see e.g. \citealt{Belotsky2019}) and have non-trivial implications on the effects of PBHs in the IGM evolution, structure formation and GW astronomy (e.g. \citealt{Desjacques2018,Bringmann2019,DeLuca2020}).}, corresponding to the isocurvature term $\delta_{\rm iso}^{0}$ (i.e. discreteness noise). We then assign velocities to each PBH assuming that they are the same as its nearest-neighbor DM particle, assuming that PBHs are formed with negligible speeds relative to DM. Finally, the mass of each DM particle is reduced by a fraction of $f_{\rm PBH}$ to keep $\Omega_{m}$ identical to the CDM runs.

Next, we implement the isocurvature term $\delta_{\rm iso}(a)$ in Equ.~\ref{e1} induced by PBHs for the overdensity of DM. Deep in the linear regime with small overdensities, the isocurvature mode and adiabatic mode in DM are uncorrelated, i.e. $\delta_{\rm iso}(a)\rightarrow \delta^{0}_{\rm iso}$ for $a\rightarrow 0$. However, at later stages such as our case with $z_{\rm ini}=300$, the two modes have mixed with each other as PBHs follow the large-scale adiabatic mode to fall into larger structures and meanwhile induce/disrupt DM structures around themselves on small scales. In this regime, $\delta_{\rm iso}(a)$ should be between $\delta_{\rm iso}^{0}$ and $\delta_{\rm PBH}(a)$, whose exact form can be complex, especially for intermediate\footnote{The mode mixing/correlation is unimportant in the `seed' limit with very small $f_{\rm PBH}$, where PBHs (and the induced DM structures around them) seldom interact with each other, and also in the opposite `Poisson' limit ($f_{\rm PBH}\rightarrow 1$), where the large-scale behavior is still dominated by the abiabatic mode, and a discreteness noise is introduced at small scales (see e.g. \citealt{Carr2018,Inman2019}). } 
PBH fractions $f_{\rm PBH}\sim 10^{-4}-0.1$. In the absence of a better theory for $\delta_{\rm iso}(a)$, we use the positions of PBHs %\footnote{By using the positions of PBHs at $z_{\rm ini}$, we assume that nonlinear interactions between PBHs and DM structures are negligible. This assumption will break for haloes containing multiple PBHs, which are expected to be rare at $z_{\rm ini}=300$ (see fig.~8 in \citealt{Inman2019}). We discuss the possible errors introduced by this assumption in Sec.~\ref{s4.1}.} 
at $z_{\rm ini}$ to calculate the (co-moving) displacement and velocity fields of DM particles induced by PBHs with the Zel'dovich approximation \citep{Zel'Dovich1970,mo2010galaxy}

\begin{align}
    \vec{\psi}(\vec{x})&=-\frac{D(a)}{4\pi G\bar{\rho}_{\rm m}a^{3}}\nabla \phi_{\rm iso}(\vec{x})=-\frac{2D(a)}{3\Omega_{m}H_{0}^{2}}\nabla \phi_{\rm iso}(\vec{x})\ ,\label{e3}\\
    \Delta \vec{v}(\vec{x})&=-\frac{\dot{D}(a)}{4\pi G\bar{\rho}_{\rm m}a^{2}}\nabla \phi_{\rm iso}(\vec{x})=\frac{a\dot{D}(a)}{D(a)}\vec{\psi}(\vec{x})\ \label{e4}.
\end{align}
In this way, the mode mixing/correlation is captured by the deviation of the PBH distribution from purely random at large scales caused by the adiabatic mode. Here $D(a)=T_{\rm iso}(a)-1$ is the growth factor of PBH-induced perturbations, $H_{0}$ is the Hubble constant, and $-\nabla \phi_{\rm iso}(\vec{x})$ is the (co-moving) acceleration field from PBHs
\begin{align}
    \nabla \phi_{\rm iso}(\vec{x})=4\pi Gm_{\rm PBH}\sum_{i}\frac{\vec{x}-\vec{x}_{i}}{|\vec{x}-\vec{x}_{i}|^{3}}\ ,\label{e5}
\end{align}
given the (co-moving) coordinates of PBH particles $\vec{x}_{i}$ at $z_{\rm ini}$. Since the isocurvature mode in DM does not grow during the radiation-dominated epoch, the growth factor can be approximated with a simple analytical expression within 1.5\% accuracy \citep{Inman2019}:
\begin{align}
    D(a)&\approx \left(1+\frac{3\gamma}{2a_{-}}s\right)^{a_{-}}-1\ ,\quad s=\frac{a}{a_{\rm eq}}\ ,\notag\\
    \gamma&=\frac{\Omega_{m}-\Omega_{b}}{\Omega_{m}}\ ,\quad a_{-}=\frac{1}{4}\left(\sqrt{1+24\gamma}-1\right)\ \label{e6}\ ,
\end{align}
where $a_{\rm eq}=1/(1+z_{\rm eq})$ is the scale factor at matter-radiation equality with $z_{\rm eq}\sim 3400$. As our simulations start in the matter-dominated era with $D(a)\propto a$ approximately (given $\gamma\approx 1$ and $a_{-}\approx 1$), we have 
\begin{align}
    \frac{\dot{D}(a)}{D(a)}\approx \frac{\dot{a}}{a}=H(a)\approx \sqrt{\Omega_{m}}a^{-3/2}\ .\label{e7}
\end{align}

Substituting Equ.~\ref{e5}-\ref{e7} into Equ.~\ref{e3} and \ref{e4}, we obtain the perturbation from PBHs on every DM particle. In this process the displacement is truncated at the average separation of DM particles $d_{\rm DM}\sim 0.3\ h^{-1}\rm kpc$ to be consistent with the Zel'dovich approximation, i.e. $\vec{\psi}=\min\left(1,d_{\rm DM}/|\vec{\psi}|\right)\vec{\psi}$. 
For a given DM particle $j$ with an initial coordinate $\vec{x}_{j}$, we have $\vec{x}_{j}=\vec{x}_{j}+\vec{\psi}(\vec{x}_{j})$ and $\vec{v}_{j}=\vec{v}_{j}+\Delta\vec{v}(\vec{\psi}(\vec{x}_{j}))$. As a conservative estimation of the strength of mode mixing, we only consider at most the 64 nearest PBH particles within $2d_{\rm PBH}$ around the DM particle when evaluating Equ.~\ref{e5}, because the isocurvature mode should only be important at small scales. Here $d_{\rm PBH}$ is the average separation between PBHs. If all PBHs are considered in Equ.~\ref{e5}, the maximum correlation with $\delta_{\rm iso}(a)=\delta_{\rm PBH}(a)$ is achieved, which will further accelerate structure formation. On the other hand, if we reduce the number/volume of PBHs that contribute to the acceleration field, the correlation will be suppressed, leading to delay of structure formation. %We find that this significantly accelerates structure formation such that formation of the first star-forming halo can be $\Delta z\sim 10$ earlier than the default case. 
We discuss in detail the dependence of our results on initial conditions in Appendix~\ref{aa}. Note that the PBH perturbations are actually \textit{nonlinear} close to the PBHs (i.e. at $k\gtrsim \rm kpc^{-1}$, see fig.~4 in \citealt{Inman2019}), which are not fully captured by our approach based on \textit{linear} perturbation theory. Since our simulations start at a relatively high redshift ($z_{\rm ini}=300$) with respect to the moment of first star formation ($z\sim 20-40$), nonlinear structures around PBHs will grow and virialize within one Hubble time, i.e. by $z\sim 100$, in our simulations, before the star-forming minihaloes start to assemble, such that the application of the Zel'dovich approximation in initial conditions should have little impact on the results. 

For gas particles, we calculate the PBH-induced displacement and velocity fields following the same procedure (Equ.~\ref{e3}-\ref{e7}). The only difference is that we have replaced $a_{\rm eq}$ in Equ.~\ref{e6} with $a_{\rm rec}=1/(1+z_{\rm rec})$ given the redshift of recombination $z_{\rm rec}\sim 1100$, assuming that isocurvature perturbations in gas only starts to grow after photon-gas decoupling due to Silk damping. Our results are not sensitive to the initial displacement field of gas particles, since gas cannot condense into dark matter structures at $z\gtrsim 100$ anyway, which are not massive enough to overcome gas pressure. However, the velocity field of gas (with respect to that of DM), i.e. streaming motion between gas and DM \citep{Tseliakhovich2010}, can play an important role in early star formation (see e.g., \citealt{maio2011impact,greif2011delay,stacy2011effect,naoz12,naoz13,fialkov2012,Hirano2018formation,Anna2019,Park2020}). Therefore, for the CDM and fiducial PBH models in Case A, we also run a simulation that includes an universal\footnote{The streaming motion is coherent over scales of a few co-moving Mpc \citep{Tseliakhovich2010}, much larger than the size of our zoom-in region, such that we can adopt a constant velocity offset throughout the zoom-in region. {Note that we have ignored the smoothing of gas density by streaming motion at $z>z_{\rm ini}$, which may lead to underestimation of the delay of collapse caused by streaming motion \citep{Park2020}. However, this will not affect Pop~III star formation at the halo scale and change the general trends that we are concerned with.}} initial velocity offset between gas and DM in a random direction with a magnitude of $v_{\rm ini}=v_{\rm b\chi}a_{\rm rec}/a$. Here we choose $v_{\rm b\chi}=0.8\sigma_{\rm b\chi}$ as a typical value around which the impact on overall structure formation is largest \citep{Anna2019}, where $\sigma_{\rm b\chi}=30\ \rm km\ s^{-1}$ is the root-mean-square streaming velocity at recombination. {The effects of gas-DM streaming in PBH cosmologies have been explored in \citet{Kashlinsky2021}, finding that they are weaker than in $\rm \Lambda CDM$ due to the density perturbations induced by PBHs. As further discussed in Appendix~\ref{ab}, our results are consistent with those in \citet{Kashlinsky2021}.} 

%Since the isocurvature mode is only important at scales smaller than the average spacing between PBHs, we generate a grid 

%

\subsection{Black hole physics}
\label{s2.3}
Limited by the scope of the Tree+PM gravity solver \citep{springel2005cosmological}, our simulations do not treat PBHs as point masses. Instead, gravity from PBHs is still softened by a \textit{physical} softening length of $\epsilon_{\rm BH,phy}=10^{-3}\ h^{-1}\ \rm pc$, and the softening kernel for the gravitational potential has a fixed size of $\epsilon_{\rm g,BH}=2.8\epsilon_{\rm BH,phy}\sim 4\times 10^{-3}\ \rm pc$, much smaller than the $\sim 0.1-1\ \rm pc$ extent of star-forming clouds in minihaloes. %When two BH particles get closer than $2\epsilon_{\rm g,BH}$ and are meanwhile gravitationally bound to each other, they are combined into one BH particle, which then represents a BH binary or multiple system whose internal evolution is beyond our resolution. 
We also include a sub-grid model for dynamical friction (DF) of PBHs by DM (see Sec.~\ref{s2.3.1} below). In this way, although our simulations cannot capture the dynamics within close binaries and multiple systems of PBHs, the overall dynamics of PBHs in star-forming minihaloes is well resolved. 
Beside dynamical effects, feedback from BH accretion can also play an important role in the evolution of primordial gas clouds. We implement sub-grid models for BH accretion and feedback based on \citet{springel2005modelling,tremmel2015off,tremmel2017romulus,Takhistov2022}, as described in Sec.~\ref{s2.3.2} and Sec.~\ref{s2.3.3}.

\subsubsection{Dynamical friction}
\label{s2.3.1}
%In our simulations, the masses of Pop~III BH particles are usually comparable to that of stellar particles ($m_{\mathrm{BH}}\sim m_{\star}$), and much smaller than the masses of gas and DM particles ($m_{\mathrm{BH}}\sim 0.1 m_{\mathrm{gas}}\sim 0.02 m_{\mathrm{DM}}$). Under this condition, the corresponding gravitational softening lengths of background particles must be large enough to avoid spurious collisionality, which meanwhile will suppress dynamical friction (DF) by preventing close encounters with BH particles. Thus, DF of BHs by background objects is not naturally simulated with the gravity solver on scales smaller than the background gravitational softening length, which is a common problem in cosmological simulations with limited mass/spatial resolution. Since dynamical interactions are crucial for BH mergers, we adopt the sub-grid model from \citet{tremmel2015off} to better simulate DF of BHs by background stars\footnote{We only apply the sub-grid DF model to stars because they are the dominant source for DF, and our resolution for gas and DM particles is too low for the sub-grid model to work.}. 

Since our simulations have very high mass resolution of gas relative to the masses of BHs, i.e. $m_{\rm BH}/m_{\rm gas}\gtrsim 120\gg 1$, DF of BHs through gas is naturally captured by the gravity solver. For DM, on the other hand, the resolution is only marginally sufficient ($m_{\rm BH}/m_{\rm DM}\sim 15$), and small-scale effects may be underestimated. %DF can be important for the dynamics of BH in minihalos. 
We therefore further adopt the sub-grid model from \citet{tremmel2015off} to better simulate DF of BHs by DM.

For each BH particle, the additional acceleration from the sub-grid DF model is \citep{tremmel2015off}
\begin{align}
	\vec{a}_{\mathrm{DF}}=-4\pi G^{2}m_{\mathrm{BH}}\rho_{\rm DM}(<v_{\mathrm{BH}})\ln\Lambda\frac{\vec{v}_{\mathrm{BH}}}{v_{\mathrm{BH}}^{3}}\ ,
\end{align}
where $\vec{v}_{\mathrm{BH}}$ is the velocity of the BH relative to the local background centre of mass (COM), $\rho_{\rm DM}(<v_{\mathrm{BH}})$ is the mass density of DM particles with velocities relative to the COM smaller than $v_{\mathrm{BH}}$, and $\ln\Lambda$ is the Coulomb logarithm. The local COM velocity is defined with the 64 nearest gas particles around the BH enclosed by the radius $h_{\mathrm{BH}}$. In our case, $\rho_{\rm DM}(<v_{\mathrm{BH}})$ is estimated with
\begin{align}
	\rho_{\rm DM}(<v_{\mathrm{BH}})=\frac{M_{\rm DM}(<v_{\mathrm{BH}})}{[4\pi h_{\mathrm{BH}}^{3}/3]}\ ,
\end{align}
where $M_{\rm DM}(<v_{\mathrm{BH}})$ is the total mass of DM particles within $h_{\mathrm{BH}}$ around the BH, whose velocities relative to the COM are smaller than $v_{\mathrm{BH}}$. %We set $r_{\mathrm{DF}}=h_{\mathrm{BH}}$. 
The Coulomb logarithm is\footnote{We have $\ln\Lambda \sim 10$ typically in our simulated minihaloes, given $m_{\mathrm{BH}}\simeq 33\ \mathrm{M}_{\odot}$, $v_{\mathrm{BH}}\sim 10\ \mathrm{km\ s^{-1}}$ and $\epsilon_{\mathrm{g, DM}}\sim 1$~pc.}
\begin{align}
\begin{split}
	\ln\Lambda &= \ln(1+b_{\max}/b_{\min})\ ,\\
	b_{\max}&=\epsilon_{\mathrm{g, DM}}\ ,\quad b_{\min}=\frac{Gm_{\mathrm{BH}}}{v_{\mathrm{BH}}^{2}}\ .
\end{split}
\end{align}
Here we use $b_{\max}=\epsilon_{\mathrm{g, DM}}\equiv 2.8a\epsilon_{\rm DM}$ and multiply the acceleration $\vec{a}_{\mathrm{DF}}$ by a factor $1/[1+m_{\mathrm{BH}}/(5m_{\rm DM})]\sim 0.1-0.25$ to avoid double counting the frictional forces on resolved (larger) scales.

\subsubsection{Black hole accretion}
\label{s2.3.2}
We use a modified Bondi-Hoyle formalism developed by \citet{tremmel2017romulus} to calculate the BH accretion rate $\dot{m}_{\mathrm{acc}}$, which takes into account the angular momentum of gas. For each BH particle, we first estimate the characteristic rotational velocity of surrounding gas $\epsilon_{\mathrm{g,BH}}$ away from the BH as $v_{\theta}=j/\epsilon_{\mathrm{g,BH}}$, where $j$ is the specific angular momentum of gas particles in the radius range $(3/4)h_{\mathrm{BH}}-h_{\mathrm{BH}}$. Then we compare $v_{\theta}$ with the characteristic bulk motion velocity $v_{\mathrm{bulk}}$, approximated by the smallest relative velocity between the BH and gas particles within $h_{\mathrm{BH}}$. When $v_{\theta}\le v_{\mathrm{bulk}}$, the effect of angular momentum is negligible, so that the original Bondi-Hoyle accretion formula is used:
\begin{align}
	\dot{m}_{\mathrm{acc}}=\frac{4\pi (G m_{\mathrm{BH}})^{2}\rho_{\mathrm{\rm gas}}}{\tilde{v}^{3}}=\frac{4\pi (G m_{\mathrm{BH}})^{2}\rho_{\mathrm{\rm gas}}}{(c_{s}^{2}+v_{\mathrm{\rm gas}}^{2})^{3/2}}\ ,\label{e11}%\text{ if }v_{\theta}\le v_{\mathrm{bulk}}\ ,
\end{align}
where $\rho_{\mathrm{\rm gas}}$ is the gas density computed from the hydro kernel at the position of the BH, $c_{s}$ is the sound speed and $v_{\mathrm{\rm gas}}$ the velocity dispersion of gas particles with respect to the BH. Here $c_{s}$ is calculated with the mass-weighted average temperature of surrounding gas. While for $v_{\theta}> v_{\mathrm{bulk}}$, a rotation-based formula is adopted \citep{tremmel2017romulus}:
\begin{align}
	\dot{m}_{\mathrm{acc}}=\frac{4\pi (G m_{\mathrm{BH}})^{2}\rho_{\mathrm{\rm gas}}c_{s}}{\left(c_{s}^{2}+v_{\theta}^{2}\right)^{2}}\ ,\label{e12}%\text{ if }v_{\theta}> v_{\mathrm{bulk}}\ .
\end{align}
%We also place an upper limit on $\dot{M}_{\mathrm{acc}}$ as the Eddington accretion rate (for primordial gas with $\mu\simeq 1.22$)\footnote{Throughout our simulations, accretion is always highly sub-Eddington, even when BH feedback is turned off. We include this upper limit for completeness, although it is never invoked here.}

Once $\dot{m}_{\mathrm{acc}}$ is known, we increase the BH mass at each timestep with $\delta m_{\mathrm{BH}}=\dot{m}_{\mathrm{acc}}\delta t$. The dynamical masses of BH particles are also updated smoothly. However, the masses of surrounding gas particles are not reduced in this continuous fashion. Instead, we adopt the algorithm from \citealt{springel2005modelling} (see their equ.~35), in which BH particles swallow nearby gas particles stochastically\footnote{Different from the original scheme in \citet{springel2005modelling}, in our case the BH mass is no longer increased when a gas particle is swallowed, as it has already been updated (smoothly).}. This implies that mass conservation is not explicitly enforced at each timestep in our simulations, but overall mass conservation still holds\footnote{The stochastic effect is negligible since the average fraction of accreted mass in BH mass is less than one percent throughout our simulations, and BHs only make up a small fraction of DM.}. 
%\footnote{This implies that mass conservation is not explicitly enforced in our simulations. Nevertheless, the effect is negligible since $M_{\mathrm{BH,res}}\sim m_{\mathrm{gas}}$, and the total number of BH particles is much smaller than the total number of gas particles. Besides, the fraction of accreted mass in BH mass is typically less than one percent in our simulations.} 
%with $\delta m_{\mathrm{BH}}=\delta M_{\mathrm{BH}}$ for $M_{\mathrm{BH}}<M_{\mathrm{BH,res}}=10^{4}\ \mathrm{M}_{\odot}\sim m_{\mathrm{gas}}$. While for $M_{\mathrm{BH}}\ge M_{\mathrm{BH,res}}$, 
We also apply drag forces from accretion on BH particles according to momentum conservation, following \citet{springel2005modelling}.

\subsubsection{Black hole feedback}
\label{s2.3.3}

For simplicity, we only consider thermal feedback from BH accretion in terms of photoionization heating. To save computational resources, BH feedback is only turned on at $z<100$, as we are mainly concerned with the effects of PBHs in potential star-forming minihaloes with $M_{\rm h}\gtrsim 10^{5}\ \rm M_{\odot}$, formed at late stages ($z\lesssim 40$). Actually, our initial conditions do not fully capture the nonlinear structures around PBHs at small scales, which will form and virialize within a Hubble time after the simulation starts (i.e. $z\sim 100$). It is therefore reasonable to turn on BH feedback thereafter. Note that we have neglected the X-ray background produced by BH accretion at $z\gtrsim 100$, which can increase the electron abundance and, therefore, increase the $\rm H_{2}$ abundance to $x_{\rm H_{2}}\sim 10^{-5}-10^{-4}$ in the IGM at $z\lesssim 100$ for $m_{\rm PBH}\sim 100\ \rm M_{\odot}$, $f_{\rm PBH}\gtrsim 10^{-4}$ \citep{Ricotti2008}. It is found in our simulations and previous studies that formation of $\rm H_{2}$ during virialization is more important than in the diffuse IGM, such that $x_{\rm H_{2}}\sim 10^{-4}-10^{-3}$ in minihaloes hosting collapsing primordial gas clouds, regardless of the background $\rm H_{2}$ abundance. We expect the $\rm H_{2}$ abundance in the IGM to have little impact on our results\footnote{Our simulations do produce $x_{\rm H_{2}}\sim 10^{-5}-10^{-4}$ in the IGM at $z\lesssim 100$ rather than the standard value $x_{\rm H_{2}}\sim 10^{-6}$. The reason is that we have ignored the reactions between CMB photons and $\rm H_{2}$, $\rm H^{-}$ and $\rm H_{2}^{+}$, which are non-negligible at $z\gtrsim 100$. Therefore, instead of underestimating $x_{\rm H_{2}}$ in the PBH runs, we actually overestimate $x_{\rm H_{2}}$ in the CDM runs. We have checked that this leads to slightly ($\lesssim 10\ \rm Myr$) earlier collapse for the CDM case, but will not change the trends seen in our simulations (see Sec.~\ref{s3.1}).}.

%\footnote{}
%\footnote{We have verified that turning on the BH feedback at the beginning ($z_{\rm ini}=300$) only has minor impact on the results.}. 

We adopt the sub-grid model in \citet{springel2005modelling} that implements the thermal feedback as energy injection into the gas particles within a hydro kernel of size $h_{\mathrm{BH}}$ for each BH particle. The total amount of energy to be injected over a timestep $\delta t$, is $\delta E=\epsilon_{r}L_{\mathrm{BH}}\delta t$, where $\epsilon_{r}$ is the efficiency of radiation-thermal coupling, and $L_{\mathrm{BH}}=\epsilon_{\mathrm{EM}}\dot{m}_{\mathrm{acc}}c^{2}$ is the luminosity from BH accretion. Instead of using a fixed radiation efficiency $\epsilon_{\mathrm{EM}}$, we here use the method in \citet{negri2017black} to calculate $\epsilon_{\mathrm{EM}}$ as
\begin{align}
\epsilon_{\mathrm{EM}}=\frac{\epsilon_{0}A\eta}{1+A\eta}\ ,\quad \eta\equiv \dot{m}_{\mathrm{acc}}/\dot{m}_{\mathrm{Edd}}\ ,\label{epsilonEM}
\end{align}
where $A=100$ and $\epsilon_{0}=0.057$ is the radiative efficiency for non-rotating Schwarzschild BHs (assuming negligible spins of PBHs), and $\dot{m}_{\mathrm{Edd}}$ is the Eddington accretion rate
\begin{align}
	\dot{m}_{\mathrm{Edd}}=2.7\times 10^{-7}\ \mathrm{M_{\odot}\ yr^{-1}}\ \left(\frac{m_{\mathrm{BH}}}{100\ \mathrm{M_{\odot}}}\right)\left(\frac{\epsilon_{0}}{0.1}\right)^{-1}\ .\label{e14}
\end{align}
This model is meant to capture the transition from optically thick and geometrically thin, radiatively efficient accretion discs, to optically thin, geometrically thick, radiatively inefficient advection dominated accretion flows (ADAFs). The only free parameter is the coupling efficiency $\epsilon_{r}$, which can be further written as $\epsilon_{r}\equiv f_{h}f_{\rm abs}$, where $f_{\rm abs}$ is the fraction of radiation energy absorbed by the ISM, and $f_{h}$ is the fraction of energy deposited as heat. 

\begin{figure}
    \centering
    \includegraphics[width=1\columnwidth]{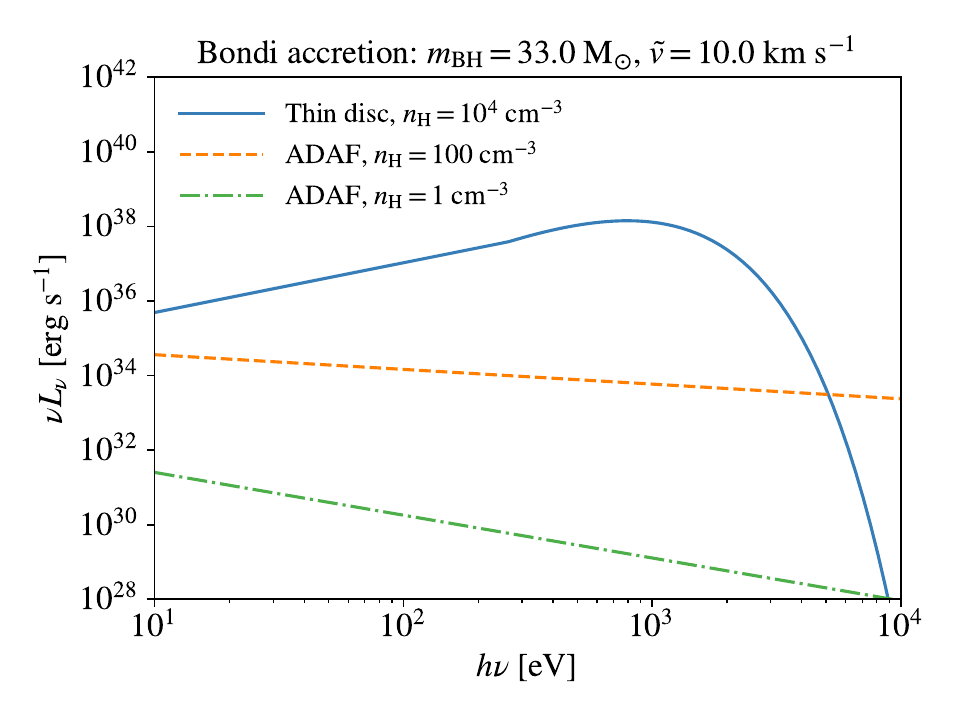}
    \vspace{-25pt}
    \caption{Spectra of BH accretion discs at 3 densities $n_{\rm H}=10^{4}$ (solid), 100 (dashed) and 1~$\rm cm^{-3}$ (dashed-dotted) for Bondi accretion (Equ.~\ref{e11}) around a BH of $m_{\rm BH}=33\ \rm M_{\odot}$ and $\tilde{v}=10\ \rm km\ s^{-1}$, based on the thin disc and ADAF models in \citet{Takhistov2022}. } % However, the discrepancy can be as large as a factor $\sim 10$ in the LHAF regime. Nevertheless, this regimes only occupies a small density range ($n_{\rm H}\sim 100-1000\ \rm cm^{-3}$) and therefore, is expected to be unimportant for our simulation results. }
    \label{bhspec}
\end{figure}

Following \citet{Takhistov2022}, we adopt $f_{h}=1/3$. To further determine $f_{\rm abs}$ and $\epsilon_{r}$, we define the ISM heating efficiency as $\epsilon_{\rm heat}=\epsilon_{r}\epsilon_{\rm EM}$ and compare the efficiency derived from Equ.~\ref{epsilonEM} as a function of surrounding gas density with that obtained from detailed calculations of the BH accretion disc spectra and radiative transfer based on \citet{Takhistov2022}. For simplicity, we only consider the thin disc regime and the standard ADAF regime under a constant cooling efficiency parameter $A_{c}=1.1$ in our calibration\footnote{We have ignored the `electron' ADAF and luminous hot accretion flow (LHAF) regimes. In principle, the `electron' ADAF regime has a lower efficiency of inverse-Compton cooling and the LHAF regime is dominated by ion-electron collisional heating rather than viscous electron heating, such that the dependence of electron temperature on accretion rate is different from the standard ADAF case \citep{Takhistov2022}. Considering these two regimes leads to up to a factor of 10 discrepancies in $\epsilon_{\rm heat}$ between our sub-grid model and more detailed radiative transfer calculations at $n_{\rm H}\sim 1-1000\ \rm cm^{-3}$. As shown in Appendix~\ref{ac}, varying $\epsilon_{\rm r}$ (and $\epsilon_{\rm heat}$) by a factor of 10 does not change our results significantly. So we expect the discrepancies here to have little impact on our conclusions.}. For instance, Fig.~\ref{bhspec} shows the spectra of BH accretion discs at 3 densities $n_{\rm H}=10^{4}$, 100 and 1~$\rm cm^{-3}$ for $m_{\rm BH}=33\ \rm M_{\odot}$ and $\tilde{v}=10\ \rm km\ s^{-1}$ under Bondi accretion (Equ.~\ref{e11}). For $n_{\rm H}=10^{4}\ \rm cm^{-3}$, we are in the thin disc regime dominated by UV (ionizaing) photons, while the other two cases are in the ADAF regime, which is more common in our simulations (see Sec.~\ref{s3}). Ionizing photons ($h\nu>13.6\ \rm eV$) are produced with much lower efficiencies in a ADAF disc by inverse Compton scattering (of synchrotron radiation), which has a power-law spectrum \citep{Takhistov2022}. %In the latter case, the spectrum is dominated by photons of lower energies (synchrotron and free-free radiation). Since these photons are hardly absorbed by the ISM

\begin{figure}
    \centering
    \includegraphics[width=1\columnwidth]{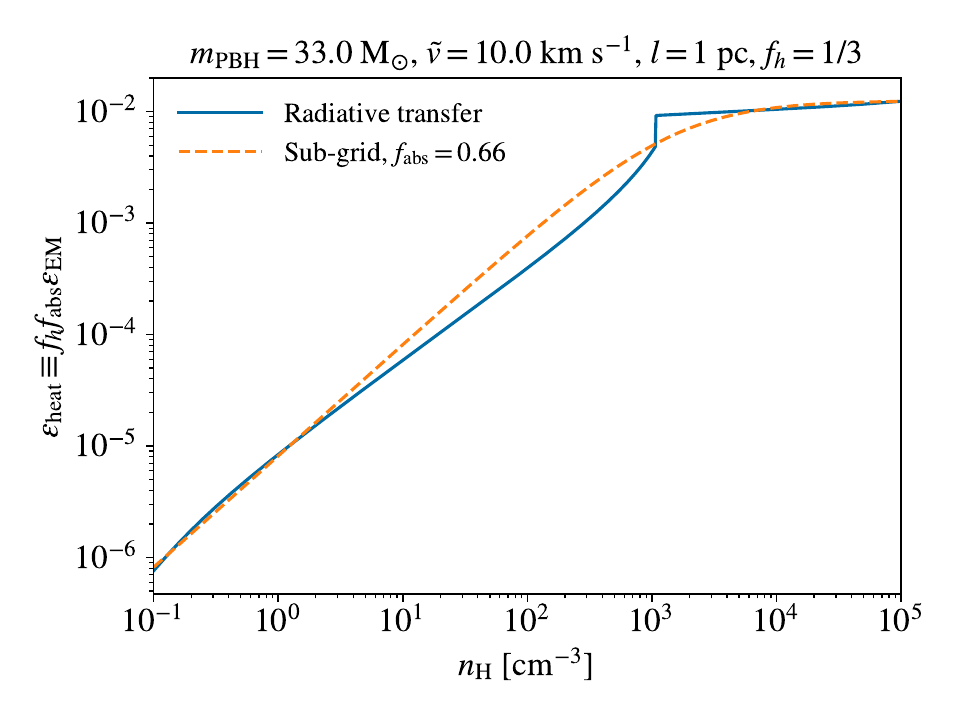}
    \vspace{-25pt}
    \caption{ISM heating efficiency as a function of gas density. The solid curve shows the results from a simplified version of the BH accretion disc spectrum model and radiative transfer in \citet[see the main text of Sec.~\ref{s2.3.3}]{Takhistov2022}. The dashed curve shows the predictions of our sub-grid model (Equ.~\ref{epsilonEM}) with $f_{\rm abs}=0.66$, which agrees with the model in \citet{Takhistov2022} within a factor of 2.} % However, the discrepancy can be as large as a factor $\sim 10$ in the LHAF regime. Nevertheless, this regimes only occupies a small density range ($n_{\rm H}\sim 100-1000\ \rm cm^{-3}$) and therefore, is expected to be unimportant for our simulation results. }
    \label{eheat}
\end{figure}

It is found that in typical primordial (star-forming) gas clouds with a size of $l\sim 1\ \rm pc$ and $\tilde{v}\sim 5-10\ \rm km\ s^{-1}$, the sub-grid model is generally consistent with the more complex model based on \citet{Takhistov2022} within a factor of $\sim 3$ given $f_{\rm abs}\simeq 0.66$ (i.e. $\epsilon_{r}=0.22$), for the hydrogen density range $n_{\rm H}\sim 0.1-10^{5}\ \rm cm^{-3}$ and BH masses $m_{\rm BH}\sim 30-100\ \rm M_{\odot}$ relevant to our work. For example, Fig.~\ref{eheat} shows $\epsilon_{\rm heat}$ as a function of $n_{\rm H}$ for $m_{\rm BH}=33\ \rm M_{\odot}$ and $\tilde{v}=10\ \rm km\ s^{-1}$. For this specific case, the difference between our sub-grid model and the detailed calculation based on \citet{Takhistov2022} is within a factor 2. %, except for the regime of luminous hot accretion flow (LHAF)\footnote{In the LHAF regime, ion-electron collisional heating exceeds viscous electron heating to become the dominant heating mechanism, which changes the dependence of electron temperature on accretion rate and therefore changes the trend in heating efficiency as the inverse-Compton scattering spectrum is very sensitive to electron temperature \citep{Takhistov2022}.}, which only occurs at $n_{\rm H}\sim 100-1000\ \rm cm^{-3}$ and has little impact on ISM heating by BHs. 
In light of this, we adopt $\epsilon_{r}=0.22$ as the fiducial value and explore several cases in the range $\epsilon_{r}\sim 0-1$. In general, reducing (increasing) $\epsilon_{r}$ (i.e. the feedback strength) will accelerate (delay) the collapse of primordial gas clouds. %For the fiducial PBH model with $m_{\rm PBH}=33\ rm M_{\odot}$, 
In the PBH models considered in this paper ($m_{\rm PBH}\sim 30-100\ \rm M_{\odot}$, $f_{\rm PBH}\sim 10^{-4}-0.1$), the effects are rather minor at $\epsilon_{r}\gtrsim 0.02$ and even the strongest feedback cannot stop the collapse of gas but only delay it. %and increase the temperature of dense gas (from $\sim 200\ \rm K$ to $500\ \rm K$). 
Therefore, in the main body of the paper, we only show the results in the fiducial case ($\epsilon_{r}=0.22$) and the extreme case with $\epsilon_{r}=0$ (no feedback), while a detailed analysis of how feedback strength affects simulation outcomes is given in Appendix~\ref{ac}. 
%We have verified that varying $\epsilon_{r}$ by a factor of a few does not change the trends in our results. %, which implies that at least for $f_{\rm abs}\gtrsim 0.1$ the BH feedback is self-regulated and the amount of ISM heating from PBHs is not sensitive to $f_{\rm abs}$.
%\citep{negri2017black}. 
%We set $\epsilon_{r}=0.02$ as a conservative choice, based on the calibration with SMBHs in \citet{tremmel2017romulus}. \citep{Takhistov2022}

\section{Simulation results}
\label{s3}

\begin{figure*}
    \centering
    \includegraphics[width=1.9\columnwidth]{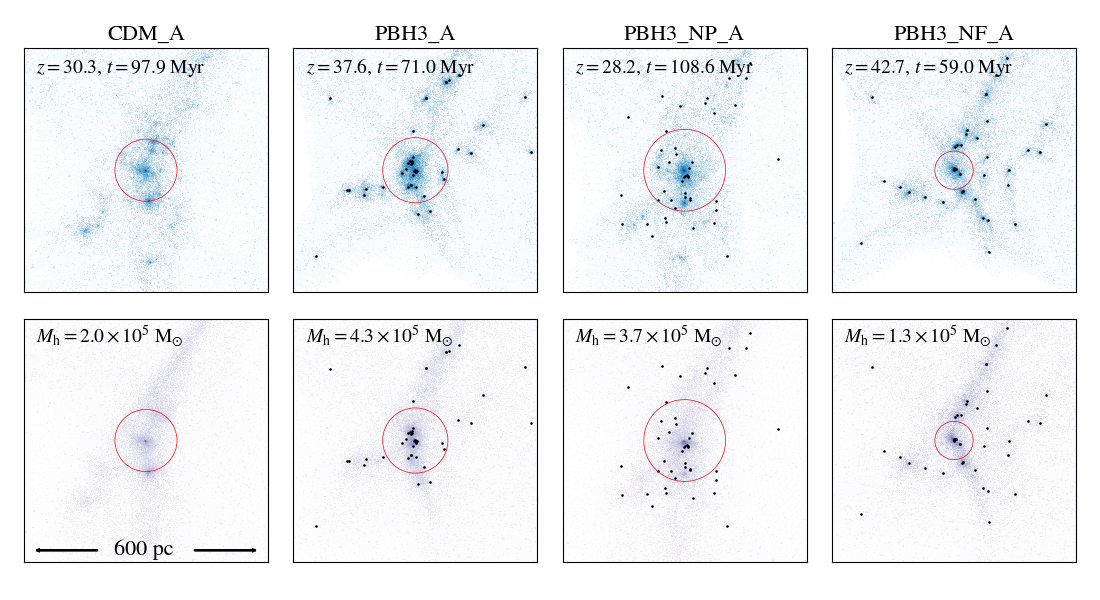}
    \vspace{-15pt}
    \caption{Matter field at the moment of cloud collapse in Case A, as projected distributions of DM (\textit{top}) and gas (\textit{bottom}) particles, for the reference CDM simulation (\texttt{CDM\_A}), fiducial PBH ($m_{\rm PBH}=33\ \rm M_{\odot}$, $f_{\rm PBH}=10^{-3}$) simulation (\texttt{PBH3\_A}) and two extreme situations, one (\texttt{PBH3\_NP\_A}) without PBH perturbations and the other (\texttt{PBH3\_NF\_A}) without BH feedback. The data slice has a (physical) extent of $600\ \rm pc$ and a thickness of $300\ \rm pc$. PBHs are plotted with black dots and the circles show the halo virial radii.}
    \label{cswb_a}
\end{figure*}

\begin{figure}
    \centering
    \includegraphics[width=\columnwidth]{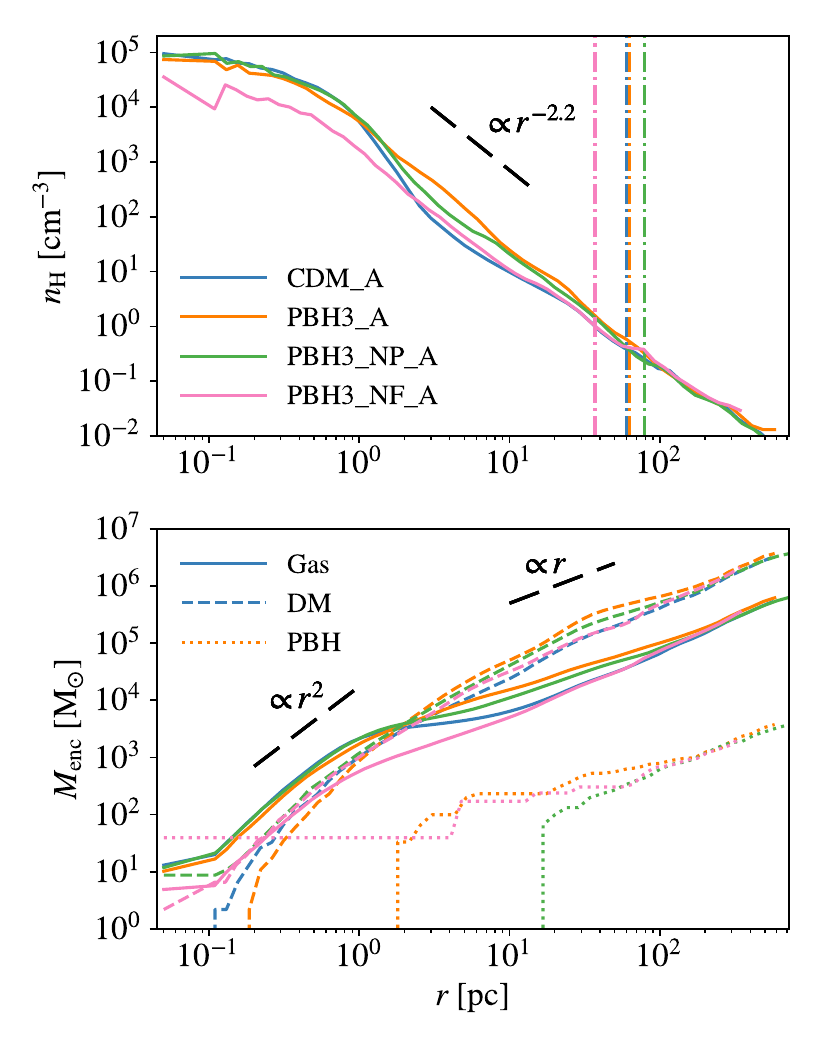}
    \vspace{-20pt}
    \caption{Internal structure of the collapsing cloud and host halo in Case A, for the reference CDM simulation (blue, \texttt{CDM\_A}), fiducial PBH ($m_{\rm PBH}=33\ \rm M_{\odot}$, $f_{\rm PBH}=10^{-3}$) simulation (orange, \texttt{PBH3\_A}) and two extreme simulations without PBH perturbations (green, \texttt{PBH3\_NP\_A}) and without BH feedback (pink, \texttt{PBH3\_NF\_A}). \textit{Top}: hydrogen number density profile, where the virial radii of host haloes are shown with vertical dashed-dotted lines. \textit{Bottom}: enclosed mass profiles for gas (solid), DM (dashed) and PBHs (dotted). }%The results for \texttt{CDM\_A}, \texttt{PBH3\_A}, \texttt{PBH3\_NP\_A} and \texttt{PBH3\_NF\_A} are shown. }
    \label{dpro_a}
\end{figure}

\begin{figure}
    \centering
    \includegraphics[width=1\columnwidth]{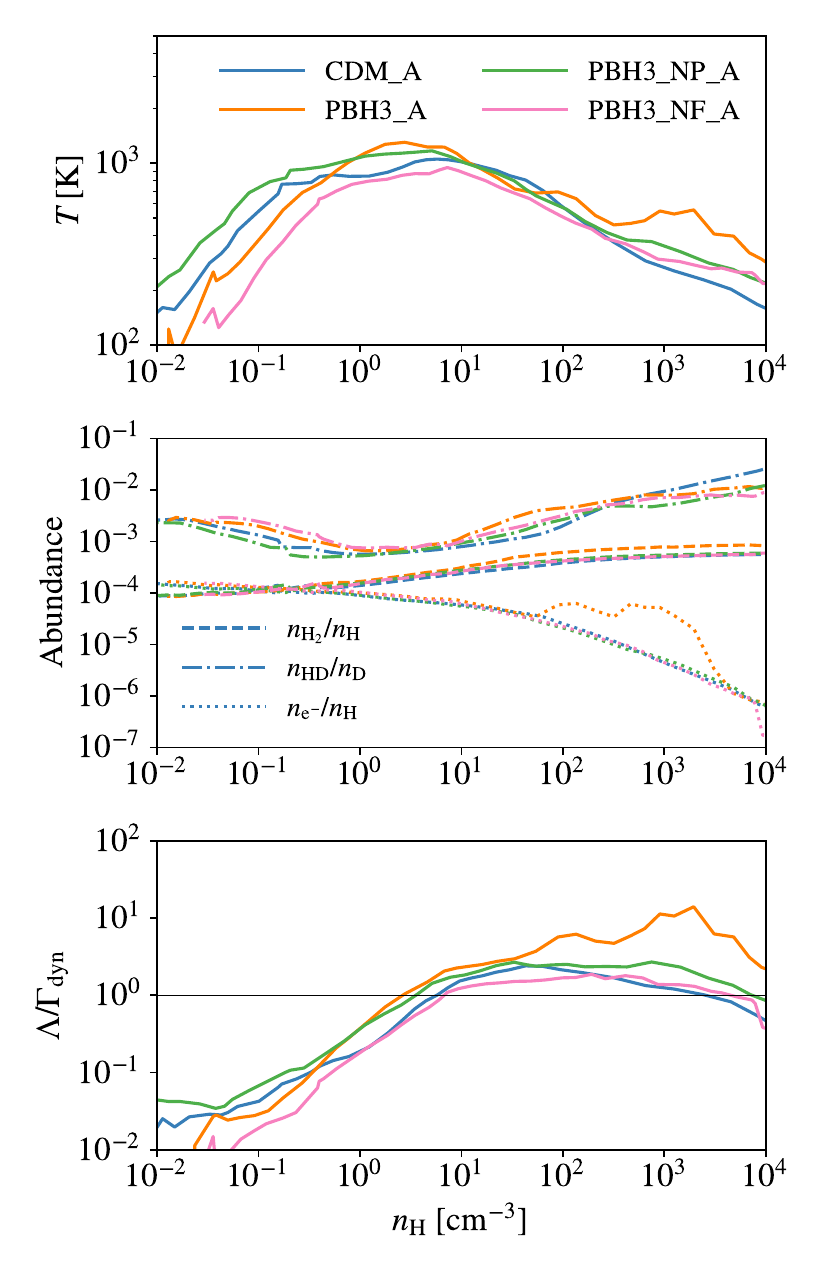}
    \vspace{-20pt}
    \caption{Phase diagrams of the collapsing cloud in Case A, for the reference CDM simulation (blue, \texttt{CDM\_A}), fiducial PBH ($m_{\rm PBH}=33\ \rm M_{\odot}$, $f_{\rm PBH}=10^{-3}$) simulation (orange, \texttt{PBH3\_A}) and two extreme simulations without PBH perturbations (green, \texttt{PBH3\_NP\_A}) and without BH feedback (pink, \texttt{PBH3\_NF\_A}). \textit{Top}: temperature-density diagram. \textit{Middle}: abundances of $\rm H_{2}$ (dashed), $\rm HD$ (dashed-dotted) and $\rm e^{-}$ (dotted) as functions of density. \textit{Bottom}: ratio of cooling and dynamical heating rates as a function of density.}
    \label{cpro_a}
\end{figure}

\begin{comment}
\begin{figure*}
    \centering
    \includegraphics[width=1.9\columnwidth]{cswb_01_B.png}
    \vspace{-15pt}
    \caption{Same as Fig.~\ref{cswb_a} but for Case B.}
    \label{cswb_b}
\end{figure*}
\end{comment}

Our simulations terminate when the maximum hydrogen number density reaches $10^{5}\ \rm cm^{-3}$. At this moment (denoted by $z_{\rm col}$ and $t_{\rm col}$), a dense ($n_{\rm H}\gtrsim 10^{4}\ \rm cm^{-3}$) cold ($T\lesssim 10^{3}\ \rm K$) gas clump of a few $10^{3}\ \rm M_{\odot}$ has formed at the central parsec of the halo by run-away collapse under efficient molecular cooling. This is the typical condition of Pop~III star formation, and it is met in all cases considered in our study (see Table~\ref{t1}), implying that \textit{the standard picture of Pop~III star formation is not changed by the presence of stellar-mass PBHs with $m_{\rm PBH}\sim 30-100\ \rm M_{\odot}$ and $f_{\rm PBH}\lesssim 0.1$}. 

Besides, growth of PBHs via accretion is highly inefficient in our simulations. Even without BH feedback, PBHs can only grow by up to 10\% in mass and the average Eddington ratio is $\sim 10^{-3}$. When feedback is considered, PBHs can hardly grow by more than 0.1\% and the Eddington ratio is $\sim 10^{-4}\ll 1$ on average. This is consistent with previous simulations \citep{johnson2007aftermath,alvarez2009accretion,hirano2014one,smith2018growth,liu2020gw}, showing that light seeds ($\lesssim 100 \rm M_{\odot}$) hardly grow in most cases, unless by super-Eddington accretion under special conditions {\citep{Alexander2014,madau2014super,volonteri2015case,pezzulli2016super,
inayoshi2016hyper,Takeo2018,toyouchi2019super}. More massive ($m_{\rm PBH}\gtrsim 100\ \rm M_{\odot}$) PBHs are more likely to be seeds of supermassive BHs \citep{Cappelluti2022}. }

Nevertheless, we find that PBHs do alter the timing of collapse, host halo mass, as well as chemical and thermal properties of gas during collapse. The thermal history of the IGM is also significantly affected in extreme cases ($f_{\rm PBH}\gtrsim 0.01$). In this section, we use the data from the final snapshots (i.e. at $z_{\rm col}$) to demonstrate the effects of PBHs on primordial star-forming clouds. The key information of our simulations is summarised in Table~\ref{t1}.

\subsection{Fiducial PBH model: perturbation vs. feedback}
\label{s3.1}

We first focus on the fiducial PBH model (\texttt{PBH3}) with $m_{\rm PBH}=33\ \rm M_{\odot}$ and $f_{\rm PBH}=10^{-3}$ in comparison with the reference CDM case and two extreme situations, one (\texttt{PBH3\_NP}) without (isocurvature) perturbations from PBHs (see Sec.~\ref{s2.2}) and the other (\texttt{PBH3\_NF}) without BH feedback (Sec.~\ref{s2.3.3}). In addition to these two extreme cases, we also consider select in-between situations for different strengths of PBH perturbations and feedback, as discussed in Appendices~\ref{aa} and \ref{ac}, respectively.

For Case A (without streaming motion between gas and the dark sector), the cloud collapses at $z_{\rm col}=30.3$ ($t_{\rm col}=97.9\ \rm Myr$) in a halo of $M_{\rm h}=2\times 10^{5}\ \rm M_{\odot}$ without PBHs (\texttt{CDM\_A}). With fiducial PBHs, collapse is \textit{accelerated} by $\sim 30\ \rm Myr$ to $z_{\rm col}=37.6$ ($t_{\rm col}=71.0\ \rm Myr$) in a more massive halo with $M_{\rm h}=4.3\times 10^{5}\ \rm M_{\odot}$ (\texttt{PBH3\_A}). When feedback is turned off (\texttt{PBH3\_NF\_A}), the acceleration is more significant (by $\sim 40\ \rm Myr$) with $z_{\rm col}=42.7$ ($t_{\rm col}=58.9\ \rm Myr$). However, when perturbations from PBHs are not considered (\texttt{PBH3\_NP\_A}), collapse is delayed by $\sim 10\ \rm Myr$ to $z_{\rm col}=28.2$ ($t_{\rm col}=108.6\ \rm Myr$). The trend in Case B is similar: We have $z_{\rm col}=22.3$ ($t_{\rm col}=152.7\ \rm Myr$), $M_{\rm h}=3.8\times 10^{5}\ \rm M_{\odot}$ in \texttt{CDM\_B}, and collapse is accelerated by $\sim 10$ and $50\ \rm Myr$ in \texttt{PBH3\_B} and \texttt{PBH3\_NF\_B}, but delayed by $\sim 10\ \rm Myr$ in \texttt{PBH3\_NP\_B} (see Table~\ref{t1}). The biggest difference is that contrary to Case A, the host halo mass at $t_{\rm col}$ is smaller in \texttt{PBH3\_B} than in \texttt{CDM\_B}. The reason is that in the CDM case collapse of gas happens after a major merger between two haloes, while in \texttt{PBH3\_B}, the merger is delayed due to the Poisson noise introduced by PBHs (see Sec.~\ref{s4.1}), but structure formation is accelerated at smaller scales such that collapse happens in one of the two haloes before merger with a smaller mass.  %, as shown in Fig.~\ref{cswb_b}. 
%We also find that the overall effect is weaker in Case B (at lower redshifts in less over-dense regions) compared with Case A. The reason is that the isocurvature perturbations from PBHs are more important at smaller scales and higher redshifts.

In general, enhancement of density perturbations and BH accretion feedback/heating are two competing effects of PBHs that regulate the formation site and timing of Pop~III stars. For our fiducial PBH model with $m_{\rm PBH}=33\ \rm M_{\odot}$ and $f_{\rm PBH}=10^{-3}$, the former wins over the latter in the simulated over-dense regions, such that although heating tends to delay star formation by increasing the mass threshold above which efficient molecular cooling is activated (see Sec.~\ref{s4.2}), structure formation proceeds much faster under the perturbations of PBHs (see Sec.~\ref{s4.1}) and forms massive haloes with efficient cooling still earlier than in the CDM case\footnote{This outcome is sensitive to the implementation of perturbations from PBHs in the initial matter field at $z_{\rm ini}=300$. As shown in Appendix~\ref{aa}, with weaker perturbations, collapse happens later and can be close to that in the CDM case.}. 

Using Case A as an example\footnote{The results for Case B are similar.}, to better evaluate the aforementioned two effects, we show the (projected) distribution of DM, gas and PBHs in Fig.~\ref{cswb_a}. When perturbations from PBHs are included in the initial condition, by the time of $z_{\rm col}$, DM haloes first form around individual PBHs and the host halo of collapsing gas is assembled by mergers of such PBH-induced structures. The presence of PBHs facilitates formation of filaments and nodes, increasing the clustering strength of DM at small scales. This is consistent with the simulations in \citet{Inman2019} for PBHs of similar masses at $z\gtrsim 99$ (e.g. their fig.~4). The large-scale structure around the host halo is not significantly affected since PBHs follow the same adiabatic mode on large scales. However, without initial PBH perturbations, PBHs behave like test particles, and the DM structures are not modified at almost all scales. This implies that our simulations are very sensitive to the initial condition at $z_{\rm ini}=300$, when perturbations of PBHs have already grown significantly (see Appendix~\ref{aa}). 

We also plot the density profiles of gas, DM and PBHs around the densest gas particle in Fig.~\ref{dpro_a}. The evolution of temperature, chemical composition and the ratio of cooling and dynamical heating rates with gas density (i.e. phase diagrams) are presented in Fig.~\ref{cpro_a}. The density profiles of different models seem similar\footnote{Excluding \texttt{PBH3\_NF\_A}, the slight difference between the density profiles in the other three cases including BH feedback can be explained by their assembly stages. For instance, the snapshot of \texttt{CDM\_A} captures the ongoing merger of two haloes, such that the outer part ($r\sim 2-100\ \rm pc$) of the star-forming halo experiences the gravity from the other halo. }, especially at the central few parsecs\footnote{As the physical softening length of DM and gas particles is $\sim 0.3- 1\ \rm pc$ at $z\sim 20-40$ in our simulations, the core-like feature at $r\lesssim 0.1\ \rm pc$ is a numerical artifact. We expect this to have little impact on our conclusions since cloud collapse is governed by dynamics and cooling at larger scales. }, except for the PBH model without BH feedback (\texttt{PBH\_NF\_A}). In this rather unphysical case, cold gas in the central region condenses rapidly onto two PBHs, reaching the density threshold $n_{\rm H}=10^{5}\ \rm cm^{-3}$ very close ($\lesssim 10^{-3}\ \rm pc$) to the BHs when the cloud at larger scales ($r\sim 10\ \rm pc$) has not collapsed as far as it should be in the standard picture. In order words, BHs accelerate central collapse. When feedback is turned on, gas cannot condense onto BHs due to heating and the densest particle is $\gtrsim 1$ (10)~pc away from the nearest BH with (without) PBH perturbations. The phase diagrams in the four simulations also look similar, except for the fiducial PBH model (\texttt{PBH3\_A}) in which BHs can penetrate into the central $\sim 10\ \rm pc$ %(due to initial perturbations of DM around PBHs)
and meanwhile heat/ionize the relatively dense ($n_{\rm H}\gtrsim 10\ \rm cm^{-3}$) gas therein. This leads to higher temperatures and cooling rates at $n_{\rm H}\gtrsim 10^{2}\ \rm cm^{-3}$ and slightly enhanced $\rm H_{2}$ abundances at $n_{\rm H}\gtrsim 10\ \rm cm^{-3}$. The electron abundance is also increased by a factor of $\sim 10$ around $n_{\rm H}\sim 10^{3}\ \rm cm^{-3}$ by the nearest BHs around the density peak. %Similar trends have been discussed in the early work by \citep{Ricotti2008} for the IGM, where efficient accretion of PBHs can enhance the ionized fraction of hydrogen and $\rm H_{2}$ abundance. %The evolution of $\rm HD$ abundance can be explained by the temperature and electron abundance profiles, as the $\rm HD$ abundance becomes higher with lower temperature and higher electron abundance. 

\begin{comment}
\begin{figure}
    \centering
    \includegraphics[width=\columnwidth]{denspro_B.pdf}
    \vspace{-20pt}
    \caption{}
    \label{dpro_b}
\end{figure}

\begin{figure}
    \centering
    \includegraphics[width=1\columnwidth]{coolpro_B.pdf}
    \vspace{-20pt}
    \caption{}
    \label{cpro_b}
\end{figure}

\end{comment}

\begin{figure*}
    \centering
    \includegraphics[width=1.9\columnwidth]{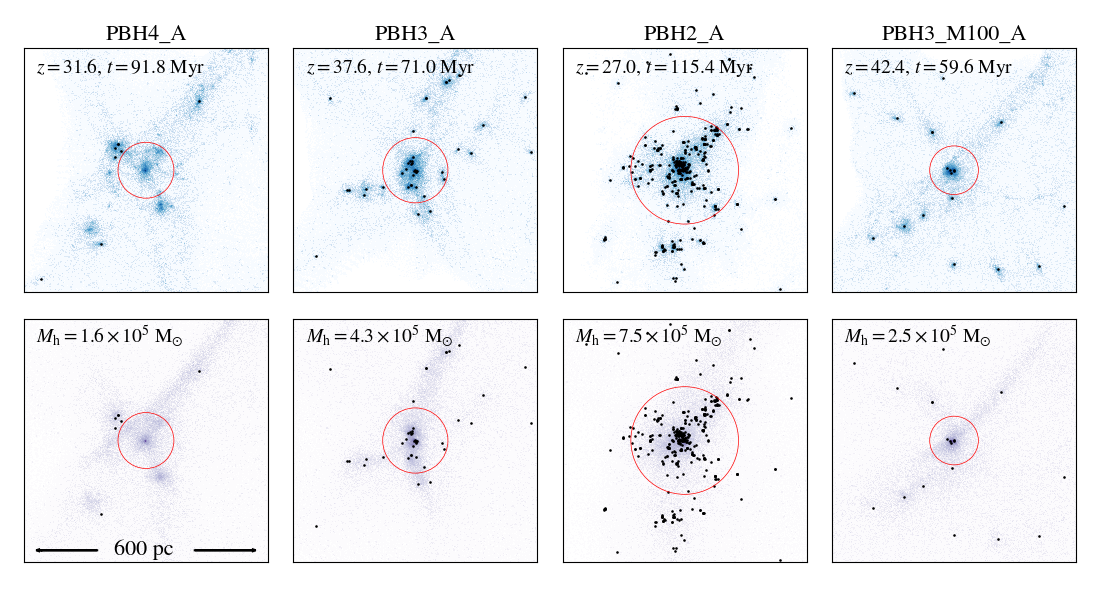}
    \vspace{-15pt}
    \caption{Same as Fig.~\ref{cswb_a} but for PBH models with $m_{\rm PBH}=33\ \rm M_{\odot}$, $f_{\rm PBH}=10^{-4}$, $10^{-3}$, 0.01 and $m_{\rm PBH}=100\ \rm M_{\odot}$, $f_{\rm PBH}= 10^{-3}$.}
    \label{cswb_pbha}
\end{figure*}

\begin{figure}
    \centering
    \includegraphics[width=\columnwidth]{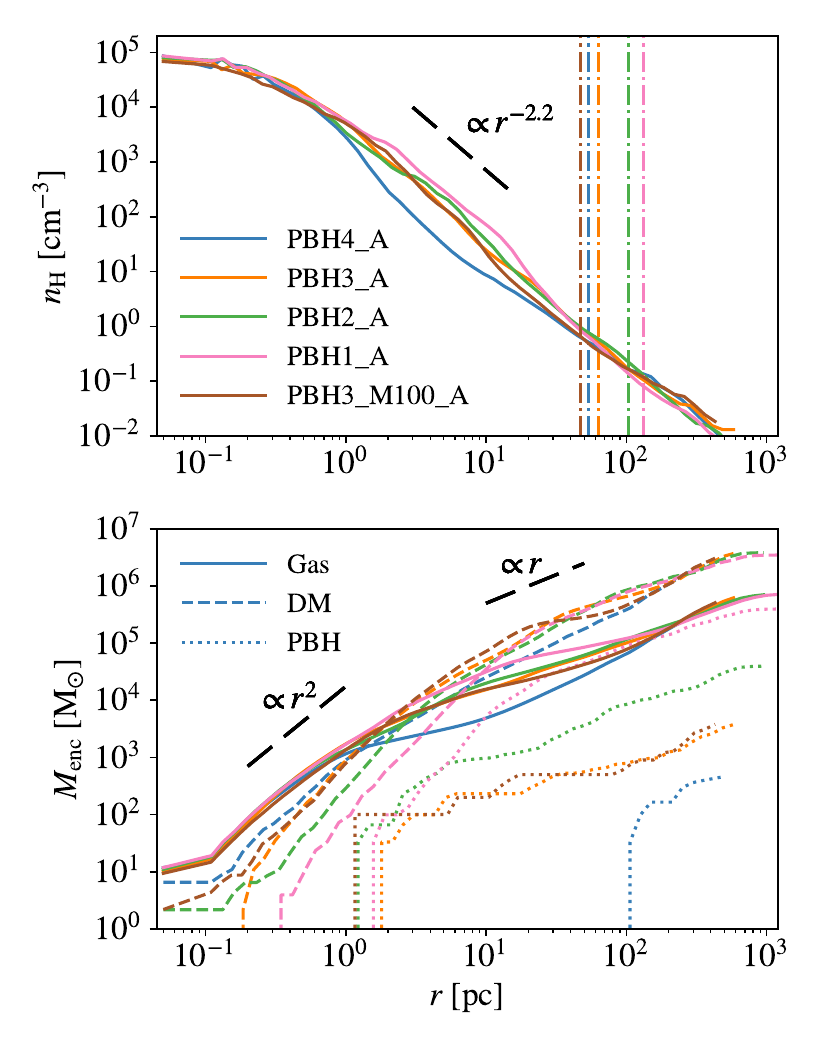}
    \vspace{-20pt}
    \caption{Same as Fig.~\ref{dpro_a} but for PBH models with $m_{\rm PBH}=33\ \rm M_{\odot}$, $f_{\rm PBH}=10^{-4}$ (blue), $10^{-3}$ (orange), 0.01 (green), 0.1 (pink) and $m_{\rm PBH}=100\ \rm M_{\odot}$, $f_{\rm PBH}= 10^{-3}$ (brown).}
    \label{dpro_pbha}
\end{figure}

\begin{figure}
    \centering
    \includegraphics[width=1\columnwidth]{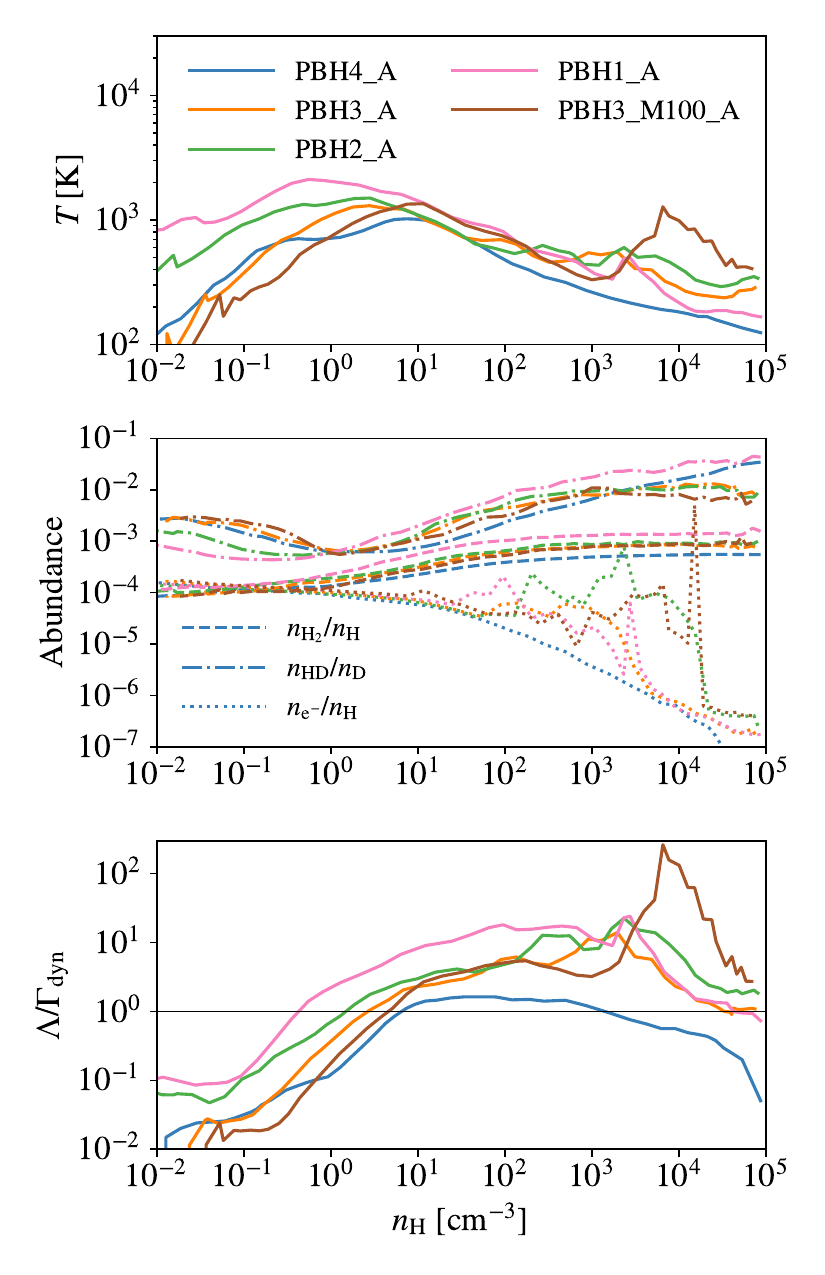}
    \vspace{-20pt}
    \caption{Same as Fig.~\ref{cpro_a} but for PBH models with $m_{\rm PBH}=33\ \rm M_{\odot}$, $f_{\rm PBH}=10^{-4}$ (blue), $10^{-3}$ (orange), 0.01 (green), 0.1 (pink) and $m_{\rm PBH}=100\ \rm M_{\odot}$, $f_{\rm PBH}= 10^{-3}$ (brown).}
    \label{cpro_pbha}
\end{figure}

\begin{figure*}
    \centering
    \includegraphics[width=1.9\columnwidth]{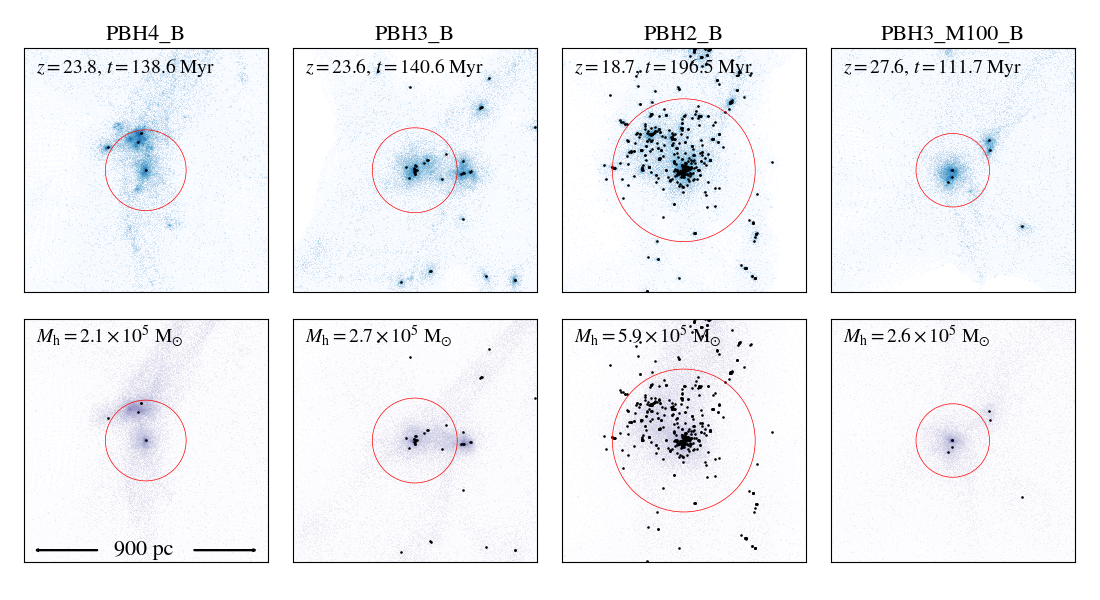}
    \vspace{-20pt}
    \caption{Same as Fig.~\ref{cswb_a} but for PBH models with $m_{\rm PBH}=33\ \rm M_{\odot}$, $f_{\rm PBH}=10^{-4}$, $10^{-3}$, 0.01 and $m_{\rm PBH}=100\ \rm M_{\odot}$, $f_{\rm PBH}= 10^{-3}$ in Case B, and the data slice has a (physical) extent of $900\ \rm pc$ and a thickness of $450\ \rm pc$.}
    \label{cswb_pbhb}
\end{figure*}

\begin{figure}
    \centering
    \includegraphics[width=\columnwidth]{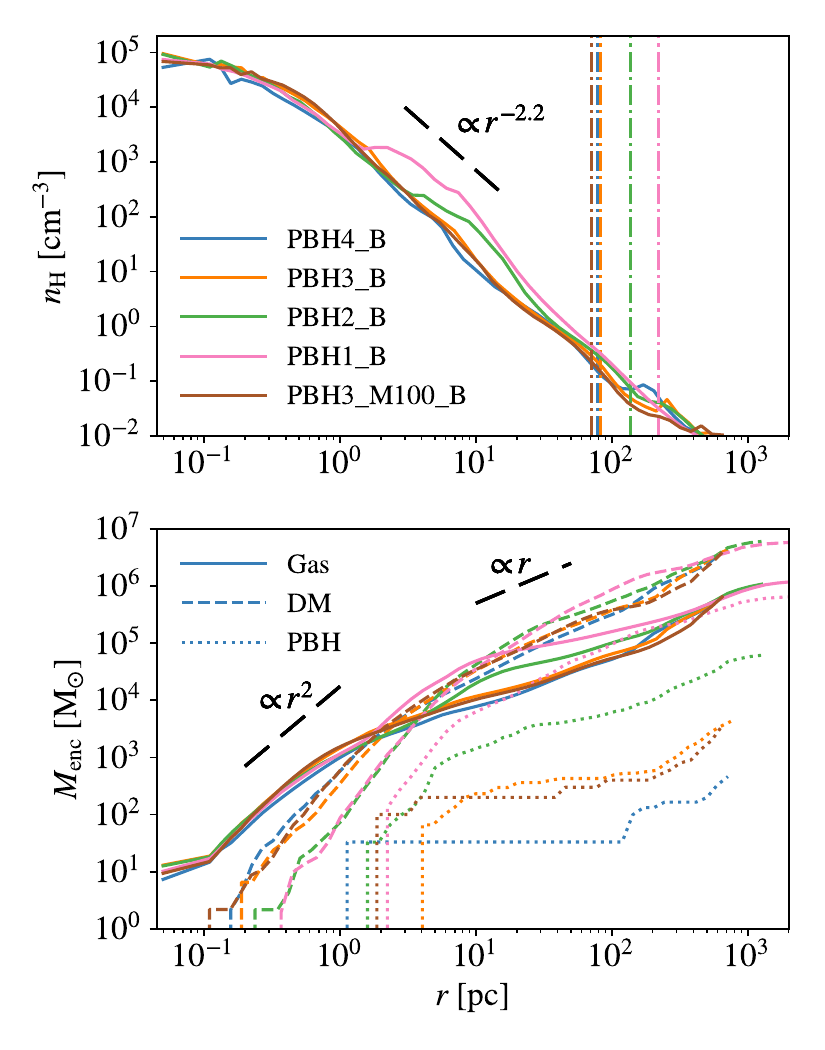}
    \vspace{-20pt}
    \caption{Same as Fig.~\ref{dpro_a} but for PBH models with $m_{\rm PBH}=33\ \rm M_{\odot}$, $f_{\rm PBH}=10^{-4}$ (blue), $10^{-3}$ (orange), 0.01 (green), 0.1 (pink) and $m_{\rm PBH}=100\ \rm M_{\odot}$, $f_{\rm PBH}= 10^{-3}$ (brown) in Case B.}
    \label{dpro_pbhb}
\end{figure}

\begin{figure}
    \centering
    \includegraphics[width=1\columnwidth]{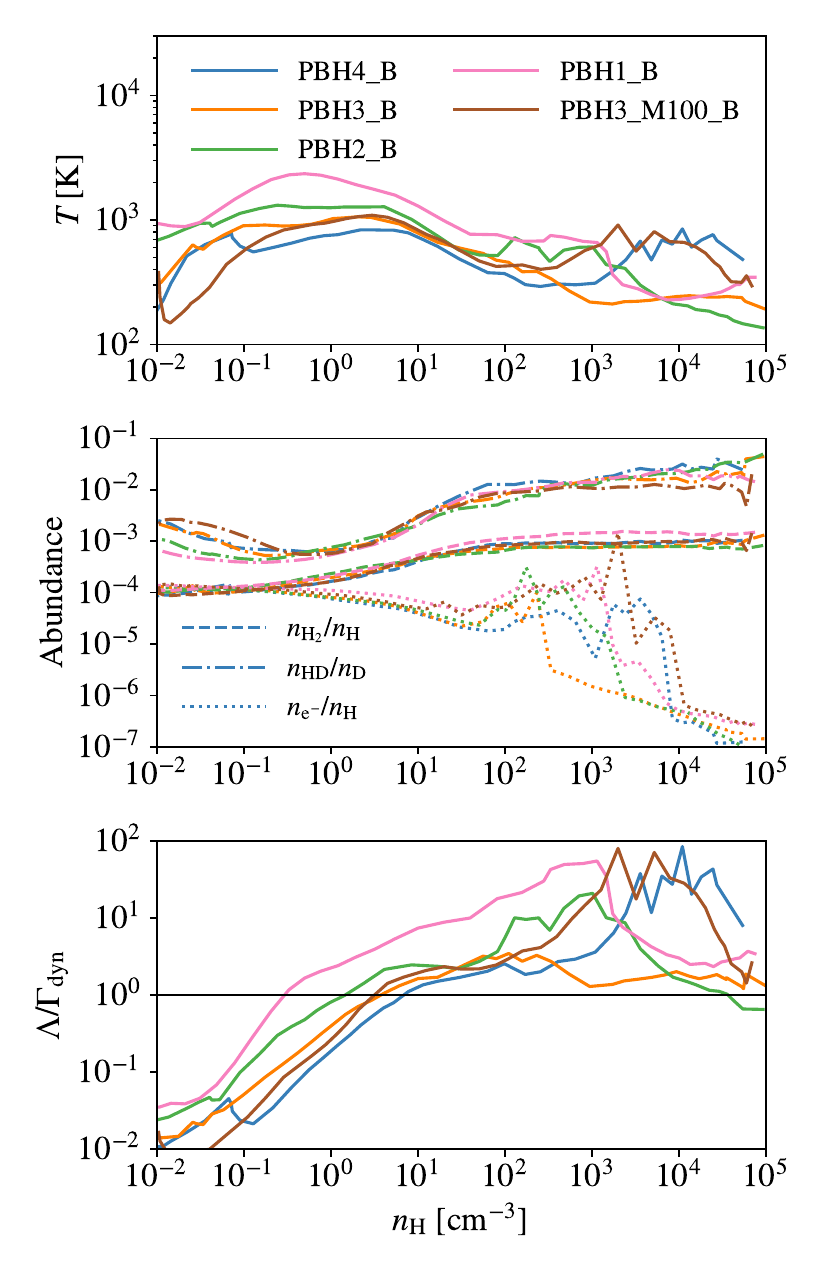}
    \vspace{-20pt}
    \caption{Same as Fig.~\ref{cpro_a} but for PBH models with $m_{\rm PBH}=33\ \rm M_{\odot}$, $f_{\rm PBH}=10^{-4}$ (blue), $10^{-3}$ (orange), 0.01 (green), 0.1 (pink) and $m_{\rm PBH}=100\ \rm M_{\odot}$, $f_{\rm PBH}= 10^{-3}$ (brown) in Case B.}
    \label{cpro_pbhb}
\end{figure}

\subsection{Parameter dependence}
\label{s3.2}
Next, we explore how the properties of star-forming clouds depend on PBH parameters with 4 models in addition to the fiducial model for both Case A and B: $m_{\rm PBH}=33\ \rm M_{\odot}$ with $f_{\rm PBH}=10^{-4}$ (\texttt{PBH4}), $0.01$ (\texttt{PBH2}), 0.1 (\texttt{PBH1}) and $m_{\rm PBH}=100\ \rm M_{\odot}$ with $f_{\rm PBH}=10^{-3}$ (\texttt{PBH3\_M100}). The matter field, density profiles and phase diagrams for Case A and B are shown in Fig.~\ref{cswb_pbha}-\ref{cpro_pbha} and Fig.~\ref{cswb_pbhb}-\ref{cpro_pbhb}, respectively. 

With $m_{\rm PBH}=33\ \rm M_{\odot}$ fixed, we have $t_{\rm col}=91.8$ (138.6), 71.0 (140.6), 115.5 (196.5), 150.2 (239.2) Myr for $f_{\rm PBH}=10^{-4},\ 10^{-3},\ 0.01$ and 0.1 in Case A (B), compared with the reference CDM value $t_{\rm col}=97.9$ (152.7) Myr (see Table~\ref{t1}). That is to say, collapse is accelerated by PBHs for $f_{\rm PBH}\le 10^{-3}$, but delayed for $f_{\rm PBH}\ge 0.01$. %The  although the collapse time is not necessarily a monotonic function of $f_{\rm PBH}$: 
$t_{\rm col}$ always increases with $f_{\rm PBH}$ in Case B, implying that the effect of BH heating is enhanced more rapidly with increasing $f_{\rm PBH}$ than the effect of PBH perturbations. However, in Case A, this trend only holds for $f_{\rm PBH}\ge 10^{-3}$, while the $f_{\rm PBH}=10^{-4}$ model (\texttt{PBH4\_A}) is very similar to the CDM case. The reason is that the host halo of collapsing gas contains no PBHs (within the virial radius) in the $f_{\rm PBH}=10^{-4}$ model of Case A (see Fig.~\ref{cswb_pbha} and \ref{dpro_pbha}), while in Case B, the host halo contains one PBH (see Fig.~\ref{cswb_pbhb} and \ref{dpro_pbhb}) that is sufficient to provide density perturbations overcoming the heating effect for slightly earlier collapse compared with the fiducial model ($f_{\rm PBH}=10^{-3}$). The host halo mass also increases with $f_{\rm PBH}$, from $1.6\ (2.1)\ \times 10^{5}\ \rm M_{\odot}$ for $f_{\rm PBH}=10^{-4}$ to $9.2\ (17)\ \times 10^{5}\ \rm M_{\odot}$ for $f_{\rm PBH}=0.1$ in Case A (B), which is a natural consequence of BH feedback/heating (see Sec.~\ref{s4.2}). When $f_{\rm PBH}=10^{-3}$ is fixed, increasing the PBH mass to $100\ \rm M_{\odot}$ accelerates collapse by $\sim 10$ (30)~Myr in Case A (B). The reason is that isocurvature perturbations from PBHs are more sensitive to $m_{\rm PBH}$ than BH feedback, given the fact that $f_{\rm PBH}$ is fixed and BH feedback tends to be self-regulated locally. %(see Sec.~\ref{s4} and Appendix~\ref{ac}). 
Besides, the density, chemical and thermal structures of the host halo for $m_{\rm PBH}=100\ \rm M_{\odot}$ are very similar to the fiducial model, showing that PBH feedback is mostly sensitive to $f_{\rm PBH}$. Therefore, below we focus on the dependence on $f_{\rm PBH}$.

In both Case A and B, the gas density profiles look very similar in the inner parsec (with $n_{\rm H}\gtrsim 10^{4}\ \rm cm^{-3}$) for all models considered here, indicating that the condition of first star formation at $n_{\rm H}\sim 10^{4}-10^{5}\ \rm cm^{-3}$ is not changed by PBHs. However, the outer part of the gas distribution ($1\ \mathrm{pc}\lesssim r\lesssim R_{\rm vir}$) becomes more clumpy, and the density profile is shallower with increasing $f_{\rm PBH}$, which may be caused by the heating from BHs and/or substructures around PBHs that can slow down the collapse. The gas density profile eventually converges to the power-law $n_{\rm H}\propto r^{-2.2}$ at $r\rightarrow R_{\rm vir}$, consistent with previous studies (e.g. \citealt{gao2007first,hirano2015primordial}). For DM, the density in the centre ($r\lesssim 1\ \rm pc$) becomes lower with increasing $f_{\rm PBH}$ and the density profile is also generally shallower. The reason is that substructures around PBHs are more tightly bound and therefore more difficult to destroy during virialization compared with their BH-less counterparts. That is to say, it is more difficult for DM to concentrate at the centre with more PBHs. Actually, it is also seen in the simulations by \citet{Inman2019} that halo profiles are cuspy around isolated PBHs but significantly less so for haloes containing multiple PBHs. The distribution of PBHs also %is approximately isothermal at large scales ($r\gtrsim 10\ \rm pc$), while at smaller scales, it 
becomes shallower with increasing $f_{\rm PBH}$, especially in the central region ($r\lesssim 10\ \rm pc$) for $f_{\rm PBH}\gtrsim 0.01$, such that only up to a few PBHs can reach $r\lesssim 2\ \rm pc$ regardless of the value of $f_{\rm PBH}$. This can be interpreted with the lower central DM density with higher $f_{\rm PBH}$, the collisional nature of the BH system and the survivorship bias for gas condensation, as we define the halo centre with the location of the densest gas particle, and gas can only condense when not significantly heated by nearby BHs.

\begin{figure}
    \centering
    \includegraphics[width=1\columnwidth]{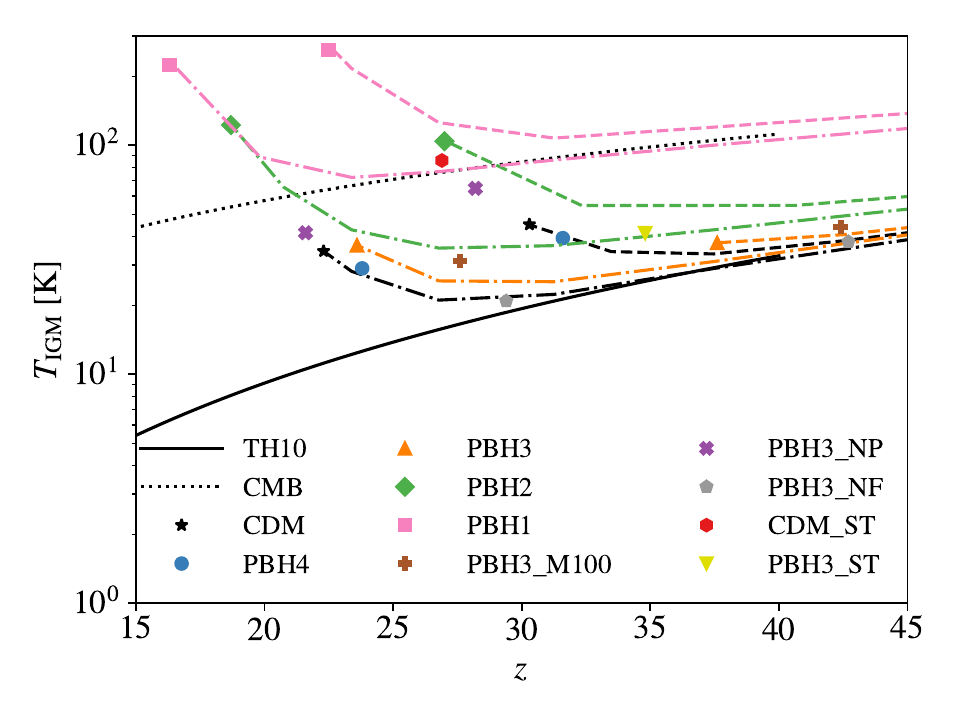}
    \vspace{-20pt}
    \caption{IGM temperature evolution. %estimated with the volume-weighted temperature of gas in the density range of $\rho_{\rm gas}\sim 0.1-10\bar{\rho}_{\rm gas}$ given $\bar{\rho}_{\rm gas}$ as the cosmic average gas density. 
    Symbols denote the values at $z_{\rm col}$ (see Table~\ref{t1}) from our simulations. We also plot evolutionary tracks for the CDM case and PBH models with $f_{\rm PBH}=10^{-3}$ (\texttt{PBH3}), 0.01 (\texttt{PBH2}) and 0.1 (\texttt{PBH1}) in dashed and dashed-dotted curves for Case A and B, respectively. For comparison, the IGM temperature in the standard $\Lambda$CDM cosmology (ignoring the effects of structure formation) from the fitting formula in \citet[TH10]{Tseliakhovich2010} is shown with the solid curve, and the CMB temperature with the dotted curve. }
    \label{TIGM_z}
\end{figure}

In the phase diagrams, it is shown that BH heating is generally stronger with higher $f_{\rm PBH}$, increasing gas temperature and cooling rates throughout the halo. The effect is particularly strong at halo outskirts $n_{\rm H}\lesssim 1\ \rm cm^{-3}$, where cooling is relatively inefficient. In the central region $n_{\rm H}\gtrsim 10^{2}\ \rm cm^{-3}$ ($r\lesssim 10\ \rm pc$), the temperature and electron abundance profiles exhibit large (temporal) variations (with respect to the standard CDM evolution track), such that the dependence on $f_{\rm PBH}$ is less clear. The reason is that this regime is sensitive to the detailed stochastic gaseous environments around a small number of PBHs. The $\rm H_{2}$ abundance at $n_{\rm H}\gtrsim 1\ \rm cm^{-3}$ generally increases with $f_{\rm PBH}$, which results from the fact that $\rm H_{2}$ formation rates become higher in regions processed by the heating and ionization of BH feedback.  %, especially for $n_{\rm H}\lesssim 1\ \rm cm^{-3}$ and $n_{\rm H}\gtrsim $

Finally, we consider the effects of PBHs on IGM temperature, which is estimated with the volume-weighted temperature of gas in the density range of $\rho_{\rm gas}\sim 0.1-10\bar{\rho}_{\rm gas}$ from our simulations, where $\bar{\rho}_{\rm gas}$ is the cosmic average gas density. The results are shown in Fig.~\ref{TIGM_z} for eight cases. We find significant heating of the IGM by $\sim 10-100\ \rm K$ (up to the CMB temperature) from PBHs with $f_{\rm PBH}\gtrsim 0.01$, which can have a great impact on the global 21-cm signal. Similar trends are also seen in previous (semi-analytical) studies \citep{Hektor2018,Mena2019,Yang2021}, which have shown that such extreme models with $f_{\rm PBH}\gtrsim 0.01$ will be ruled out if the detection of the global 21-cm absorption signal at $z\sim 17$ by EDGES \citep{nature21cm} is confirmed\footnote{Whether this signal has an astrophysical origin is still in debate (see e.g. \citealt{Singh2022}).}. For models with lower abundances of PBHs, the effects of PBHs are rather minor. Note that the IGM temperature measured from our simulations at late stages, i.e. $z\lesssim 30$ (20) for Case A (B), deviates from the standard $\Lambda$CDM case with an upturn even without PBHs. This is caused by the fact that our simulations target over-dense regions that will collapse in the end, such that virialization shocks can heat up low-density gas when nonlinear structures grow at scales comparable to the whole zoom-in region. Nevertheless, we can still see the effect of PBH heating on top of shocks. The effect is relatively weaker when perturbations of PBHs are included, which accelerate collapse at small scales and terminate the simulation earlier. This is why in some PBH models, the IGM temperature at the end of the simulation is lower than that of the CDM case.

%\subsection{IGM heating by PBHs}
%\label{s3.3}

\section{Cosmological Context}
\label{s4}
To better understand the simulation results in Sec.~\ref{s3} and their implications, we use semi-analytical models to reproduce the trends seen in our simulations. In Sec.~\ref{s4.1}, we adopt the Press-Schechter (PS) formalism \citep{Press_1974,mo2010galaxy} to calculate halo mass functions (HMFs), which are compared with simulation data to shed light on the effect of PBHs on structure formation. In Sec.~\ref{s4.2}, we derive the mass threshold above which molecular cooling is efficient for cloud collapse by comparing the cooling and free-fall timescales with a one-zone model based on \citet{trenti2009formation}, in comparison with host halo masses at $z_{\rm col}$ from simulations. The mass thresholds are then combined with HMFs to evaluate the impact of PBHs on the cosmic star formation history of Pop~III stars. Finally, we discuss the possible effects of PBHs on star formation in later (than simulated) stages in Sec.~\ref{s4.3}.

\subsection{Structure formation with PBHs}
\label{s4.1}
To calculate the HMF, $dn_{\rm h}/dM_{\rm h}$, with the PS formalism, we need to know the linear power spectrum (extrapolated to $z=0$) of DM density perturbations, which, according to the formalism in Sec.~\ref{s2.2} that includes the isocurvature perturbations of PBHs, can be written as
\begin{align}
    P(k)&=P_{\rm\Lambda CDM}(k) + P_{\rm iso}(k)\ ,\notag\\
    P_{\rm iso}(k) &= \left[f_{\rm PBH}D_{0}\right]^{2}/\bar{n}_{\rm PBH} + T^{2}_{\rm mix}(k)P_{\rm \Lambda CDM}(k)\ ,\label{e15}
\end{align}
where $P_{\rm\Lambda CDM}(k)$ is the standard power spectrum (for the adiabatic mode) in $\rm \Lambda CDM$ cosmology\footnote{We use the $\rm \Lambda CDM$ power spectrum measured by \citet{planck} from the \textsc{python} package \href{https://bdiemer.bitbucket.io/colossus/index.html}{\textsc{colossus}} \citep{colossus}.},  $T_{\rm mix}(k)$ is the transfer function meant to capture the effect of mode mixing, $\bar{n}_{\rm PBH}=f_{\rm PBH}\frac{3H_{0}^{2}}{8\pi G}(\Omega_{m}-\Omega_{b})/m_{\rm PBH}$ is the cosmic (co-moving) number density of PBHs, and $D_{0}=D(a=1)=\left[1+3\gamma/(2a_{-}a_{\rm eq})\right]^{a_{-}}-1$ is the growth factor of isocurvature perturbations evaluated at $z=0$ (see Equ.~\ref{e6}). 

\begin{figure}
    \centering
    \includegraphics[width=1\columnwidth]{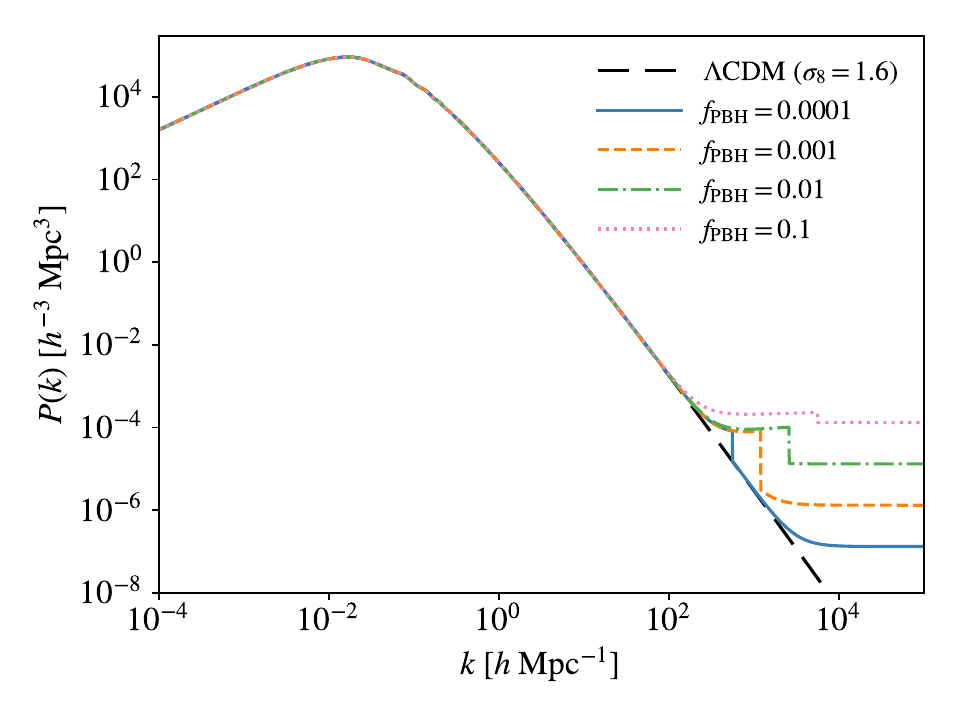}
    \vspace{-20pt}
    \caption{Power spectra of the DM density field for 4 PBH models with $m_{\rm PBH}=33\ \rm M_{\odot}$, $f_{\rm PBH}=10^{-4}$ (solid), $10^{-3}$ (dashed), 0.01 (dashed-dotted) and 0.1 (dotted) based on the formalism in Equ.~\ref{e15} and \ref{e16}, in comparison with the standard $\Lambda$CDM power spectrum (long-dashed) measured by \citet{planck}, which is enhanced by a factor of $\sim 4$ to be consistent with Case B simulations.}
    \label{powspec}
\end{figure}

Next, we need to evaluate $T_{\rm mix}(k)$. In principle, the effect of mode mixing is only important at intermediate scales where the gravitational fields of PBHs do affect clustering of DM but are not strong enough to completely disrupt the structures generated by the adiabatic mode, while larger (smaller) scales will be dominated by the adiabatic mode (discreteness noise), i.e. $T^{2}_{\rm mix}(k)\rightarrow 0$ as $k\rightarrow 0$ or $k\rightarrow \infty$. Heuristically, we find that  
\begin{align}
T^{2}_{\rm mix}(k)=\begin{cases} f_{\rm PBH}D_{0}^{2}D_{\rm ad,0}^{-1}(k/k_{\rm PBH})^{3}\ ,\quad &k\le 3k_{\rm PBH} \\
0\ ,\quad &k>3k_{\rm PBH}
\end{cases}\label{e16}
\end{align}
can well reproduce the trends seen in our simulations (of default initial conditions), where $k_{\rm PBH}=(2\pi^{2}\bar{n}_{\rm PBH})^{1/3}$ denotes the characteristic scale below which the isocurvature mode dominates and $D_{\rm ad,0}=D_{\rm ad}(a=1)/D_{\rm ad}(a=a_{\rm eq})$, given $D_{\rm ad}(a)$ as the growth factor of the adiabatic mode \citep{mo2010galaxy}. Substituting the expression for $T_{\rm mix}(k)$ in Equ.~\ref{e15}, we obtain the power spectra for 4 PBH models with $m_{\rm PBH}=33\ \rm M_{\odot}$, $f_{\rm PBH}=10^{-4}$, $10^{-3}$, 0.01 and 0.1, as shown in Fig.~\ref{powspec}, together with the standard $\Lambda$CDM power spectrum. Here we have increased $P_{\rm \Lambda CDM}(k)$ by a factor of $(1.6/0.8159)^{2}\sim 4$ to be consistent with our Case B simulations, where adiabatic perturbations are enhanced to $\sigma_{8}=1.6$, compared to the cosmic mean $\sigma_{8}=0.8159$ \citep{planck}.

\begin{figure}
    \centering
    \includegraphics[width=1\columnwidth]{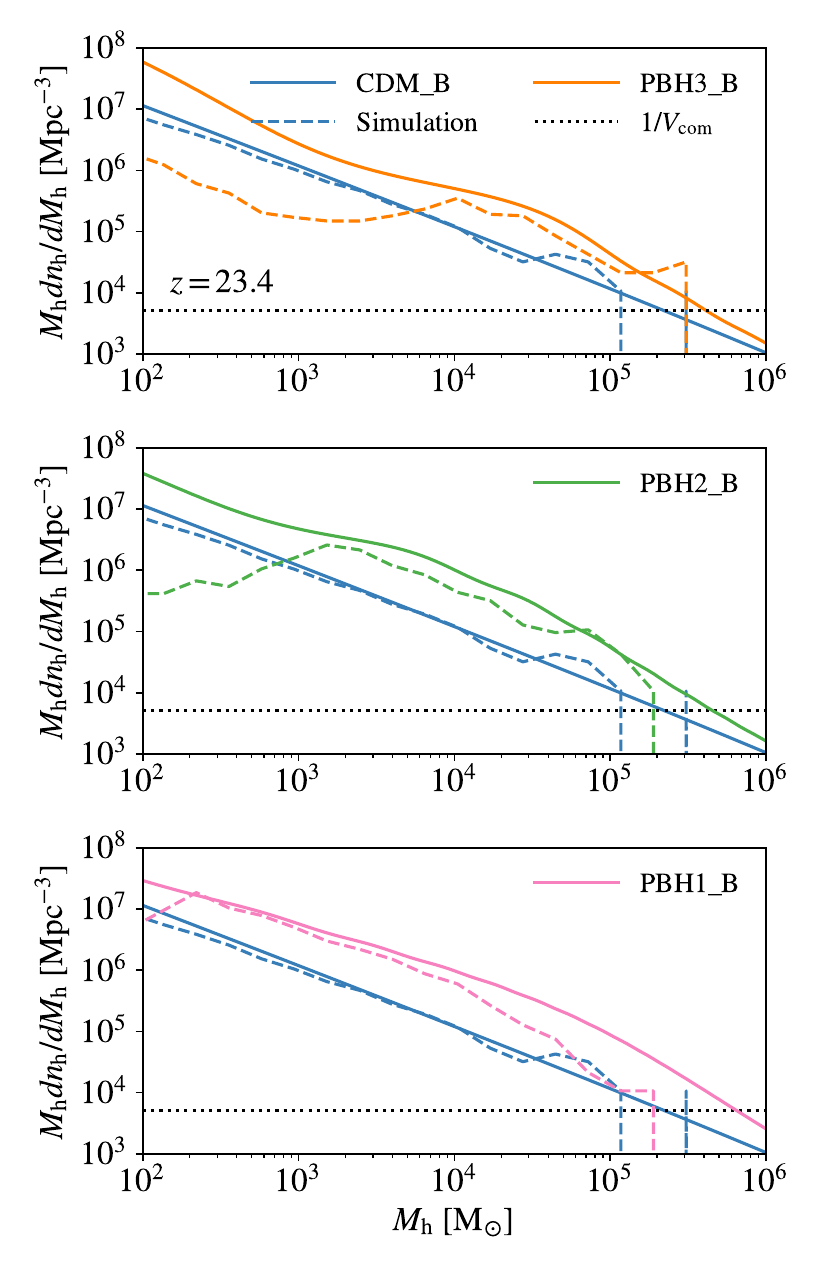}
    \vspace{-20pt}
    \caption{Halo mass functions at $z=23.4$, for the reference CDM model (blue) and PBH models with $m_{\rm PBH}=33\ \rm M_{\odot}$, $f_{\rm PBH}=10^{-3}$ (orange, top), 0.01 (middle, green) and 0.1 (bottom, pink). Predictions from the PS formalism based on Equ.~\ref{e15} and \ref{e16} with (enhanced) power spectra (see Fig.~\ref{powspec}) are shown with solid curves, while those measured in the relevant Case B simulations are shown with dashed curves. The horizontal dotted line is $1/V_{\rm com}$, whose intersection with the HMF curve approximately corresponds to the maximum halo mass expected to show up in the simulation. }
    \label{hmf}
\end{figure}

From the power spectrum, we derive the HMFs under the influence of PBHs with the PS formalism that includes corrections for ellipsoidal dynamics \citep{mo2010galaxy}. As an example, the results at $z=23.4$ are shown in Fig.~\ref{hmf} for the reference CDM case and PBH models with $m_{\rm PBH}=33\ \rm M_{\odot}$, $f_{\rm PBH}=10^{-3}$, 0.01 and 0.1, which are compared with the HMFs calculated from Case B simulation data based on the \href{https://bitbucket.org/gfcstanford/rockstar/src}{\textsc{rockstar}} halo finder \citep{behroozi2012rockstar}. The semi-analytical predictions agree well with simulations for the CDM case. {However, in PBH models, the agreement is only marginally good (within a factor of 2) for haloes of $M_{\rm h}\sim 10^{4}-10^{5},\ 10^{3}-10^{5},\ \text{and } 200-10^{4}\ \rm M_{\odot}$ in the simulations for $f_{\rm PBH}=10^{-3}$, 0.01 and 0.1, respectively, while the abundances of low-mass haloes are significantly underestimated. Besides, the abundance of massive ($M_{\rm h}\gtrsim 10^{4}\ \rm M_{\odot}$) haloes is also underestimated in the case of $f_{\rm PBH}=0.1$. These features} can be understood with the fact that at $z\lesssim 100$, individual PBHs have already been surrounded by tightly-bound DM haloes before falling into larger structures. On the one hand, these BH-induced haloes can disrupt their BH-less counterparts of smaller or comparable masses, reducing the abundance of low-mass haloes. This mechanism only works at small scales where on average each halo contains less then one BH, which explains why the effect becomes weaker with increasing $f_{\rm PBH}$. On the other hand, as shown in Sec.~\ref{s3.2}, substructures around BHs can impede the assembly of large haloes containing multiple BHs, an effect that becomes stronger with increasing $f_{\rm PBH}$. In other words, the large-scale flows can be disrupted by Poisson noise from PBHs when $f_{\rm PBH}$ is high (i.e. approaching the `Poisson' limit).

\begin{figure}
    \centering
    \includegraphics[width=1\columnwidth]{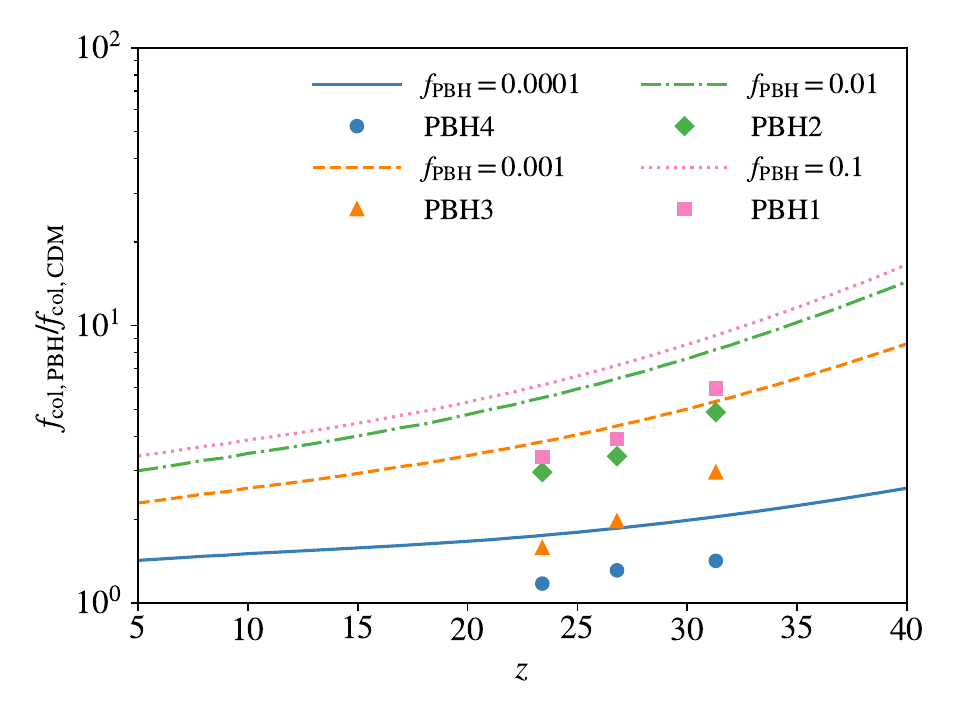}
    \vspace{-20pt}
    \caption{Ratio of the collapsed mass in the halo mass range $M_{\rm h}\sim 64\ \mathrm{M_{\odot}}-M_{\rm mol}$ in PBH and CDM models, where $M_{\rm mol}$ is the mass threshold for efficient molecular cooling in the standard CDM case (with no baryon-DM streaming motion) from \citet[TS09]{trenti2009formation}. Predictions from the PS formalism under (enhanced) power spectra (see Fig.~\ref{powspec}) for PBH models with $m_{\rm PBH}=33\ \rm M_{\odot}$, $f_{\rm PBH}=10^{-4}$, $10^{-3}$, 0.01 and 0.1 are shown with the solid, dashed, dashed-dotted and dotted curves respectively. The relevant values obtained from Case B simulations at $z=23.4$, 26.8 and 31.3 are shown with filled circles, triangles, diamonds and squares for $f_{\rm PBH}=10^{-4}$, $10^{-3}$, 0.01 and 0.1.}
    \label{fcol}
\end{figure}

We also calculate the collapsed mass fraction of haloes in the mass range $M_{\rm h}\sim 64\ \mathrm{M_{\odot}}-M_{\rm mol}$, where our simulations of limited volumes have {marginally} good statistics of haloes. Here $M_{\rm mol}$ is the mass threshold for efficient molecular cooling in the standard CDM case from \citet[see their equ.~9]{trenti2009formation}. In Fig.~\ref{fcol}, we present our results in terms of the ratio of the collapsed mass fractions in PBH and CDM models for $m_{\rm PBH}=33\ \rm M_{\odot}$, $f_{\rm PBH}=10^{-4}$, $10^{-3}$, 0.01 and 0.1, where the results from Case B simulations are also shown for comparison. The semi-analytical approach and simulations produce similar trends that the collapsed mass fraction is increasingly enhanced by PBHs at higher redshifts and with higher $f_{\rm PBH}$, and the effect of PBHs tends to saturate at $f_{\rm PBH}\gtrsim 0.01$. This is consistent with intuition and previous studies (see e.g. fig.~2 in \citealt{Cappelluti2022}). However, the ratio for $M_{\rm h}\sim 64\ \mathrm{M_{\odot}}-M_{\rm mol}$ is overestimated by a factor of $\sim 2$ with the PS formalism compared with simulations, likely due to the aforementioned nonlinear effects of haloes surrounding individual BHs. %For haloes with $M_{\rm h}\sim 1-10 \times M_{\rm mol}$ that potentially form Pop~III stars, the semi-analytical model predicts enhancement in the collapsed mass fraction by a factor of $\sim 2-10$ at $z\sim 10-40$ for the fiducial PBH model with $f_{\rm PBH}=10^{-3}$, and up to $\sim 30$ for $f_{\rm PBH}=0.1$. This indicates that PBH models allowed by existing observational constraints can significantly enhance early star formation. Nevertheless, the mapping from collapsed mass fraction to star formation rate density and observables of the first stars can be complex, considering that feedback from BHs may change the mass threshold and efficiency of star formation, as well as the properties of stars. We address these issues in the following subsections.

{
Given the mass functions, we further evaluate the radiation backgrounds produced by PBH accretion at $z> 6$ in collapsed structures for $m_{\rm PBH}=33\ \rm M_{\odot}$. We focus on the Lyman-Werner (LW, $h\nu\sim 11.2-13.6\ \rm eV$) and X-ray photons. The former can dissociate $\rm H_{2}$ and $\rm HD$, and, therefore, reduce cooling and delay star formation (e.g. \citealt{safranek2012star,Schauer2021}). The latter can heat and ionize the IGM, which may not affect Pop~III star formation significantly \citep{Hummel2015,Park2021} but can have a great impact on the 21-cm signal (e.g. \citealt{Mirocha2019,fialkov2019,schauer2019constraining,21cm2020}). In this calculation we assume isothermal distributions of gas and BHs\footnote{ It is found in simulations that the gas distribution in high-$z$ atomic-cooling haloes follows approximately $\rho\propto r^{-2}$ at $r\gtrsim 0.003\ \rm pc$ (see equ. 2 in \citealt{Safarzadeh2020}). Similar density profiles are also seen in our simulations for molecular-cooling minihaloes. Therefore, isothermal distribution is a good approximation. Here we have also truncated the BH distribution at $r=1\ \rm pc$ for conservative estimates, since no BHs travel into the central parsec in our simulations (see Sec.~\ref{s3.2}).} and Bondi accretion with $\tilde{v}\sim \sqrt{GM_{\rm h}/R_{\rm vir}}$ (see Equ.~\ref{e22}). The radiation output from any halo with a given mass and redshift can be obtained by integrating the BH spectra times BH density profile over the halo volume, where the spectra are computed based on the model from \citet{Takhistov2022} as explained in Sec.~\ref{s2.3.3}. The outputs from individual haloes are then combined with the halo mass function to produce the radiation background (see e.g. \citealt{schauer2019constraining,Cappelluti2022}). We only consider haloes containing at least one PBH ($M_{\rm h}\gtrsim m_{\rm PBH}/f_{\rm PBH}$) with virial temperatures above 100~K and $M_{\rm h}\lesssim 10^{10}\ \rm M_{\odot}$, as the abundance and gas properties of smaller haloes are uncertain and more massive haloes are rare at $z\gtrsim 6$ whose density structures can also be complex (with e.g. central massive BHs).

We find that the background intensity of LW radiation $J_{\rm 21,bg}$ (in units of $10^{-21}\ \rm erg\ s^{-1}\ cm^{-2}\ Hz^{-1}\ sr^{-2}$) produced by PBHs is always below\footnote{We have $J_{\rm 21, bg}\lesssim 10^{-7},\ 10^{-4}\ \text{and }0.6$ for $f_{\rm PBH}=10^{-4},\ 10^{-3}$ and 0.01 at $z\sim 6-40$.} the critical value $J_{\rm 21}\sim 1$ at which the destruction rate of $\rm H_{2}$ equals the formation rate, except for the extreme case with $f_{\rm PBH}=0.1$ where $J_{\rm 21, bg}$ is above 1 at $z\gtrsim 25$, reaching $\sim 8$ at $z\sim 40$. Therefore, the LW background from PBH accretion in collapsed haloes only has minor impact on Pop~III star formation in stellar-mass PBH models allowed by current observational constraints. However, the signals in X-rays are stronger. The cumulative X-ray background intensity from PBHs in the $0.5-2$~keV band (for an observer at $z=0$) reaches $J_{[0.5-2\ \rm keV]}\sim 5.5\times 10^{-14}$, $4.4\times 10^{-12}$, $1.1\times 10^{-10}$ and $1.6\times 10^{-9}\ \rm erg\ s^{-1}\ cm^{-2}\ deg^{-2}$ at $z=6$ for $f_{\rm PBH}=10^{-4}$, $10^{-3}$, 0.01 and 0.1, respectively. When compared with the observed extragalactic (unresolved) cosmic X-ray background (CXB) $\sim 8\ (3) \times 10^{-12}\ \rm erg\ s^{-1}\ cm^{-2}\ deg^{-2}$ \citep{Cappelluti2017}, our results can rule out PBH models with $f_{\rm PBH}\gtrsim 10^{-3}$. %(for $m_{\rm PBH}\sim 30\ \rm M_{\odot}$). 
This implies that PBH accretion in collapsed structures can contribute significantly to the CXB in addition to accretion in the IGM (e.g. \citealt{Hasinger2020,Cappelluti2022}) and the CXB may place stronger constraints on PBH parameters even than Galactic X-ray observations \citep{Inoue2017,Manshanden2019}. We will investigate the effects of X-ray emission from PBHs in more detail in future work. %(see e.g. \citealt{Park2021}).
}

\subsection{Mass threshold for star formation}
\label{s4.2}
\begin{comment}
BH heating rate (thermal energy deposited by BHs per unit time per baryon):
\begin{align}
    \Gamma_{\rm BH}&=4\pi\int_{r_{1}}^{r_{2}}n_{\rm BH}(r)\epsilon_{\rm heat}(\dot{m})c^{2}\dot{m}r^{2}dr\times \mu m_{\rm H}/M_{\rm c} \\&= 4\pi R_{vir}^3\int_{0.01}^{0.1}n_{\rm BH}(\frac{r}{R_{vir}})\epsilon_{\rm heat}(\dot{m})c^{2}\dot{m}(\frac{r}{R_{vir}})^{2}d(\frac{r}{R_vir})\times \frac{\mu m_{\rm H}}{M_{\rm c}},\\ \text{given} \notag \\
    \dot{m}&\equiv\dot{m}(r)=\dot{m}(n_{\rm gas}(r),v_{\rm BH}(r),m_{\rm PBH})\ ,\notag\\
    M_{\rm c}&=4\pi\int_{0}^{r_{2}}n_{\rm gas}(r)\times\mu m_{\rm H}r^{2}dr\approx \frac{r_{2}\Omega_{b}}{R_{\rm vir}\Omega_{m}} M_{\rm h}\ ,\notag\\
    n_{\rm BH}&\propto r^{-2}\ ,\quad 4\pi\int_{r_{1}}^{R_{\rm vir}}n_{\rm BH}(r)r^{2}dr\approx \frac{f_{\rm PBH}M_{\rm h}}{m_{\rm PBH}}\ ,\notag
\end{align}
where $r_{1}=0.01 R_{\rm vir}$, $r_{2}=0.1R_{\rm vir}$, $v_{\rm BH}(r)\sim \sqrt{GM_{\rm h}/R_{\rm vir}}$, $\mu\simeq 1.22$ is the mean molecular weight of primordial gas, and $m_{\rm H}$ is proton mass.
\end{comment}

We use the Rees-Ostriker-Silk cooling criterion $t_{\mathrm{cool}}\le t_{\mathrm{ff}}$ \citep{rees1977cooling, silk1977} to determine the mass threshold $M_{\rm mol}$ for efficient molecular cooling (see also \citealt{Sullivan2018,Liu2019}). For simplicity, we only consider $\rm H_{2}$ as the dominant coolant, whose maximum abundance in a halo of a virial temperature $T_{\rm vir}$ is approximately \citep{trenti2009formation}
\begin{align}
    \hat{x}_{\rm H_{2},\max}(T_{\rm vir})\simeq 3.5\times 10^{-4}\left(\frac{T_{\rm vir}}{1000\ \rm K}\right)^{1.52}\ ,
\end{align}
in the CDM case. We further consider the (mild) enhancement of $\rm H_{2}$ abundance by PBH heating and ionization, with a simple power-law fit to Case A simulation results for $m_{\rm PBH}=33\ \rm M_{\odot}$, $f_{\rm PBH}\sim 10^{-4}-0.1$ (see Fig.~\ref{cpro_pbha}):
\begin{align}
    \frac{x_{\rm H_{2},\max}(T_{\rm vir},f_{\rm PBH})}{\hat{x}_{\rm H_{2},\max}(T_{\rm vir})}=\max\left[1,3\left(\frac{f_{\rm PBH}}{0.1}\right)^{0.15}\right].
\end{align}
Now, given the halo mass $M_{\rm h}$ and redshift $z$, the cooling and free-fall timescales can be written as
\begin{align}
    t_{\rm cool}&=\frac{(3/2)k_{B}T_{\rm vir}}{\Lambda(T_{\rm vir},x_{\rm H_{2}},n) - \Gamma (m_{\rm PBH}, f_{\rm PBH}, \rho_{\rm gas}, \tilde{v})}\ ,\\
    t_{\rm ff}&=\sqrt{\frac{3\pi}{32G\rho_{\rm gas}}}\ ,\quad \rho_{\rm gas}=\mu m_{\rm H}n=\Delta\bar{\rho}_{\rm gas}\ ,
\end{align}
where $\mu\simeq 1.22$ is the mean molecular weight of primordial gas, $m_{\rm H}$ is proton mass, $\Delta$ is a free parameter that sets the characteristic density of gas in units of the cosmic average $\bar{\rho}_{\rm gas}$, $\Lambda$ is the cooling rate (per baryon), and $\Gamma$ is the BH heating rate
\begin{align}
    \Gamma=\frac{f_{\rm PBH}\mu m_{\rm H}(\Omega_{m}-\Omega_{b})}{m_{\rm PBH}\Omega_{b}}P(m_{\rm PBH}, \rho_{\rm gas}, \tilde{v})\ ,
\end{align}
in which $P\equiv\epsilon_{r}L_{\rm BH}$ is the heating power per BH, given the coupling efficiency $\epsilon_{r}$ and BH luminosity $L_{\rm BH}$ derived from the same BH accretion and feedback model used in the simulations (see Sec.~\ref{s2.3.2} and \ref{s2.3.3}), and we estimate the characteristic velocity between PBHs and gas as
\begin{align}
    \tilde{v}\sim \sqrt{\frac{GM_{\rm h}}{R_{\rm vir}}}\sim {5.4\ \rm km\ s^{-1}} \left(\frac{M_{\rm h}}{10^{6}\ \rm M_{\odot}}\right)^{\frac{1}{3}}\left(\frac{21}{1+z}\right)^{\frac{1}{2}}\ .\label{e22}%\notag
\end{align}
At last, we set the overdensity parameter as $\Delta=1300$ to reproduce the results in \citet[see their equ.~9]{trenti2009formation} for the CDM case ($f_{\rm PBH}$=0). The mass threshold $M_{\rm mol}$ is defined by $t_{\mathrm{cool}}= t_{\mathrm{ff}}$.

\begin{figure}
    \centering
    \includegraphics[width=1\columnwidth]{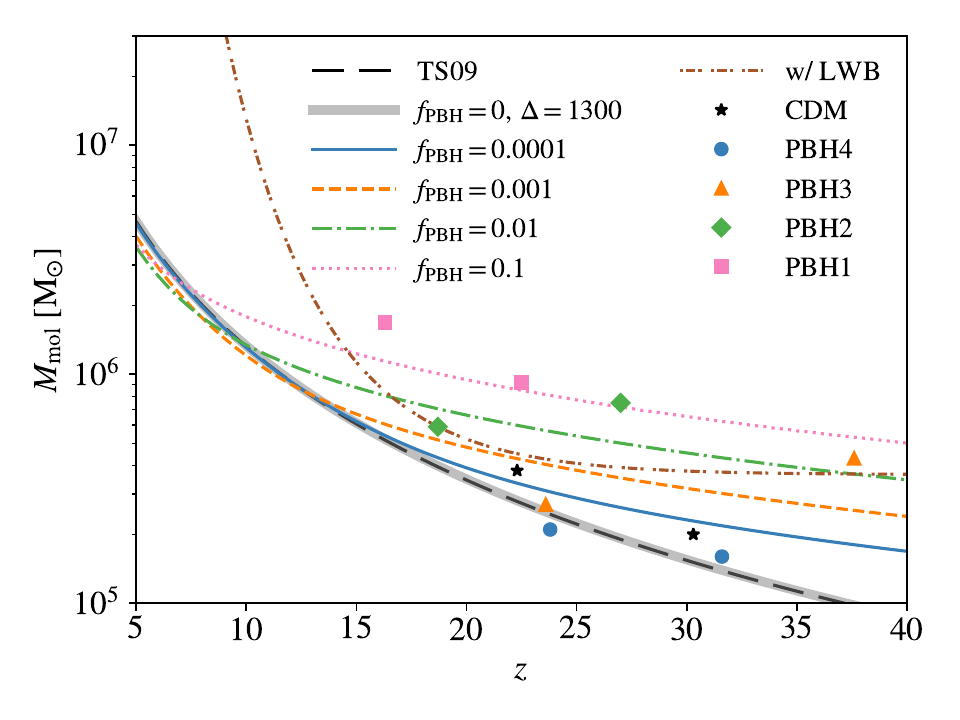}
    \vspace{-20pt}
    \caption{Mass threshold of efficient cooling, predicted by the one-zone model, for PBH models with $m_{\rm PBH}=33\ \rm M_{\odot}$, $f_{\rm PBH}=10^{-4}$ (solid), $10^{-3}$ (dashed), 0.01 (dashed-dotted) and 0.1 (dotted curves), respectively. The corresponding halo masses at the moment of collapse from simulations are shown with filled circles, triangles, diamonds and squares. For comparison, we plot 2 versions of the mass threshold in the CDM case: the prediction of our one-zone model (thick gray), ignoring the effect of Lyman-Werner radiation background (LWB), which agrees perfectly with \citet[TS09, long-dashed]{trenti2009formation}, and the mass threshold regulated by a LW background from \citet{Greif2006}, based on the fitting formula in \citet[dotted-dashed-dotted]{Schauer2021}.}
    
    %in comparison with the mass threshold for efficient molecular cooling in the standard CDM case (with no baryon-DM streaming motion) from \citet[TS09]{trenti2009formation}.  Note: We may show some predictions from the top-hat model here.}
    \label{mh_z}
\end{figure}

Fig.~\ref{mh_z} shows the mass threshold as a function of $z$ for $m_{\rm PBH}=33\ \rm M_{\odot}$, $f_{\rm PBH}=0,\ 10^{-4},\ 10^{-3}$, 0.01 and 0.1, compared with the host halo masses at the moment of collapse in simulations. Clearly, the threshold mass increases with $f_{\rm PBH}$, {up to $\sim 3-5$ times the CDM value at $z\sim 20-40$} for $f_{\rm PBH}=0.1$. The increase is more significant at higher redshifts and the threshold mass converges to the CDM case at low $z$. The host halo masses in Case A simulations with $f_{\rm PBH}\ge 10^{-3}$ are all above the predicted threshold masses. However, haloes collapse at lower masses than the threshold in Case B simulations with $f_{\rm PBH}\le 0.01$ and the Case A run for $f_{\rm PBH}=10^{-4}$. The former can be explained by the specific assembly history of the zoom-in region (see Sec.~\ref{s3}). The latter can be understood with the fact that there is no PBH in the host halo in that run. Given the small sample size of simulated haloes, it is difficult to quantitatively evaluate the accuracy of our semi-analytical model for $M_{\rm mol}$, which does reproduce the general trend in our simulations.

\begin{figure}
    \centering
    \includegraphics[width=1\columnwidth]{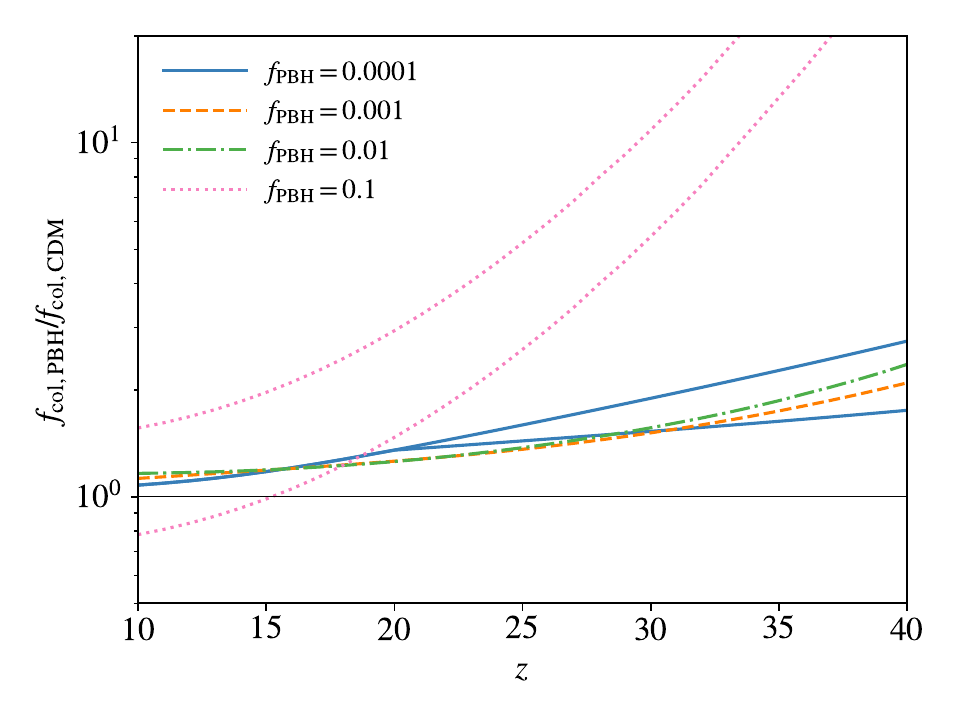}
    \vspace{-20pt}
    \caption{Ratio of the collapsed mass for haloes in the mass range $M_{\rm h}\sim M_{\rm mol}-10^{8}\ \rm M_{\odot}$ that typically host Pop~III stars. Here the mass threshold for efficient cooling $M_{\rm mol}$ depends on PBH parameters. For the $f_{\rm PBH}=0.1$ model, we show a conservative estimate (the lower curve) that takes into account the delayed assembly of large haloes containing multiple BHs (see Sec.~\ref{s4.1}). For $f_{\rm PBH}=10^{-4}$, we also provide a lower limit considering the fact that only a fraction of star-forming haloes contain PBHs at high-$z$. }
    \label{fcol_sf}
\end{figure}

Once $M_{\rm mol}\equiv M_{\rm mol}(z,f_{\rm PBH},m_{\rm PBH})$ is known, we can calculate the collapsed mass fraction in haloes with $M_{\rm h}\sim M_{\rm mol}-10^{8}\ \rm M_{\odot}$ as typical hosts %\footnote{The upper bound ($10^8 \ \rm M_\odot$) is the typical halo mass for the first galaxies above which haloes are mostly metal enriched and thus do not form Pop~III stars \citep[e.g.][]{bromm2011first}.} 
of Pop~III stars, which, to the first order, is proportional to the star formation rate (or stellar mass) density of Pop~III stars:
\begin{align}
    f_{\rm col}=\int_{M_{\rm mol}}^{10^{8}\ \rm M_{\odot}}\frac{dn_{\rm h}}{dM_{\rm h}}M_{\rm h}dM_{\rm h}/\bar{\rho}_{m}\ ,\label{e23}
\end{align}
where $\bar{\rho}_{m}$ is the average density of matter in the Universe. In Fig.~\ref{fcol_sf}, we show the resulting ratio of the collapsed mass fractions in PBH and CDM models with $m_{\rm PBH}=33\ \rm M_{\odot}$, $f_{\rm PBH}=10^{-4}$, $10^{-3}$, 0.01 and 0.1, now for the cosmic average with $\sigma_{8}=0.8159$ (not enhanced). In addition to the direct predictions of Equ.~\ref{e23}, we also estimate the lower limits for the $f_{\rm PBH}=0.1$ model in which the collapse mass fraction is reduced by a factor of 2 to take into account the nonlinear dynamics that delays the assembly of large haloes containing multiple BHs (see Sec.~\ref{s4.1}), and {for the $f_{\rm PBH}=10^{-4}$ model, where the collapse mass fraction is $f_{\rm col,PBH}=f_{\rm col,raw}P_{\rm PBH}+1-P_{\rm PBH}$, given $f_{\rm col,raw}$ as the raw output of Equ.~\ref{e23} and $P_{\rm PBH}=\min\{1,\ f_{\rm PBH}M_{\rm mol}(\Omega_{m}-\Omega_{b})/\Omega_{m}/m_{\rm PBH}]\}$ as the fraction of star-forming haloes containing PBHs, which is less than 1 at high-$z$. }

\begin{figure}
    \centering
    \includegraphics[width=1\columnwidth]{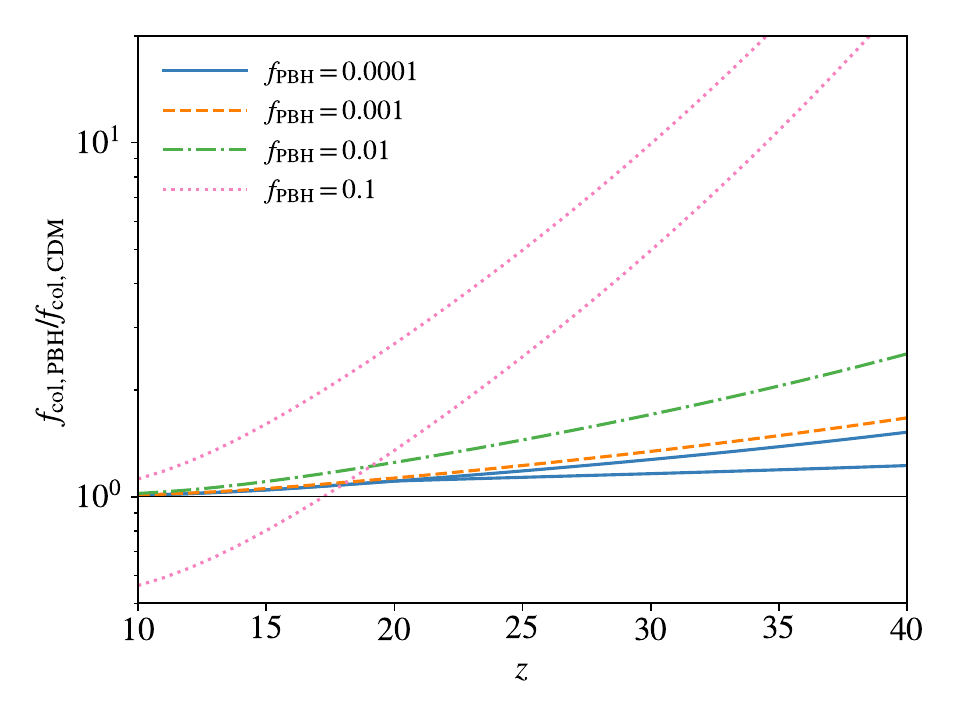}
    \vspace{-20pt}
    \caption{Same as Fig.~\ref{fcol_sf}, but now including the effects of streaming motion and LW background on the mass threshold.}
    \label{fcol_sf_ext}
\end{figure}

{Interestingly, when the increase of $M_{\rm mol}$ by BH feedback is taken into account, the collapsed mass fraction in the PBH model is similar (within a factor of $\sim 2$)\footnote{For for $f_{\rm PBH}\sim 10^{-4}-0.01$, if we ignore mode mixing in the power spectrum (Equ.~\ref{e15}), the fraction of collapse mass will be significantly reduced at $z\sim 15-40$ by up to {a factor of $\sim 6$}. On the other hand, if we ignore the increase of $M_{\rm mol}$ by BH feedback, the ratio will be much higher, reaching $\sim 3-10$ at $z\sim 15-30$ for $f_{\rm PBH}\gtrsim 10^{-3}$.} to that of the CDM case for $f_{\rm PBH}\sim 10^{-4}-0.01$ and also insensitive to $f_{\rm PBH}$. If we take the lower limit of the $f_{\rm PBH}=10^{-4}$ model, $f_{\rm col}$ slightly increases with $f_{\rm PBH}$ at $z\gtrsim 30$, which is different from the trend seen in our zoom-in simulations that $z_{\rm col}$ decreases with $f_{\rm PBH}$. 
%The collapsed mass fraction is increased by $\lesssim 60\%$ with $f_{\rm PBH}=10^{-3}- 0.01$, and even slightly ($\lesssim 10\%$) reduced at $z\sim 20-35$ for $f_{\rm PBH}=0.1$. We also find that $f_{\rm col}$ decreases with $f_{\rm PBH}$ at $z\gtrsim 15$, consistent with the trend seen in our simulations that $t_{\rm col}$ generally increases with $f_{\rm PBH}$. %This is inconsistent with the trend seen in our zoom-in simulations of over-dense regions for $f_{\rm PBH}=10^{-4}$ and $10^{-3}$. The reason is that structure formation is accelerated more in regions with higher initial overdensities (from the adiabatic mode) and stronger mode mixing. %In the case of $f_{\rm PBH}\sim 10^{-4}-0.01$, star formation is accelerated by PBHs in the simulated over-dense regions, but slightly delayed on cosmic average as predicted by the semi-analytical model. 
Besides, in the extreme case with $f_{\rm PBH}=0.1$, the collapsed mass fraction is significantly higher than the CDM value, by up to a factor of $\sim 10$ at $z\lesssim 30$. That is to say, the semi-analytical model predicts significant acceleration of star formation for $f_{\rm PBH}= 0.1$, while star formation is delayed in the corresponding zoom-in simulations. 
%, but slightly below 1 at $z\lesssim 20$. %Without the correction for nonlinear dynamics, the ratio is always above 1. 
%This is also different from simulations that show delayed star formation. 
Such different trends from the semi-analytical model and simulations can be understood with the fact that structure formation is accelerated more in regions with higher initial overdensities and stronger mode mixing. Our simulation volumes are too small to capture larger-scale perturbations that lead to higher initial overdensities and increases the abundance of haloes above $M_{\rm mol}$. }

In the above calculation, we have ignored higher order and external effects that also regulate the host haloes of Pop~III stars, such as halo assembly history, mass and redshift dependence of star formation efficiency, metal enrichment, radiative feedback, cosmic rays and streaming motion between DM and gas (see e.g. {\citealt{yoshida2003simulations,oshea2008,stacy2011effect,Fialkov2013,Johnson2013FBY,Smith2015,Hummel2015,Hummel2016,Anna2019,Schauer2021,Park2021}}). For instance, it will be shown in Appendix~\ref{ab} that streaming motion also delays star formation in the presence of PBHs. To take such effects into account, as an upper limit, we replace $M_{\rm mol}$ with $\max(M_{\rm mol},M_{\rm crit})$, where $M_{\rm crit}$ is the minimum mass of star-forming haloes under a typical streaming motion velocity $v_{\rm b\chi}=0.8\sigma_{\rm b\chi}$ and a LW background $J_{21}=10^{2-z/5}$ \citep{Greif2006}, based on the fitting formulae in \citet[see thier equ.~11-13]{Schauer2021}. The results are presented in Fig.~\ref{fcol_sf_ext}. Now we find that for $f_{\rm PBH}\sim 10^{-4}-0.01$, the ratio always increases with redshift and $f_{\rm PBH}$ and remains above 1, but still below 2 at $z\lesssim 30$. If mode mixing is ignored, i.e. $T_{\rm mix}(k)\equiv 0$ in Equ.~\ref{e15}, the ratio becomes 1 for $f_{\rm PBH}\lesssim 10^{-3}$ and is reduced by a factor of 3 for $f_{\rm PBH}=0.01$.

%In general, we find that PBHs may not necessarily enhance Pop~III star formation as assumed in previous studies, and the impact of PBHs on the cosmic Pop~III star formation history is likely small for stellar-mass PBHs ($m_{\rm PBH}\sim 10-100\ \rm M_{\odot}$) allowed by existing observational constraints with $f_{\rm PBH}\lesssim 10^{-3}-0.01$ (see e.g. \citealp{Ali-Haimoud2017,Poulin2017,Hektor2018,Carr2021,Hutsi2021}). However, PBHs do shift Pop~III star formation to more massive haloes with possibly stronger spacial clustering.

\subsection{Fate of collapsing clouds}
\label{s4.3}
As shown in Sec.~\ref{s3}, our simulations indicate that the presence of stellar-mass PBHs with $m_{\rm PBH}\sim 30-100\ \rm M_{\odot}$ and $f_{\rm PBH}\lesssim 0.1$ cannot prevent primordial gas clouds from collapse in minihaloes with efficient cooling. Then one remaining question is what will happen eventually in such collapsing clouds, regarding the impact of PBHs in the later stages.

At the end of the simulation when the cloud has collapsed to reach $n_{\rm H}\sim 10^{5}\ \rm cm^{-3}$, there is a central dense ($n_{\rm H}\gtrsim 10^{4}\ \rm cm^{-3}$) clump of  size $R\sim 1\ \rm pc$ and mass $M\sim 10^{3}\ \rm M_{\odot}$ in (mostly) gas. The free-fall timescale of this central clump is $t_{\rm ff}\sim 0.5\ \rm Myr$, {which can be regarded as the upper limit of} the timescale of protostar formation. Considering the accretion timescale of Pop~III protostars $t_{\rm acc}\sim 0.01-0.1\ \rm Myr$ (see e.g. \citealt{greif2012formation,stacy2013constraining,susa2014mass,machida2015accretion,stacy2016building,hirano2017formation,sugimura2020birth,Latif2022}), we estimate that the star formation process should finish within $t_{\rm SF}\lesssim t_{\rm ff}+t_{\rm acc}\lesssim 0.6\ \rm Myr$ after the termination criterion of our simulations is met. If PBHs can sink into the centre within a dynamical friction (DF) timescale smaller than $t_{\rm SF}$ and reach the star-forming disc, they may significantly affect the properties of stars or even prevent star formation by heating and disrupting the disc. 

For a BH of mass $m_{\rm BH}$ with initial apocentric distance $r$ and velocity $v_{\rm BH}$ with respect to the centre, the DF timescale can be estimated with Chandrasekhar's formula as \citep{bt2008}
\begin{align}
    \frac{t_{\rm DF}}{\mathrm{Myr}}\simeq \frac{340}{\ln\Lambda}\left(\frac{r}{\rm3\ pc}\right)^{2}\left(\frac{v}{\rm 10\ km\ s^{-1}}\right)\left(\frac{m_{\rm BH}}{100\ \rm M_{\odot}}\right)^{-1}\ , \label{e24}
\end{align}
where $\ln\Lambda\sim 10$ is the Coulomb logarithm. In our case with $m_{\rm BH}\sim 30-100\ \rm M_{\odot}$ and $v\sim 5-10\ \rm km\ s^{-1}$, we always have $r\gtrsim 1\ \rm pc$ as long as feedback from BHs (with $\epsilon_{r}>0.02$) is considered, such that we have\footnote{We also calculate the DF timescale with the updated $t_{\rm DF}$ formula in \citet{arca2015henize}, and obtain $t_{\rm DF}\sim 2-5\ \rm Myr$ for the DM core with $R\sim 1\ \rm pc$, $M\sim 10^{3}\ \rm M_{\odot}$ and a slope of density profile $\gamma_{\rm DM}\sim1$, assuming circular orbits of BHs.} $t_{\rm DF}\gtrsim 2-10\ {\rm Myr}\gg t_{\rm SF}$. This indicates that \textit{PBHs with $m_{\rm PBH}\sim 30-100\ \rm M_{\odot}$ and $f_{\rm PBH}\le 0.1$ are unlikely to sink into Pop~III star-forming discs and, therefore, not likely to significantly change the properties of Pop~III stars (at least at birth). }

Note that the above conclusion relies on the condition $r\gtrsim 1\ \rm pc$, which is inferred from the 19 simulations including BH feedback. It is possible that in rare cases not captured by our limited sample of minihaloes, PBHs can sink into the cloud centre efficiently with $r<1\ \rm pc$. The detailed statistics of BH orbits in Pop~III star-forming clouds can only be obtained with cosmological simulations of larger volumes, beyond the scope of this paper. Moreover, $t_{\rm DF}$ is actually comparable to the lifetimes of Pop~III stars and relaxation timescales of Pop~III star clusters \citep{liu2021binary}. Therefore, PBHs may still affect Pop~III stellar evolution and dynamics of Pop~III star clusters, which can have interesting implications on binary and multiple systems involving Pop~III stars/remnants and PBHs and their GW signals. We defer the investigation of this aspect to future studies. 

{ If a PBH does sink into the star-forming disc, the final outcome depends on how deep it reaches, since the accretion rate is sensitive to gas density. If the BH sits at the centre within $r\lesssim r_{\rm B}\sim 10^{-3}\ \rm pc$, where $r_{\rm B}$ is the Bondi radius for $m_{\rm BH}\sim 30\ \rm M_{\odot}$ in ionized gas with $\tilde{v}\sim 10\ \rm km\ s^{-1}$, the (spherically averaged) gas density at the Bondi radius is $n_{\rm H}\sim 10^{9}\ \rm cm^{-3}$. This leads to hyper-Eddington accretion ($\sim 0.01\ \rm M_{\odot}\ yr^{-1}$) unimpeded by radiation feedback at least initially \citep{inayoshi2016hyper,Takeo2018}. Under such a high density, the BH mass will grow by a factor of 10 within $\sim 0.01\ \rm Myr$. The production rate of ionizing photons from the BH follows $Q_{\rm BH}\sim 2\times 10^{50}\ {\rm s^{-1}}(m_{\rm BH}/30\ \rm M_{\odot})^{2}$, which is comparable or even higher than that from a typical group/cluster of Pop~III stars with a total stellar mass of $\sim 100-1000\ \rm M_{\odot}$. We expect the radiation feedback from the BH to rapidly evaporate the (low-density) gas outside the disc such that the BH can grow at most to $\sim 1000\ \rm M_{\odot}$ by devouring all dense gas in the disc. However, if the BH never sinks very deeply within $t_{\rm acc}\sim 0.01-0.1\ \rm Myr$, the BH growth and feedback can be much weaker. For instance, at $r\sim 0.01\ \rm pc$ with $n_{\rm H}\sim 10^{7}\ \rm cm^{-3}$, the BH mass will increase by $\sim 30$\% percent in 0.1~Myr, and the radiation feedback ($Q_{\rm BH}\sim 10^{49}\ \rm s^{-1}$) is similar to that from a Pop~III star of the same mass ($\sim 30\ \rm M_{\odot}$). In this case, the BH will have minor impact on disc evolution, since the feedback is not strong enough to stop inflows and the \HII\ region can even be trapped within the disc (see e.g. \citealt{sugimura2020birth,Jaura2022} for detailed simulations of ionization feedback in Pop~III star forming clouds). }

%The Lyman-Werner (LW) photons ($h\nu\sim 11.2-13.6\ \rm eV$) from BH accretion may also play a role, as they can dissociate $\rm H_{2}$ and $\rm HD$, and, therefore, reduce cooling and even suppress fragmentation of star-forming discs. 
{Although the LW background produced by PBH accretion is unimportant for first star formation in most cases, as discussed in Sec.~\ref{s4.1}, the local LW radiation (from PBHs within the same halo) may still play a role.
Since LW feedback is not included in our simulations, we here estimate their effects by post-processing. %effects of LW photons are ignored. Here we assume that they are relatively unimportant compared with the ionizing photons and focus on their late-stage effects. 
Again, based on the BH spectrum model in \citet[see Sec.~\ref{s2.3.3}]{Takhistov2022} and assuming isothermal distributions of gas and BHs, we find that on average each BH contributes $J_{\rm 21}\sim 0.6 (m_{\rm PBH}/{33\ \rm M_{\odot}})^{1.9} $ to the intensity of LW radiation (in units of $10^{-21}\ \rm erg\ s^{-1}\ cm^{-2}\ Hz^{-1}\ sr^{-2}$) at the centre of a halo with $M_{\rm h}\sim 10^{6}\ \rm M_{\odot}$ at $z\sim 20$, ignoring self-shielding (i.e. treating it as an `external' background). }%If we include self-shielding with an $\rm H_{2}$ abundance of $x_{\rm H_{2}}\sim 10^{-3}$ based on the method in \citet{wolcott2011photodissociation}, we obtain $J_{\rm 21}\sim 4\times 10^{-5} (m_{\rm PBH}/{33\ \rm M_{\odot}})^{2}$. 
For the fiducial PBH model ($m_{\rm PBH}= 33\ \rm M_{\odot}$, $f_{\rm PBH}=10^{-3}$) consistent with current observational constraints, star-forming minihaloes typically contain $\sim 10-100$ PBHs. We thus expect that in most minihaloes the total LW intensity from PBHs cannot reach the critical value $J_{21}\sim 10^{3}$ \citep{Sugimura2014}, required to sufficiently suppress $\rm H_{2}$ cooling and fragmentation, leading to the formation of massive ($\sim 10^{4}-10^{6}\ \rm M_{\odot}$) direct-collapse BHs (DCBHs, reviewed by e.g. \citealt{Latif2019, Haemmerle2020}). Note that the power of LW radiation from BH accretion is highly sensitive to gas density and the above values can only be achieved when the cloud has collapsed to reach {$n_{\rm H}\gtrsim 10^{3}\ \rm cm^{-3}$} around BHs in the centre. While for a typical average gas density $n_{\rm H}\sim 1\ \rm cm^{-3}$ in minihaloes, we have {$J_{21}\sim 0.1(r/{\rm pc})^{-2}$} (without self-shielding) for $m_{\rm PBH}=33\ \rm M_{\odot}$, and the BH sources will be even fainter in the IGM\footnote{Accreting stellar-mass BHs are much weaker sources of LW photons compared with stars (see e.g. \citealt{jeon2014radiative}).}. {Therefore, we conclude that LW feedback from stellar-mass PBHs cannot change the standard picture of Pop~III star formation in (molecular-cooling) minihaloes, at least for $f_{\rm PBH}\lesssim 0.01$}, although it may further delay star formation %to more massive haloes 
in addition to the effect of photoionization heating (e.g. \citealt{safranek2012star,Schauer2021}).

{However, in more massive atomic-cooling haloes ($M_{\rm h}\gtrsim 10^{8}\ \rm M_{\odot}$) containing $\gtrsim 1000$ PBHs, the LW radiation from BHs can be strong enough to suppress fragmentation and form DCBHs. Actually, the LW intensity from PBHs at the halo centre as a function of halo mass follows $J_{21}\sim 5\times 10^{4}f_{\rm PBH}(M_{\rm h}/10^{7}\ \rm M_{\odot})^{0.56}(m_{\rm PBH}/{33\ \rm M_{\odot}})^{0.9}$ within a factor of 2 errors for $m_{\rm PBH}\sim 10-100\ \rm M_{\odot}$ and $M_{\rm h}\sim 10^{6}-10^{10}\ \rm M_{\odot}$ at $z\sim 10-30$. This implies that haloes with $M_{\rm h}\gtrsim 3.3\times 10^{7}\left( f_{\rm PBH}/0.01\right)^{-1.8}(m_{\rm PBH}/{33\ \rm M_{\odot}})^{-1.6}\ \rm M_{\odot}$ will meet the criterion $J_{21}\gtrsim 10^{3}$ for DCBH formation.}
In this scenario, DCBHs can be more common than in the CDM case where they only form in rare sites with strong external radiation fields or high inflow rates (e.g. \citealt{Visbal2014,wise2019formation}). We plan to apply our numerical framework to more massive haloes and include the LW feedback from PBHs in future simulations. 

\section{Summary and Conclusions}
\label{s5}
We use cosmological hydrodynamic zoom-in simulations to study the effects of stellar-mass PBHs on first star formation, which for the first time self-consistently take into account the enhancement of initial density perturbations (by the isocurvature mode introduced by PBHs) and heating of gas by the accretion feedback from PBHs. The two effects compete with each other, as the former accelerates structure formation, while the latter increases the halo mass threshold $M_{\rm mol}$ above which stars can form by efficient cooling. 
%The initial adiabatic perturbations are enhanced to accelerate structure formation, and the simulations are stopped when the maximum gas density reaches $n_{\rm H}=10^{5}\ \rm cm^{-3}$, denoted by the collapse time (redshift) $t_{\rm col}$ ($z_{\rm col}$). 
We also build semi-analytical models to calculate the halo mass functions and $M_{\rm mol}$ under the influence of PBHs, which well reproduce the trends seen in our simulations (for over-dense regions) and are used to evaluate the effects of PBHs on Pop~III star formation at larger scales. Focusing on PBH models with a monochromatic mass function peaked at $m_{\rm PBH}=33\ \rm M_{\odot}$ and PBH fractions in DM $f_{\rm PBH}=10^{-4},\ 10^{-3}$, 0.01 and 0.1, we infer the following features of first star formation in the presence of PBHs at the cloud, halo and cosmic scales.
\begin{itemize}
    \item At the end of a simulation (denoted by $z_{\rm col}$ and $t_{\rm col}$), a dense ($n_{\rm H}\gtrsim 10^{4}\ \rm cm^{-3}$) cold ($T\lesssim 10^{3}\ \rm K$) gas clump of a few $10^{3}\ \rm M_{\odot}$ has formed at the centre ($r\lesssim 1\ \rm pc$) of the target halo by run-away collapse under efficient molecular cooling, regardless of PBH parameters. The clump is expected to form stars in $t_{\rm SF}\lesssim 0.6\ \rm Myr$, and the properties of gas within it are very similar in all PBH models considered and the CDM case. No PBHs reach the central parsec in the 19 simulations including BH feedback, and we estimate the dynamical friction timescale for the nearest BH to sink into the centre as $t_{\rm DF}\sim 2-10\ {\rm Myr}\gtrsim t_{\rm SF}$, such that PBHs are unlikely to interact with star-forming discs and affect protostellar evolution. This indicates that \textit{the standard picture of Pop~III star formation is not changed by PBHs at the scales of star-forming clouds}. %However, 
    
    \item In the simulated over-dense regions, the collapse time generally increases with $f_{\rm PBH}$, such that with respect to $\rm \Lambda CDM$, star formation is accelerated by up to $\sim 30\ \rm Myr$ with PBHs of $f_{\rm PBH}\sim 10^{-4}-10^{-3}$, but delayed by $\sim 20-90\ \rm Myr$ for $f_{\rm PBH}\sim 0.01-0.1$. This implies that the effect of heating is enhanced more rapidly with increasing $f_{\rm PBH}$ than the effect of PBH perturbations. It is predicted by our semi-analytical model and generally confirmed in simulations that the mass threshold of efficient cooling increases with $f_{\rm PBH}$, up to $\sim 5$ (2) times the CDM value {at $z\lesssim 40$} for $f_{\rm PBH}=0.1\ (10^{-3})$. The internal structure of haloes are also affected by PBHs, which tend to reduce the density of DM at the centre and produce shallower density profiles (when the halo contains multiple BHs), because it is more difficult to destroy substructures of DM around PBHs. At the outer part of the halo ($r\sim 1\ {\rm pc}-R_{\rm vir}$), the gas density profile also becomes shallower with increasing $f_{\rm PBH}$ due to the heating from PBHs and/or the structures around PBHs that may disrupt the collapse process.  
    
    \item {For PBH models with $f_{\rm PBH}\sim 10^{-4}-0.01$, our semi-analytical models predict that the cosmic collapsed mass fraction $f_{\rm col}$ in typical haloes hosting Pop~III stars in the mass range $M_{\rm h}\sim M_{\rm mol}-10^{8}\ \rm M_{\odot}$ is similar (within a factor of 2) to that of the CDM case. $f_{\rm col}$ is insensitive to $f_{\rm PBH}$ for $f_{\rm PBH}\sim 10^{-4}-0.01$. When external effects (e.g. LW background and streaming motion between gas and DM) are considered that further increase the mass threshold for star formation, $f_{\rm col}$ increases with $f_{\rm PBH}$ but remains below 2 times the CDM value at $z\lesssim 30$. %This is not consistent with the trend seen in our zoom-in simulations targeting over-dense regions for $f_{\rm PBH}=10^{-4}$ and $10^{-3}$. The reason is 
In the extreme case with $f_{\rm PBH}=0.1$, $f_{\rm col}$ is significantly higher than in the CDM case, by a factor of $\gtrsim 10$ at $z\gtrsim 30$.} Considering the different trends seen in our simulations for over-dense regions, we find that structure formation is accelerated more in regions with higher initial overdensities (from the adiabatic mode). %and stronger correlation between the adiabatic and isocurvature modes (see Sec.~\ref{s2.2} and \ref{s4.1}).
    %In the case of $f_{\rm PBH}\sim 10^{-4}-0.01$, star formation is accelerated by PBHs in the simulated over-dense regions, but slightly delayed on cosmic average as predicted by the semi-analytical model.
\end{itemize}
 
Several caveats in our models may render aspects of our aforementioned results uncertain:
\begin{itemize}
    \item Our implementation of the perturbations from PBHs in the initial matter field assumes a truncation scale $\sim 2d_{\rm PBH}$ for the correlation/mixing between the adiabatic and isocurvature modes (see Sec.~\ref{s2.2} and \ref{s4.1} for details), where $d_{\rm PBH}$ is the average separation between PBHs. This correlation/mixing arises from the fact that PBHs follow the large-scale adiabatic mode to fall into larger structures and meanwhile induce/disrupt DM structures around themselves at small scales, which is particularly important for the intermediate values of $f_{\rm PBH}$ between the `seed' and `Poisson' limits \citep{Carr2018,Inman2019}. 
    Structure formation will be enhanced more with stronger mode mixing under a larger truncation scale. As shown in Appendix~\ref{aa}, changing the scale between $d_{\rm PBH}$ and $2d_{\rm PBH}$ leads to variations in the collapse time of $\Delta t_{\rm col}\sim 15\ \rm Myr$, corresponding to $\Delta z_{\rm col}\sim 7$ (1.5) at $z\sim 30$ (20). The semi-analytical model is also sensitive to the treatment of mode mixing. For instance, when mode mixing is ignored, the collapsed mass fraction of Pop~III hosts can be reduced by up to a factor of a few at $z\lesssim 30$ (see Sec.~\ref{s4.2}). 
    
    \item In our simulations, we only consider thermal feedback from BH accretion (i.e. heating of gas by ionizing photons) with a sub-grid model calibrated to more detailed calculations of BH accretion disc spectra and radiative transfer based on \citet[see Sec.~\ref{s2.3.3}]{Takhistov2022}. In this model, the uncertainty in the radiation-thermal coupling efficiency $\epsilon_{r}$ (=0.22 by default) can be up to one order of magnitude. However, as shown in Appendix~\ref{ac}, varying $\epsilon_{r}$ between 0.02 and 1 only has minor effects on our results. We also ignored the LW feedback from BHs, which, as discussed at the end of Sec.~\ref{s4.3}, will not change the main findings of this work. 
    
    %\item In the semi-analytical calculation of the collapsed mass fraction 
\end{itemize}

In general, we find that the effects of stellar-mass PBHs ($m_{\rm PBH}\sim 10-100\ \rm M_{\odot}$) on Pop~III star formation in molecular-cooling minihaloes are small at the scales of star-forming clouds and also for the cosmic star formation history at $z\gtrsim 10$, when PBHs make up $f_{\rm PBH}\lesssim 0.01$ of DM, allowed by existing observational constraints (see e.g. \citealp{Ali-Haimoud2017,Poulin2017,Hektor2018,Carr2021,Hutsi2021}). In particular, when the feedback of BH accretion that increases the mass threshold of efficient cooling is considered, early star formation may not be significantly enhanced by the accelerated structure formation with PBHs. %as assumed in previous studies. 

Nevertheless, PBHs tend to shift Pop~III star formation to more massive haloes whose abundance is also increased. This may change the spatial distribution of Pop~III hosts and affect the intensity map of 21-cm signal and cosmic infrared background (see e.g. \citealt{Kashlinsky2016,Gong2017,Cappelluti2022}). { The X-rays produced by PBH accretion in collapsed structures can also contribute significantly to the CXB and alter the thermal and ionization history of the IGM.} Besides, although PBHs have little influence on the star formation process, they can sink into newly born Pop~III star clusters within a dynamical friction timescale of $t_{\rm DF}\sim 2-10\ \rm Myr$, comparable to the stellar lifetimes and relaxation timescales of these systems. In this way, PBHs can affect Pop~III stellar evolution and dynamics of Pop~III star clusters, which may lead to hybrid compact object mergers between PBHs and Pop~III remnants. Such GW sources will be detectable and distinguishable by 3rd-generation GW detectors \citep{Franciolini2022}. PBHs with lower masses and higher abundances than considered in this work ($m_{\rm PBH}\sim 10^{22}-10^{26}\ \rm g$) can also concentrate inside the first stars and swallow them \citep{Bambi2009}. {Finally, although relatively unimportant in minihaloes ($M_{\rm h}\sim 10^{5}-10^{6}\ \rm M_{\odot}$), the LW feedback from PBHs may be able to sufficiently dissociate $\rm H_{2}$ and trigger DCBH formation in more massive %($M_{\rm h}\gtrsim 10^{8}\ \rm M_{\odot}$) 
atomic-cooling haloes with $M_{\rm h}\gtrsim 3.3\times 10^{7}\left( f_{\rm PBH}/0.01\right)^{-1.8}(m_{\rm PBH}/{33\ \rm M_{\odot}})^{-1.6}\ \rm M_{\odot}$ (see Sec.~\ref{s4.3}).} This internal feedback mechanism from PBHs may be more efficient than the channels of DCBH formation in $\rm \Lambda CDM$ that require very special conditions (e.g. \citealt{Visbal2014,wise2019formation}). %Such PBH-triggered DCBHs may 

PBHs have long been studied as a DM candidate or a component coexisting with particle DM. Even if not making up the entire dark sector, PBHs can play important roles in early structure/star/galaxy formation, and leave their imprints in a variety of observables at Cosmic Dawn ($z\sim 5-30$), such as GWs from binary BH mergers, radiation backgrounds produced/regulated by accretion around PBHs and high-$z$ quasars seeded by PBHs or formed under the feedback of PBHs. In the next decades, a large volume of observational data from Cosmic Dawn by multi-band space and ground based telescopes (e.g. \href{https://www.jwst.nasa.gov/}{JWST}, \href{https://www.euclid-ec.org/}{Euclid} and \href{https://www.skatelescope.org/}{SKA}), as well as 3rd-generation GW detectors (e.g. \href{http://www.et-gw.eu/}{ET}, \href{https://iopscience.iop.org/article/10.1088/1742-6596/840/1/012010/meta}{DECIGO} and \href{https://lisa.nasa.gov/}{LISA}), promises to shed light on the existence and properties of PBHs. It is therefore timely to include PBHs in state-of-the-art cosmological hydrodynamic simulations of early star and galaxy formation. This work focusing on Pop~III stars in minihaloes is an exploratory step in this direction. Future studies will consider a broader range of haloes and PBH models with improved modelling of initial conditions and BH feedback.

\section*{Acknowledgements}
The authors acknowledge the Texas Advanced Computing Center (TACC) for providing HPC resources under XSEDE allocation TG-AST120024.
%This work was supported by National Science Foundation (NSF) grant AST-1413501. %The authors acknowledge the Texas Advanced Computing Center (TACC) for providing HPC resources under XSEDE allocation TG-AST120024.
%Support for this work was provided by NASA through the NASA Hubble Fellowship grant HST-HF2-51418.001-A awarded by the Space Telescope Science Institute, which is operated by the Association of Universities for Research in  Astronomy, Inc., for NASA, under contract NAS5-26555. 

\section*{Data availability}
The data and codes underlying this article will be shared on reasonable request to the corresponding authors.

%\clearpage
%\newpage
\bibliographystyle{mnras}
\bibliography{ref} 

%\clearpage
%\newpage

\appendix

\section{Dependence on initial conditions}
\label{aa}

\begin{figure}
    \centering
    \includegraphics[width=\columnwidth]{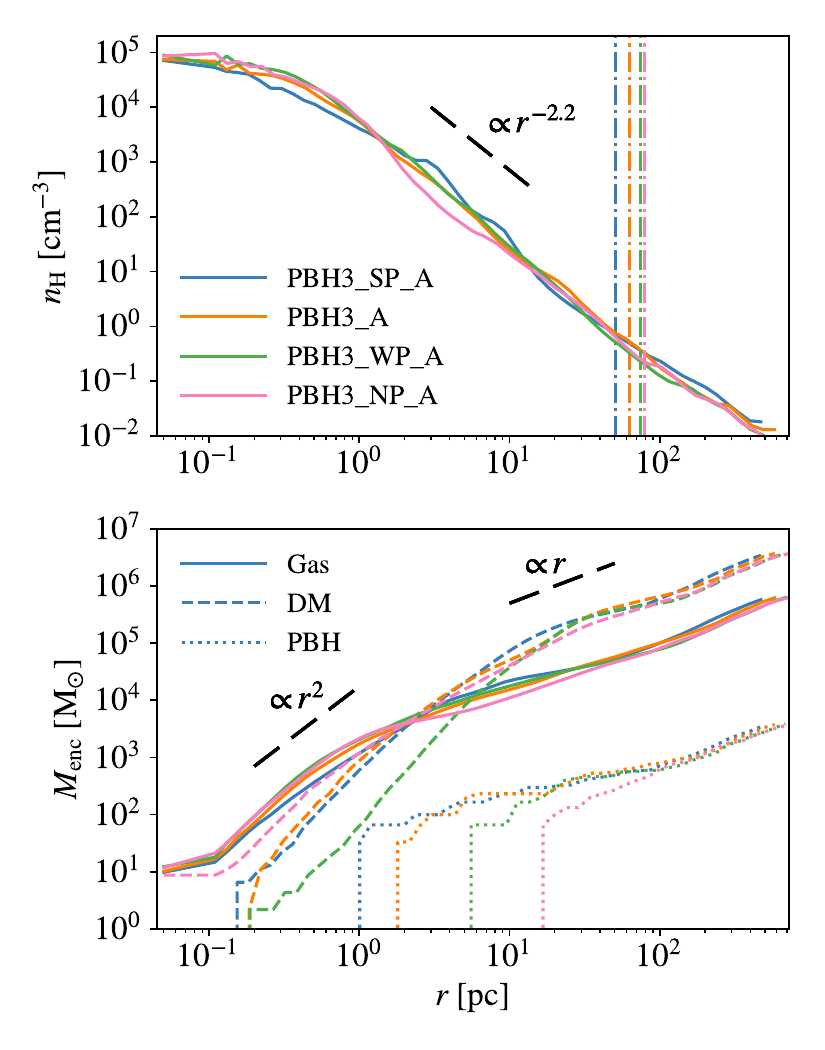}
    \vspace{-20pt}
    \caption{Internal structure of the collapsing cloud and host halo in Case A, for the fiducial PBH model ($m_{\rm PBH}=33\ \rm M_{\odot}$, $f_{\rm PBH}=10^{-3}$) with strong (blue, \texttt{PBH3\_SP\_A}), default (orange, \texttt{PBH3\_A}), weak (green, \texttt{PBH3\_WP\_A}) and no (pink, \texttt{PBH3\_SP\_A}) perturbations from PBHs on DM. \textit{Top}: hydrogen number density profile, where the virial radii of host haloes are shown with vertical dashed-dotted lines. \textit{Bottom}: enclosed mass profiles for gas (solid), DM (dashed) and PBHs (dotted). }
    \label{dpro_ic}
\end{figure}

\begin{figure}
    \centering
    \includegraphics[width=1\columnwidth]{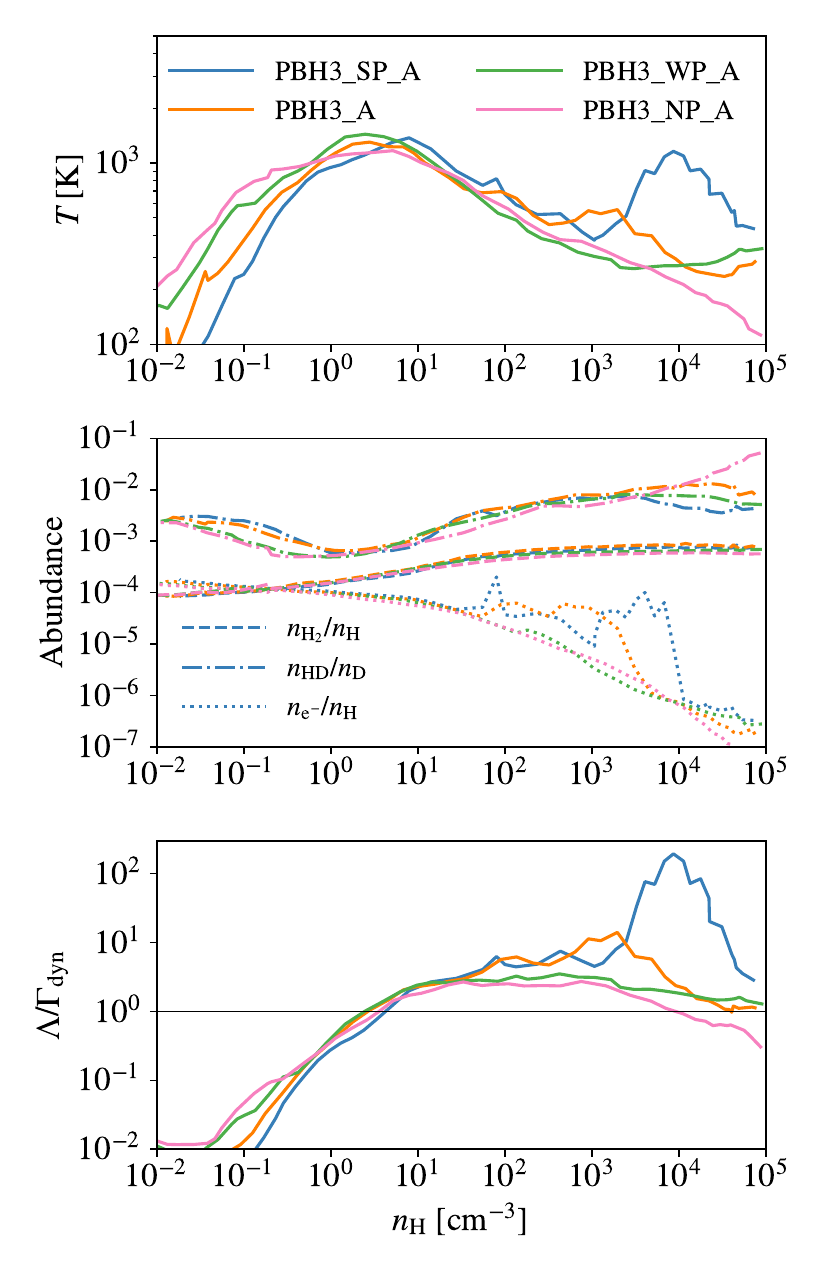}
    \vspace{-20pt}
    \caption{Phase diagrams of the collapsing cloud in Case A, for the fiducial PBH model ($m_{\rm PBH}=33\ \rm M_{\odot}$, $f_{\rm PBH}=10^{-3}$) with strong (blue, \texttt{PBH3\_SP\_A}), default (orange, \texttt{PBH3\_A}), weak (green, \texttt{PBH3\_WP\_A}) and no (pink, \texttt{PBH3\_SP\_A}) perturbations from PBHs on DM. \textit{Top}: temperature-density diagram. \textit{Middle}: abundances of $\rm H_{2}$ (dashed), $\rm HD$ (dashed-dotted) and $\rm e^{-}$ (dotted) as functions of density. \textit{Bottom}: ratio of cooling and dynamical heating rates as a function of density.}
    \label{cpro_ic}
\end{figure}

As mentioned in Sec.~\ref{s2.2}, in principle, the small-scale isocurvature perturbations on DM around PBHs grow together with the large-scale adiabatic perturbations within PBHs themselves, such that the isocurvature and adiabatic modes are mixed/correlated on intermediate scales for the PBH-induced perturbations in DM. Without an accurate theory for this mode mixing, we use the Zel'dovich approximation to generate the perturbations of DM particles by PBHs in the initial conditions, based on the positions of PBHs at $z_{\rm ini}$. By doing so, we actually assume that individual PBHs moving along the large-scale flows `carry' the induced DM structures with them all the way to $z_{\rm ini}$, and therefore enhance clustering of DM at scales larger than those dominated by the Poisson noise. Then the question is at what scale this enhancement acts, which reflects the strength of mode mixing. In our default implementation, we truncate the effects of mode mixing at the scale of $2d_{\rm PBH}$, by considering at most the 64 nearest PBH particles within $2d_{\rm PBH}$ around each DM particle for the acceleration field from PBHs (Equ.~\ref{e5}). Here, for the fiducial PBH model ($m_{\rm PBH}=33\ \rm M_{\odot}$, $f_{\rm PBH}=10^{-3}$), we further explore two cases with enhanced and suppressed mode mixing with respect to the default: In \texttt{PBH3\_SP} (strong perturbations), we consider \textit{all} PBHs for the acceleration field from PBHs (Equ.~\ref{e5}), while in \texttt{PBH3\_WP} (weak perturbations), each DM particle is only affected by the nearest PBH. Combining these two situations with the default case (\texttt{PBH3}) and the extreme model without any PBH perturbations (\texttt{PBH3\_NP}), we obtain a good coverage of the initial conditions regulated by PBHs.

Not surprisingly, collapse happens earlier with increasing strength of PBH perturbations: In Case A (B), $z_{\rm col}=28.2$, 30.4, 37.6 and 42.1 (21.6, 22.1, 23.6 and 30.6) for \texttt{PBH3\_NP}, \texttt{PBH3\_WP}, \texttt{PBH3} and \texttt{PBH3\_SP}, respectively (see Table~\ref{t1}), indicating that our results are sensitive to the initial conditions. With weak perturbations (\texttt{PBH3\_WP}) the timing of collapse is very close to that in the reference CDM run, i.e. $z_{\rm col}=30.3$ (22.3) in Case A (B), although the host halo mass is higher by up to a factor of $\sim 2$ with PBHs. This shows that first star formation may not necessarily be accelerated in our fiducial PBH model even in the simulated over-dense regions. However, if more massive halos form more stars, Pop~III star formation will still be enhanced (in over-dense regions). Excluding the extreme models \texttt{PBH3\_NP} and \texttt{PBH3\_SP}, %which are likely unphysical, 
we estimate that the error in the timing of collapse from the uncertainties in initial conditions is $\Delta t_{\rm col}\sim 15$~Myr, corresponding to $\Delta z_{\rm col}\sim 7$ (1.5) for Case A (B). %The host halo mass at the moment of collapse has little variation with the perturbation strength in the 4 cases considered here ($M_{\rm h}\sim 3-5\times 10^{5}\ \rm M_{\odot}$), in which we do not find a clear trend. 

Using Case A as an example, the density profiles and phase diagrams are shown in Fig.~\ref{dpro_ic} and Fig.~\ref{cpro_ic}. Case B results are similar. It turns out that the gas density profile at $z_{\rm col}$ is insensitive to initial conditions. Nevertheless, in the case of weak perturbations (\texttt{PBH3\_WP\_A}), DM density is reduced by a factor of $\sim 10$ in the central region ($r\lesssim 0.1 R_{\rm vir}$) compared with the other models. This is consistent with the trend seen in Sec.~\ref{s3.2} and \citet{Inman2019} that in haloes containing multiple PBHs, tightly-bound substructures around individual PBHs can hinder the concentration of DM at the centre. In \texttt{PBH3\_WP\_A}, haloes around individual PBHs will be more compact than those in the other models by construction. We also find that the distribution of PBH is more concentrated with stronger perturbations, which leads to stronger heating in dense ($n_{\rm H}\gtrsim 10^{3}\ \rm cm^{-3}$) gas at the centre, as shown in the temperature and cooling rate profiles (see Fig.~\ref{cpro_ic}). This outcome may be caused by the same mechanism that weaker mode mixing reduces the central density of DM and/or the stochastic nature of the distribution of BHs in the central region ($r\lesssim 10\ \rm pc$). The $\rm H_{2}$ abundance remains almost the same in the 4 models considered here, implying that the overall thermodynamics and chemistry of gas is insensitive to initial conditions.

\section{Effects of baryon-DM streaming motion}
\label{ab}

\begin{figure}
    \centering
    \includegraphics[width=1\columnwidth]{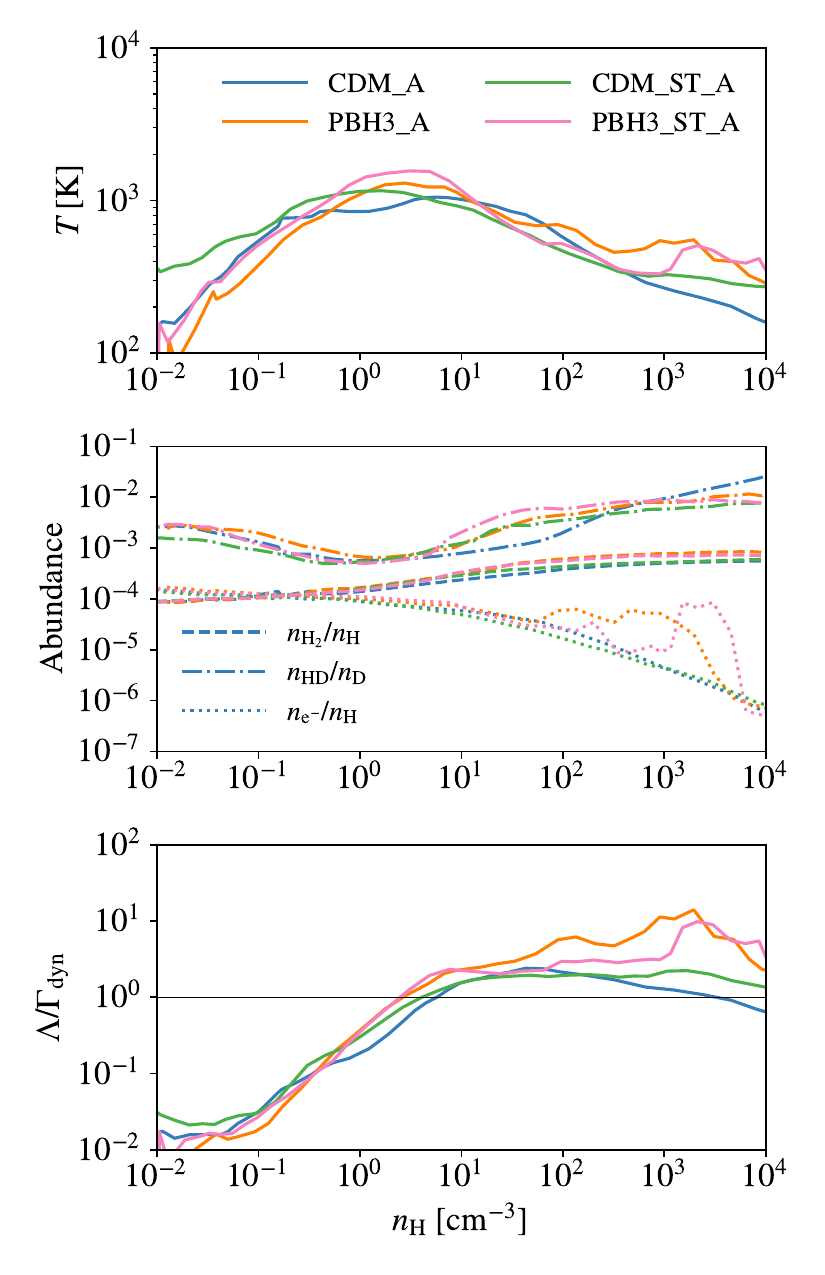}
    \vspace{-20pt}
    \caption{Same as Fig.~\ref{cpro_ic} but for the reference CDM simulation (blue, \texttt{CDM\_A}) and fiducial PBH ($m_{\rm PBH}=33\ \rm M_{\odot}$, $f_{\rm PBH}=10^{-3}$) simulation (orange, \texttt{PBH3\_A}) without streaming motion between DM and gas ($v_{\rm b\chi}=0$), and their counterparts with a typical streaming velocity $v_{\rm b\chi}=0.8\sigma_{\rm b\chi}$: \texttt{CDM\_ST\_A} (green) and \texttt{PBH3\_ST\_A} (pink). }
    \label{cpro_st}
\end{figure}

\begin{comment}
\begin{figure*}
    \centering
    \includegraphics[width=1.9\columnwidth]{cswb_01_ST_A.png}
    \vspace{-20pt}
    \caption{}
    \label{cswb_st}
\end{figure*}

\begin{figure}
    \centering
    \includegraphics[width=\columnwidth]{denspro_ST_A.pdf}
    \vspace{-20pt}
    \caption{}
    \label{dpro_st}
\end{figure}
\end{comment}

For Case A, we also investigate the effects of streaming motion between DM and gas with two simulations, \texttt{CDM\_ST\_A} and \texttt{PBH3\_ST\_A}, for CDM and the fiducial PBH model ($m_{\rm PBH}=33\ \rm M_{\odot}$, $f_{\rm PBH}=10^{-3}$) under a typical streaming velocity $v_{\rm b\chi}=0.8\sigma_{\rm b\chi}$ (at recombination), around which the contribution to overall structure formation is largest \citep{Anna2019}, given $\sigma_{\rm b\chi}=30\ \rm km\ s^{-1}$ as the root-mean-square streaming velocity. Similar to previous studies in the $\rm\Lambda CDM$ cosmology (see e.g., \citealt{maio2011impact,greif2011delay,stacy2011effect,naoz12,naoz13,fialkov2012,Hirano2018formation,Anna2019,Park2020}), collapse is delayed by streaming motion in our simulations, from $z_{\rm col}=30.3$ (27.6) to $z_{\rm col}=26.9$ (34.8) by $\sim 20$ (10)~Myr in the CDM (PBH) model, and the host halo mass at $z_{\rm col}$ is higher by a factor of $\sim 2.2$ (1.3). As shown in Fig.~\ref{cpro_st}, \ref{TIGM_z} and Table~\ref{t1}, when streaming motion is considered for CDM, temperatures are higher by a factor of $\sim 2$ at $n_{\rm H}\lesssim 0.1\ \rm cm^{-3}$ and $n_{\rm H}\gtrsim 10^{3}\ \rm cm^{-3}$, likely caused by stronger virialization shocks during more violent collapse in a more massive halo. %The IGM is heated to $T_{\rm IGM}\sim 100\ \rm K$ by virialization shocks, much higher than the $v_{\rm c\chi}=0$ case ($T_{\rm IGM}\sim 50\ \rm K$). 
While for the PBH model, the temperature-density phase diagram is almost identical with and without streaming motion, and the IGM temperature is only slightly increased (by $\sim 10\%$, see Table~\ref{t1} and Fig.~\ref{TIGM_z}). 
{Clearly, the effect of streaming motion is weaker with PBHs, implying that the perturbations from PBHs accelerate the decoupling of gas from the large-scale flow (relative to the underlying DM structures). This trend is consistent with the prediction by \citet{Kashlinsky2021} that the equalization of DM and baryonic velocity components is more efficient with the `granulation' in the density field caused by PBHs.}
%, in which case the collapse happens before virialization shocks significantly heat the IGM. %despite the fact that collapse happens earlier in the PBH model when the large-scale relative velocity between DM and gas is higher. 
%In general, considering a non-zero $v_{\rm b\chi}$ does not change the main picture that PBHs accelerate structure formation and increase the mass threshold of efficient cooling for cloud collapse.

%the isocurvature term $\delta_{\rm iso}(a)$ in the DM density field (Equ.~\ref{e1}) induced by PBHs is unclear. 

\section{Dependence on BH feedback strength}
\label{ac}
As mentioned in Sec.~\ref{s2.3.3}, the strength of BH feedback is characterised by the thermal-radiation coupling parameter $\epsilon_{r}$ in our sub-grid model, which is uncertain within a factor of 10 compared with more complex models of BH spectra and radiative transfer (\citealt{Takhistov2022}). To better evaluate the effects of the uncertainty in BH feedback, we consider two cases with weak ($\epsilon_{r}=0.02$, \texttt{PBH3\_WF\_A}) and strong feedback ($\epsilon_{r}=1$, \texttt{PBH3\_SF\_A}) in addition to the fiducial case ($\epsilon_{r}=0.22$, \texttt{PBH3\_A}) and the extreme case without any feedback ($\epsilon_{r}=0$, \texttt{PBH3\_NF\_A}), for the fiducial PBH model ($m_{\rm PBH}=33\ \rm M_{\odot}$, $f_{\rm PBH}=10^{-3}$) in the Case~A zoom-in region. Opposite to the trend with PBH perturbation strength (see Appendix~\ref{aa}), collapse is delayed by stronger feedback. The collapse redshift (time) is $z_{\rm col}=42.7$, 39.8, 37.6 and 37.0 ($t_{\rm col}=58.9$, 65.5, 71.0 and 72.6~Myr) in \texttt{PBH3\_NF\_A}, \texttt{PBH3\_WF\_A}, \texttt{PBH3\_A} and \texttt{PBH3\_SF\_A}, respectively (see Table~\ref{t1}). Note that even with the strongest feedback (\texttt{PBH3\_SF\_A}), collapse is still accelerated by PBHs with respect to the CDM case ($z_{\rm col}=30.3$). Excluding the unphysical case \texttt{PBH3\_NF\_A}, we have $\Delta t_{\rm col}\sim 7\ \rm Myr$ as the error in collapse time caused by the uncertainties in BH feedback strength, smaller than that introduced by uncertain initial conditions ($\Delta t_{\rm col}\sim 15\ \rm Myr$). This implies that our results are more sensitive to PBH perturbations in the initial conditions than BH feedback. Furthermore, we find that the density, chemical and thermal structures of the host halo is almost the same in the 3 models with $\epsilon_{r}\sim 0.02-1$. 

\begin{comment}
\begin{figure*}
    \centering
    \includegraphics[width=1.9\columnwidth]{cswb_01_FDBK_A.png}
    \vspace{-20pt}
    \caption{}
    \label{cswb_fdbk}
\end{figure*}

\begin{figure}
    \centering
    \includegraphics[width=\columnwidth]{denspro_FDBK_A.pdf}
    \vspace{-20pt}
    \caption{}
    \label{dpro_fdbk}
\end{figure}

\begin{figure}
    \centering
    \includegraphics[width=1\columnwidth]{coolpro_FDBK_A.pdf}
    \vspace{-20pt}
    \caption{}
    \label{cpro_fdbk}
\end{figure}
\end{comment}

\bsp

\label{lastpage}
\end{document}